\documentclass[prx,twocolumn,preprintnumbers,superscriptaddress,longbibliography]{revtex4-1}

\usepackage{graphicx} 
\usepackage{dcolumn}
\usepackage{bm}
\usepackage{times}

\usepackage{braket}
\usepackage{natbib}
\usepackage{floatrow}
\usepackage{comment}
\usepackage{color}
\usepackage{soul,xcolor}

\usepackage[normalem]{ulem}
\setstcolor{red}

\usepackage{amsthm} 

\definecolor{black}{rgb}{0,0,0}
\definecolor{blue}{rgb}{0,0,1}
\definecolor{green}{rgb}{0,1,0}

\definecolor{brown}{rgb}{0.4,0.2,0}
\definecolor{darkgreen}{rgb}{0,0.7,0}
\definecolor{darkblue}{rgb}{0.0,0.0,0.5}
\definecolor{red}{rgb}{1,0,0}
\definecolor{deepmagenta}{rgb}{0.8, 0.0, 0.8}

\usepackage[colorlinks,
citecolor = blue,
linkcolor = blue,
urlcolor  = blue,
anchorcolor = blue,
breaklinks = true
]{hyperref}

\usepackage{hyperref}

\usepackage{mathdots}
\usepackage{amssymb}
\usepackage{amsmath}

\begin{document}
%\begin{large}		
	
	\title{Correlated Metals, Metamagnetism, and Orbital Nematic Order in moir\'e Materials with Neural Quantum States}
	\date{\today }
    
\author{Wei Zhang}
\affiliation{
Institute for Theoretical Physics,
ETH Z\"urich, 8093 Z\"urich, Switzerland
}

\author{Ataç İmamoğlu}
\affiliation{
Institute for Quantum Electronics,
ETH Z\"urich, 8093 Z\"urich, Switzerland
}

\author{Tao Shi}
\email{tshi@itp.ac.cn}
\affiliation{
Institute of Theoretical Physics, Chinese Academy of Sciences, Beijing 100190, China
}
\affiliation{
School of Physical Sciences, University of Chinese Academy of Sciences, Beijing 100049, China
}

\author{Eugene A. Demler}
\email{demlere@ethz.ch}
\affiliation{
Institute for Theoretical Physics,
ETH Z\"urich, 8093 Z\"urich, Switzerland
}

\author{Ivan Morera}
\email{imorera@ethz.ch}
\affiliation{
Institute for Theoretical Physics,
ETH Z\"urich, 8093 Z\"urich, Switzerland
}
\affiliation{Institute for Quantum Electronics, ETH Zürich, Zürich, Switzerland}
\affiliation{\'Ecole Polytechnique F\'ed\'erale de Lausanne (EPFL), 1015 Lausanne, Switzerland}

\begin{abstract}
The interplay between kinetic energy, strong interactions, and external periodic potentials gives rise to a rich landscape of correlated quantum phases.  
In this work, we investigate this interplay in a continuum model of two-dimensional electrons interacting via long-range Coulomb forces and subject to a triangular periodic potential at filling factor $\nu=2$. Using neural quantum states based on message-passing neural networks optimized with variational Monte Carlo, we determine the ground-state phase diagram as a function of periodic potential strength and Zeeman field. 
Although the noninteracting system is expected to form a conventional band insulator, we show that strong correlations can qualitatively reshape the insulating state and, in part of the phase diagram, suppress the band insulator in favor of a correlated metallic phase, with a first-order transition separating the two regimes.
Moreover, the metal--insulator transition exhibits a strong dependence on spin polarization: the unpolarized system becomes insulating at substantially weaker periodic potentials through the formation of an $s$-wave molecular insulator, where opposite-spin electrons bind near each potential minimum, whereas the fully polarized system requires a significantly deeper potential and instead forms a symmetry-broken orbital nematic molecular crystal.
This spin-dependent molecular transition gives rise to discontinuous magnetization curves, metamagnetic transitions, and several multicritical points in the phase diagram, together with clear signatures in the density profiles, real-space correlation functions, and structure factors. 
To further elucidate the role of the long-range character of the Coulomb interaction, we also study short-range repulsive interactions, which can be realized with ultracold-atom platforms. We find that the qualitative structure of the many-body phase diagram remains unchanged, demonstrating the robustness of our results across different interaction models.
\end{abstract}

\maketitle

\section{Introduction}

Strongly correlated electrons in the presence of an external periodic potential represent a paradigmatic setting for the emergence of collective quantum phenomena. In weakly interacting systems, periodic potentials modify the single-particle spectrum and open band gaps, naturally leading to conventional band insulators. 
Strong interactions can further enhance localization and drive the formation of Mott insulating states, where charge fluctuations are suppressed even in partially filled bands to minimize the energetic cost of double occupancy\cite{ImadaMI_REV}.
However, interactions do not always favor localization. In several strongly correlated systems, interactions can instead suppress single-particle band gaps and drive a transition from a band insulator to a correlated metallic state characterized by enhanced quantum fluctuations, strong exchange effects, and multiband hybridization \cite{GargMI_2006,KancharlaMI_2007,ScalettarMI_2007}. Understanding this interaction-induced metallicity, together with the magnetic properties of the resulting correlated metal, remains a central challenge across condensed matter systems and synthetic quantum platforms.

Semiconductor moir\'e materials have recently emerged as particularly tunable platforms for addressing these questions~\cite{AndreiRev2021}. In particular, transition-metal dichalcogenide (TMD) heterostructures provide an especially clean realization due to the large energy detuning between layers and the strong spin--orbit coupling inherited from the parent semiconductors. In this limit, the low-energy physics is effectively described by electrons moving in an external periodic potential \cite{MakRev2022}. Experimental studies have already uncovered a remarkably rich landscape of correlated phases in moir\'e materials, including antiferromagnetic Mott insulators \cite{Regan2020,Tang2020a,Wang2020,Shimazaki2020}, generalized Wigner crystals \cite{Regan2020,Xu2020a,Huang2021,Tang2023}, heavy-fermion metallic behavior \cite{Zhao2023}, anomalous Hall responses \cite{Li2021}, and kinetic magnetism \cite{Ciorciaro2023,Tao2024}.

Ultracold atoms in optical lattices offer a complementary route to realizing strongly correlated fermionic systems, together with a distinct experimental toolbox. Their high degree of control enables the systematic tuning of interactions, lattice geometry, and filling, while quantum-gas microscopy and nonequilibrium protocols provide complementary probes of many-body states~\cite{Bloch2012_review,Schafer2020}. These platforms have realized key aspects of Fermi--Hubbard physics, including antiferromagnetic correlations~\cite{Mazurenko2017}, magnetic polarons~\cite{Koepsell2019,Prichard2024,Lebrat2024}, and unconventional correlated regimes~\cite{Koepsell_doping_2021,chalopin_observation_2026,kendrick2025pseudogapfermihubbardquantumsimulator}. Recent advances have also introduced atomic analogues of angle-resolved photoemission spectroscopy~\cite{Brown2020}, lattice-modulation spectroscopy~\cite{kendrick2025pseudogapfermihubbardquantumsimulator}, and two-photon Raman spectroscopy~\cite{Prichard2025}. In strongly interacting regimes, higher Bloch bands may also become relevant~\cite{Li_REV_2016}, generating multiband effects that qualitatively modify the low-energy description~\cite{Luhmann_2012}.

However, most theoretical studies of correlated phenomena in both moir\'e electron systems and ultracold fermions in optical lattices have focused on effective single-band extended Fermi--Hubbard models~\cite{PhysRevA.79.063412,PhysRevA.79.043626}. While such approaches successfully capture many qualitative features of the observed phases, recent studies suggest that they may fail to quantitatively describe the relevant energy scales and can even overlook important physical effects. The origin of this limitation is that the interaction energy may substantially exceed the confinement energy associated with the external periodic potential, leading to strong interaction-induced hybridization between different bands. Evidence for this multiband character has recently been reported in scanning tunneling microscopy experiments on moir\'e materials, where the localized orbitals were found to contain substantial contributions from higher moir\'e bands \cite{FengWang2024}. These observations motivate the development of theoretical approaches that go beyond the effective single-band approximation and explicitly incorporate interaction-driven multiband physics.

Motivated by these considerations, in this work we investigate interacting electrons in a triangular periodic potential directly in the continuum, without relying on Wannierization. We consider both Coulomb interactions and short-range interactions, allowing us to isolate the role of long-range interactions and provide predictions for both synthetic platforms and moir\'e materials. The long-range interacting electron gas has been extensively studied over the last decades \cite{Ceperley1,Ceperley2,ShiweiZhang,MPNN1,PBC_varible,Backflow1,Backflow2,Attaccalite2002,PhysRevB.79.085414}, with state-of-the-art calculations pointing toward the emergence of a Wigner crystal \cite{Wigner} at sufficiently low densities, together with unconventional magnetic states in the vicinity of the crystallization transition \cite{ShiweiZhang}. Here, however, we focus on an intermediate interaction regime that remains metallic in the absence of the external periodic potential, i.e., below the conventional Wigner crystallization transition, and investigate how the periodic potential reshapes the correlated many-body state.

Tackling the continuum problem of strongly correlated electron systems remains a central theoretical challenge. Mean-field approaches, such as Hartree-Fock, fail to capture strong correlation effects, while numerically exact techniques, such as exact diagonalization, are limited to relatively small clusters. This motivates the development of variational approaches for continuum systems capable of accurately describing strongly correlated electronic states.

Recent advances in artificial intelligence and machine learning have introduced a new class of variational wavefunctions inspired by the universal representation capabilities of neural networks, the so-called neural quantum states (NQS)~\cite{CarleoTroyer2017}. These approaches have demonstrated remarkable flexibility in representing quantum many-body wavefunctions and have been successfully applied to ground-state simulations~\cite{NQS_Hubbard,NQS_Hubbard2,NQS_Hubbard3}, thermal density matrices~\cite{PhysRevResearch.2.013284,PhysRevLett.120.240503,nys2025fermionicneuralgibbsstates}, and real-time quantum dynamics~\cite{NQSDynamics-2026,VandeWalle_2025,Nys2024}.

The introduction of backflow transformations in continuum electron systems \cite{Feynman1956,Backflow1,Backflow2} substantially improved their description while preserving computational efficiency. Combined with neural-network-based approaches for Slater determinant~\cite{FermiNet2020,MPNN1,LiangFu2025} and Pfaffian wavefunctions~\cite{MPNN2}, these variational ansätze have achieved significant success in the study of nuclear matter~\cite{NQS_nuclear,NQS_nuclear2,NQS_nuclear3}, quantum chemistry~\cite{FermiNet2020,PauliNet2020,NQS_chemistry}, and continuum electron systems~\cite{ShiweiZhang,MPNN1,LiangFu2024,Jichen2025}. 

In this work, we employ NQS to simulate a continuum electron system in the presence of an external triangular potential at filling factor $\nu=2$, corresponding to two electrons per minimum of the periodic potential. Our results show that, for both long-range and short-range interacting models, there exists a finite critical strength of the periodic potential separating a correlated metallic phase from an insulating phase for any spin polarization. The transition is first order. 
Throughout this study, we consider exclusively the Zeeman effect of the external magnetic field, neglecting orbital couplings. In the absence of this field (yielding zero spin polarization), the insulating state corresponds to an s-wave molecular crystal, in which two opposite-spin electrons form a strongly bound pair localized near each minimum of the periodic potential.
We further investigate the response of the system to an external magnetic field. Below the metal--insulator transition, the correlated metallic phase exhibits a finite spin susceptibility and undergoes a metamagnetic transition at a critical field, where the spin polarization displays a sudden jump to full polarization. As the system approaches the transition point, the spin susceptibility diverges, signaling an incipient magnetic instability. Beyond this point, the energy landscape becomes concave as a function of polarization, implying that intermediate polarization states are thermodynamically unstable. As a consequence, the polarization remains pinned to zero up to a critical magnetic field, at which it discontinuously jumps to full polarization.

In platforms where the total polarization is externally constrained, such as ultracold atoms in optical lattices with separately conserved spin populations, we expect the homogeneous states at intermediate polarization to be thermodynamically unstable. The system can lower its energy by separating into regions with different local polarizations, while the imposed global polarization is maintained through the relative volume fractions of the coexisting phases. For example, a system constrained to a global polarization $\zeta=0.5$ would consist of equal fractions of unpolarized and fully polarized regions.

In the fully polarized sector, we also observe a metal--insulator transition at a larger critical potential strength. Importantly, the resulting insulating phase is not continuously connected to the unpolarized insulator. At full polarization, the electrons instead form an orbital nematic molecular crystal characterized by reduced rotational and spatial symmetry. This distinction generates a broad intermediate regime of potential strength in which the unpolarized system is insulating while the fully polarized system remains a correlated metal.

We show how the multiple phases studied and their transitions appear in experimentally relevant
observables. The real-space density provides direct indicators of the metal--insulator transition and it can be employed to characterize the nematic character of the fully polarized insulating state.  The density and spin
correlation functions reveal the buildup of short-range molecular and
magnetic correlations. The distinction between the connected and
bare structure factors is essential: the bare structure factor contains
Bragg peaks from the static potential-induced density profile, whereas the
connected structure factor measures fluctuations about this inhomogeneous
background. The absence of a divergent connected-structure-factor peak is
consistent with a first-order transition in an explicitly pinned
system.

Finally, we introduce a short-range interaction model described by a P\"oschl--Teller potential and investigate how the many-body phases evolve with the interaction range. We find that the qualitative structure of the phase diagram remains robust over a broad range of interaction ranges, while the orientation of the orbitals in the orbital-nematic molecular crystal can be tuned by varying the interaction range.

\section{Theoretical Framework}
We consider a system of interacting spin-$1/2$ electrons in the presence of an external periodic potential, described by the Hamiltonian
\begin{equation}
H
=\sum_{i=1}^{N_e}\left(-\frac{\hbar^2}{2m^*} \nabla_i^2 + V_M(\bm r_i) \right)
+\sum_{1\le i<j\le N_e} V_{ee}(\mathbf{r}_i - \mathbf{r}_j),
\label{Eq:Hamiltonian}
\end{equation}
where $N_e$ is the number of electrons, $m^*$ denotes the effective band mass and $V_M(\bm r)$ is a triangular periodic potential modeled within the lowest-harmonic approximation,
\begin{equation}
V_M(\bm r)=-V_0\sum_{n=1}^{6}
e^{i{\bm G}_M^{(n)}\cdot \bm r},
\end{equation}
Here, $V_0$ characterizes the strength of the periodic potential and ${\bm G}_M^{(n)}=\frac{4\pi}{\sqrt{3}a_M}\left(\cos(n\pi/3),\sin(n\pi/3) \right)$ are the six shortest reciprocal lattice vectors of the triangular lattice, with $a_M$ the lattice constant.

The Hamiltonian~\eqref{Eq:Hamiltonian} provides an effective low-energy description of several experimentally relevant systems, including TMD heterobilayers with approximate $\mathcal{C}_6$ symmetry, such as $\rm MoSe_2/WSe_2$ and $\rm MoS_2/WS_2$~\cite{LiangFu2024,LiangFu2025,LiangFu2020}, as well as ultracold fermionic atoms loaded into triangular optical lattices~\cite{XuOLtriangular,Lebrat2024,Prichard2024}. 

For TMD heterobilayers, we consider long-range Coulomb interactions~\cite{Ceperley1,Ceperley2,ShiweiZhang,MPNN1,PBC_varible},
\begin{equation}
    V_{ee}(\mathbf{r})=\frac{\mathrm{e}^2}{4\pi\epsilon_0\epsilon_r} \frac{1}{|\bm r|} +V_{+-},
\end{equation}
where $\mathrm{e}$ denotes the elementary charge, $\epsilon_0$ and $\epsilon_r$ denote the vacuum and relative dielectric constants, respectively. The term $V_{+-}$ represents the interaction between the electrons and a uniform positive background charge. This standard jellium contribution guarantees global charge neutrality and removes the long-wavelength divergence of the Coulomb energy.

Throughout this work, when considering TMD heterobilayers we assume hBN encapsulation and use representative material parameters,
$a_M\sim 10.2\,\mathrm{nm}$, $\epsilon_r \sim 4.5$, and $m^*\sim 0.7 m_e$.
Here, $m_e$ denotes the bare electron mass.
We explore a broad range of periodic potential strengths, $V_0\in[0,12]\,\mathrm{meV}$.

We additionally define the dimensionless interaction parameter $r_s=a/a_B^*$, with the effective Bohr radius
$a_B^*=4\pi\epsilon_0\epsilon_r\hbar^2/(m^*\mathrm{e}^2)$ and average
interparticle spacing $a=1/\sqrt{\pi\rho_e}$ with $\rho_e$ being the electron density. For fixed dielectric environment, effective mass, and lattice constant, the interaction parameter is fully determined by the filling factor $\nu$,
\begin{equation}
r_s=
\frac{1}{a_B^*}
\left(
\frac{\pi \nu}{\sqrt{3}a_M^2/2}
\right)^{-1/2}
\propto \nu^{-1/2}.
\end{equation}
For the representative filling considered in this work $\nu=2$, we obtain $r_s \approx 11.13$, which remains well below the critical value for the transition to the Wigner crystal phase~\cite{ShiweiZhang}.

To demonstrate the broader applicability of the physical phenomena discussed in this work, we also investigate a short-range interacting model relevant to ultracold atom platforms. In particular, we consider a short-range repulsive P\"oschl--Teller (PT) interaction given by
\begin{equation} 
V_{ee}(\bm r) = V_{\rm PT} \frac{\hbar^2\kappa^2}{m^*} \frac{1}{\cosh^2(\kappa |\bm r|)} , \label{eq:poschl_teller_potential} \end{equation} 
where $V_{\rm PT}>0$ is a dimensionless parameter controlling the interaction strength and $\kappa^{-1}$ sets the interaction range. Unlike the Coulomb interaction, the PT potential remains finite at the origin and decays exponentially at large distances, thereby providing a controlled short-range interaction that captures the essential effects of strong repulsion without the long-range Coulomb tail.

\begin{figure*}[t]
\includegraphics[trim=10 60 5 0, clip,width=1.0\linewidth]{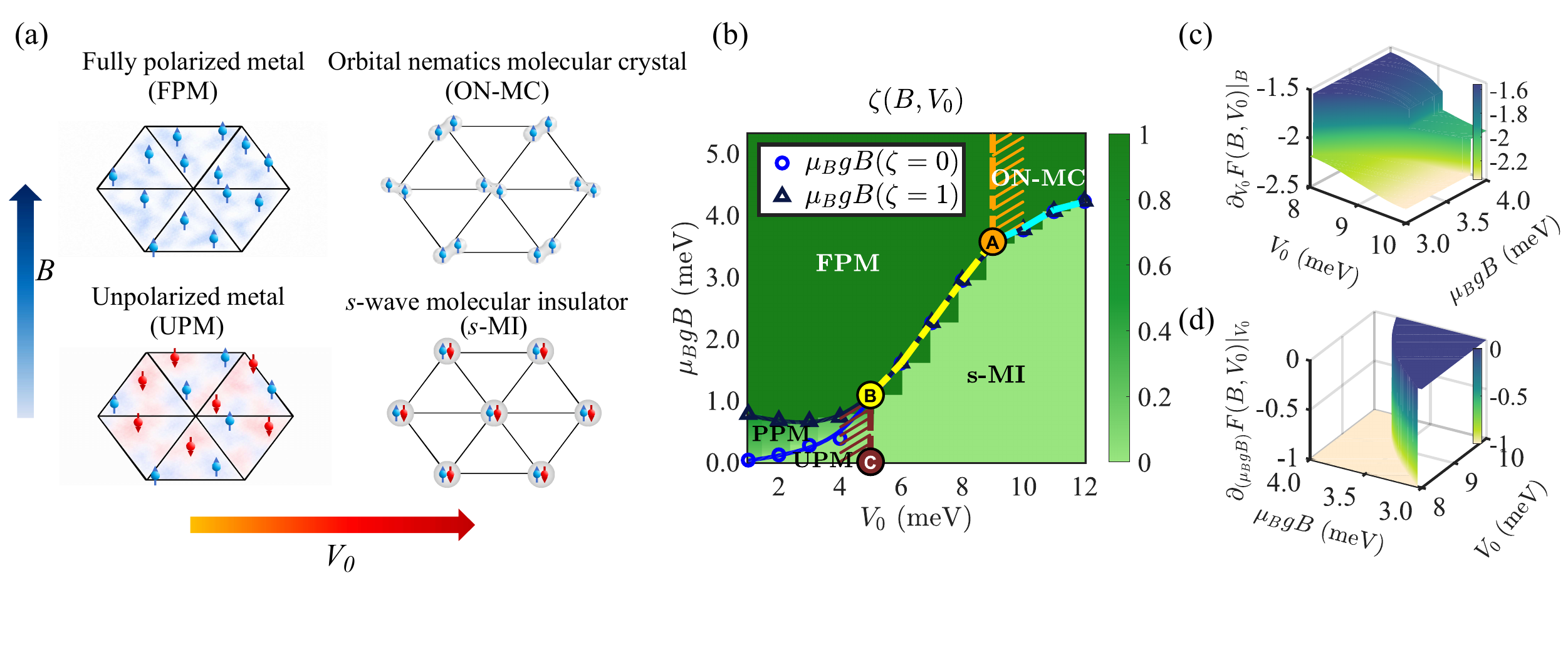}
\caption{
Many-body phase diagram and thermodynamic properties obtained for the long-range Coulomb interaction.
(a) Schematic illustration of the many-body phases encountered in this work. Upon increasing the moir\'e potential strength $V_0$, the system undergoes a metal--insulator transition. The metallic state exhibits a metamagnetic transition at finite magnetic field $B$, whereas the insulator exhibits a critical magnetic field at which it transitions to an insulator with a reduced symmetry pattern.
(b) Phase diagram of the spin polarization $\zeta$ in the $B$--$V_0$ plane. Here, UPM, PPM, and FPM denote the unpolarized ($\zeta=0$), partially polarized ($0<\zeta<1$), and fully polarized ($\zeta=1$) metallic phases, respectively. The s-MI and ON-MC phases denote the $s$-wave molecular insulator with $\zeta=0$ and the orbital-nematic molecular crystal with $\zeta=1$, respectively.
The line $\mu_B g B(\zeta=0)$ marks the onset of finite spin polarization upon increasing the magnetic field, while $\mu_B g B(\zeta=1)$ indicates the boundary beyond which the fully polarized state ($\zeta=1$) becomes energetically favorable.
The orange, yellow, cyan-blue, and brown dot-dashed lines indicate the phase boundaries separating FPM--ON-MC, FPM--s-MI, s-MI--ON-MC, and PPM/UPM--s-MI, respectively. The red, yellow, and brown points labeled A, B, and C denote phase-boundary junctions. The diagonally hatched shaded regions along the brown and orange boundaries indicate the numerical uncertainty of the transition lines arising from the finite resolution of the sampled parameter points.
(c) First derivative of $F(B,V_0)\equiv E(B,V_0)/N_e-\mu_B g B\,\zeta(B,V_0)$ with respect to $V_0$ near triple point A at fixed $B$. For the moir\'e
potential used in this work, $\partial F/\partial V_0$ is physically related to the Fourier component of the real-space density, $\tilde n(\bm G_M)$, at the moir\'e reciprocal lattice vectors.
(d) First derivative of $F(B,V_0)$ with respect to $B$ near triple point A at fixed $V_0$. Physically, $\partial F/\partial B$ corresponds to the spin
polarization $\zeta$.
}
\label{key_map}
\end{figure*}

\subsection{Variational Approach}
To simulate the quantum many-electron problem, we employ a variational ansatz optimized using variational Monte Carlo (VMC). Our ansatz consists of a Slater--Jastrow wave function augmented by neural-network backflow transformations. The variational wave function is written as
\begin{equation}
\Psi(\bm R)
=
J(\bm R)\,
\det\!\left[
\Phi_{\bm k}(\bm r_i;\delta \bm r_i)
\right],
\end{equation}
where $\bm R=(\bm r_1,\dots,\bm r_{N_e})^{T}$ denotes the full electronic configuration, $\delta\bm r_i$ are backflow coordinates that depend explicitly on the positions of all particles, and $J(\bm R)$ is a global Jastrow factor parameterized by neural networks. For $\zeta<1$, the determinant notation is shorthand for the product of the spin-resolved determinants, $\det D^\uparrow \det D^\downarrow$, as detailed in Appendix~\ref{app_MPNN}.

The single-particle orbitals are chosen to take a Bloch-like backflow form,
\begin{equation}
\Phi_{\bm k}(\bm r;\delta\bm r)
=
e^{i\bm k\cdot(\bm r+\delta\bm r)}
\,u_{\bm k}(\delta\bm r),
\end{equation}
where $\bm k$ labels the allowed simulation-cell momenta, and $u_{\bm k}$ is periodic in the simulation cell. The periodic component $u_{\bm k}$ is also represented by a neural network and separately parameterized for each $\bm k$. The skip connections in the MPNN ensure that $\delta\bm r_i$ retains information about the original coordinate $\bm r_i$.

The backflow transformation is constructed using a message-passing neural network (MPNN) architecture~\cite{MPNN1,MPNN2}. Message-passing architectures provide an efficient means of incorporating pair correlations and permutation symmetry~\cite{MPNN1}. Compared with fully connected network designs, the MPNN-based ansatz generally requires substantially fewer variational parameters while maintaining competitive accuracy. For moderate system sizes, their performance is comparable to state-of-the-art approaches such as FermiNet~\cite{FermiNet2020}. Since the system considered here has $r_s\simeq 11.13$, we build upon the neural-network backflow ansatz developed for NQS studies of the homogeneous interacting electron gas~\cite{MPNN1} and extend it by introducing the periodic factor $u_{\bm k}(\delta\bm r_i)$ to describe the moir\'e potential.

The network is constructed from one-particle and
two-particle features defined from the particle coordinates
$\{\bm r_i\}$ and relative coordinates
$\{\bm r_{ij}=\bm r_i-\bm r_j\}$.
The central idea of the method is to iteratively construct message 
variables $\bm m_{ij}^{(l)}$ from the two-particle features and use them to update both one-particle and two-particle representations through message-passing operations, where $l$ labels the message-passing layer. By iterating this procedure, the network progressively builds higher-order many-body correlations among the particles.

The final network output defines the backflow corrections $\delta\bm r_i$.
The construction is permutation equivariant: exchanging two particle coordinates, $\bm r_i \leftrightarrow \bm r_j$, simultaneously exchanges the corresponding backflow corrections, $\delta\bm r_i \leftrightarrow \delta\bm r_j$.
Thus, the backflow coordinates encode many-body correlation effects while preserving exchange symmetry and periodic boundary conditions.
Full details of the feature construction and network architecture are provided in Appendix~\ref{app_MPNN}.

The Hamiltonian in Eq.~\eqref{Eq:Hamiltonian} possesses an SU(2) spin-rotation symmetry. In our variational ansatz, however, we impose only a U(1)$\otimes$ U(1) symmetry, which fixes the number of electrons in each spin sector independently. Accordingly, we compute the lowest variational energy per particle, $E(\zeta)/N_e$, within sectors of fixed spin polarization $\zeta={(N_{\uparrow}-N_{\downarrow})}/{N_e}$, where $N_\sigma$ denotes the number of electrons with spin $\sigma$.

Performing the variational optimization requires a careful initialization of the variational parameters. For intermediate values of $V_0$, several competing states become nearly degenerate. Together with the large number of variational parameters in the neural-network ansatz, this makes direct optimization from a fully random initialization prone to becoming trapped in metastable local minima. To improve convergence, we therefore employ an annealing (transfer-learning) strategy. 
Specifically, we first obtain well-converged reference states in both the weak- and strong-moir\'e-potential regimes, where the variational landscape is comparatively smooth and easier to optimize. Starting from these optimized states, we adiabatically sweep the external potential by gradually increasing $V_0$ from the weak-potential side and decreasing $V_0$ from the strong-potential side. At each step of the sweep, the variational parameters are re-optimized using the solution from the previous step as the initialization. This procedure generates two variational branches spanning the phase diagram.
At each parameter point, the ground-state candidate is identified by comparing the energies of the two branches and selecting the lower-energy solution.
As an additional cross-check, we also use wave functions optimized for one interaction model as initial states for the optimization of another interaction model. For example, a state obtained for the Coulomb interaction can be transferred to the corresponding P\"oschl--Teller system and subsequently re-optimized. The resulting energy is then compared with that obtained from a direct optimization within the P\"oschl--Teller model. This cross-model initialization is particularly useful for the short-range interaction at large $V_0$, where our numerical simulations show that configurations containing localized particle pairs can easily become trapped in metastable local minima. By exploring several physically motivated initialization protocols and selecting the lowest-energy solution among them, we reduce the likelihood of misidentifying a metastable state as the variational ground state.

\subsection{Observables}
To characterize the phases realized by the optimized variational states, we evaluate a set of real- and momentum-space observables that probe charge ordering, spin polarization, and many-body correlations.

We first consider the local spin-resolved densities
\begin{equation}
n_{\sigma}(\bm r)=\left\langle\hat{n}_{\sigma}(\bm r)\right\rangle,
\qquad
n(\bm r)=n_\uparrow(\bm r)+n_\downarrow(\bm r),
\label{density_formula}
\end{equation}
where $\langle\cdots\rangle$ denotes the expectation value with respect to the many-body quantum state, and \(\hat{n}_{\sigma}(\bm r)=\sum_j \delta(\bm r-\bm r_{j\sigma})\).
Together with the local spin polarization
\begin{equation}
s_z(\bm r)=
\left\langle \hat{s}_z(\bm r)\right\rangle
=
n_\uparrow(\bm r)-n_\downarrow(\bm r),
\end{equation}
where \(\hat{s}_z(\bm r)=\hat{n}_\uparrow(\bm r)-\hat{n}_\downarrow(\bm r)\), these quantities provide direct information about spatial charge inhomogeneities and magnetic textures.

To further characterize the local shape of the density profile, we introduce the second-moment matrix of the density distribution around each potential minimum $\bm R_j$,
\begin{equation}
M_{ab}^{[j]}
\equiv
\frac{\int_{D_j} d^2r\, n(\bm r)\,\delta r_a^{[j]}\delta r_b^{[j]}}
{\int_{D_j} d^2r\, n(\bm r)} .
\end{equation}
Here $\delta\bm r^{[j]}=\bm r-\bm R_j$, and
$D_j=\{|\delta\bm r^{[j]}|<r_0\}$ denotes the disk over which the local density profile is sampled. Throughout this work, we choose $r_0=0.5a_M$.

The corresponding local quadrupole can be conveniently encoded in the dimensionless complex order parameter
\begin{equation}
\bm{\mathcal{Q}}_j
\equiv
\frac{(M_{xx}^{[j]}-M_{yy}^{[j]})+2iM_{xy}^{[j]}}
{M_{xx}^{[j]}+M_{yy}^{[j]}}
=
\mathcal{Q}_j e^{2i\varphi_j}.\label{Eq:quadrupole}
\end{equation}
Here $\mathcal{Q}_j$ measures the strength of the local anisotropy, while $\varphi_j$ denotes the orientation angle of the quadrupole director. Explicitly,
\begin{equation}
\begin{aligned}
\mathcal{Q}_j
&=
\frac{\sqrt{(M_{xx}^{[j]}-M_{yy}^{[j]})^2+4(M_{xy}^{[j]})^2}}
{M_{xx}^{[j]}+M_{yy}^{[j]}},
\\
\varphi_j&=\frac{1}{2}\arg \bm{\mathcal{Q}}_j .
\end{aligned}
\end{equation}
 The factor $e^{2i\varphi_j}$ sets the direction of the quadrupole, which is invariant under $\varphi_j\rightarrow\varphi_j+\pi$.

To quantify correlations beyond local expectation values, we compute the density and spin correlation functions
\begin{equation}
C_n(\bm 0,\bm r)=
\langle \hat{n}(\bm 0)\hat{n}(\bm r)\rangle,
\qquad
C_s(\bm 0,\bm r)=
\langle \hat{s}_z(\bm 0)\hat{s}_z(\bm r)\rangle.
\end{equation}
Because the periodic potential explicitly breaks translational symmetry,
all correlation functions are evaluated with one reference point fixed at
a potential minimum, chosen as $\bm 0$.

Additionally, we define the Fourier components of the density,
\begin{equation}
\tilde n(\bm k)
=
\int d^2r\,
\langle \hat{n}(\bm r)\rangle
e^{i\bm k\cdot\bm r},
\end{equation}
which allows us to identify the characteristic ordering wave vectors of density-ordered phases.

Finally, we compute the connected static structure factor
\begin{equation}
S(\bm k)
=
S_0(\bm k)
-
\frac{1}{N_e}
\left|
\left\langle
\sum_i e^{i\bm k\cdot\bm r_i}
\right\rangle
\right|^2,
\label{Eq:Sk}
\end{equation}
where
\begin{equation}
S_0(\bm k)
=
\frac{1}{N_e}
\left\langle
\left|
\sum_i e^{i\bm k\cdot\bm r_i}
\right|^2
\right\rangle
\label{Eq:S0k}
\end{equation}
is the corresponding bare structure factor. These quantities allow us
to characterize density fluctuations associated with the relevant ordering
wavevectors.

\section{Phase Diagram}
The phase diagram exhibits a rich interplay between correlation effects and spin polarization, giving rise to several metallic and insulating phases with distinct magnetic and orbital character, as shown in Fig.~\ref{key_map}(b). To determine the magnetic polarization of the ground state, we minimize the free energy per particle in the presence of an external magnetic field,
\begin{equation}
F(B,V_0)=\frac{E(\zeta,V_0)}{N_e}-\mu_B g B \zeta,
\end{equation}
where $E(\zeta,V_0)$ is the variational total ground-state energy at spin polarization $\zeta$ and potential strength $V_0$, $\mu_B$ is the Bohr magneton, $g$ is the Land\'e g-factor, and $B$ denotes the applied magnetic field. 

According to the spin polarization and charge ordering pattern, several distinct metal and insulator phases appear in the phase diagrams. Their nomenclature is summarized in the caption of Fig.~\ref{key_map}.

Let us emphasize that minimizing the free-energy functional yields the thermodynamically stable solution. Nevertheless, our ground-state calculations also allow us to identify solutions at finite polarization $\zeta$ that do not necessarily correspond to the global minimum of the free energy, see Fig.~\ref{E_zeta} (a) for a phase diagram for different polarization sectors. These states may represent metastable configurations that could be realized experimentally, for instance in ultracold-atom optical lattice setups where the polarization can be externally fixed.

Throughout this work we neglect orbital magnetic effects and retain only the Zeeman contribution. This approximation is justified because, for the parameter regime considered, the magnetic flux penetrating a moir\'e unit cell remains small. Taking $g\sim 10$, magnetic fields up to $B\sim 10\,\mathrm{T}$, and a moir\'e lattice constant $a_M\sim 10\,\mathrm{nm}$, we estimate an effective flux ratio of
$\Phi = \sqrt{3} a_M^2 B/2 \sim 0.2\Phi_0$, where $\Phi_0=h/e$ is the magnetic flux quantum. 

\subsection{Zero Magnetic Field}
We have verified that the zero-polarization state remains the ground state for all potential strengths at zero applied magnetic field. Let us now examine the nature of this state in the presence of the potential. At the single-particle level, the periodic potential opens a gap proportional to $V_0$, driving the system from a metal to a band insulator for any finite value of $V_0$. However, once correlation effects are included, this gap can be closed, leading instead to a strongly correlated metallic phase.

This behavior is illustrated in Fig.~\ref{key_map}(b), where we find that a minimum potential strength of approximately $V_0\sim 5\,\mathrm{meV}$ is required to stabilize the insulating phase. The insulating state is characterized by strongly bound opposite-spin electron pairs localized at each potential minimum. We therefore refer to this phase as an $s$-wave molecular insulator (s-MI).

The density profile associated with these molecular states exhibits strong hybridization among different Bloch bands, reflecting the substantial band mixing induced by Coulomb interactions. In particular, we
observe a strongly suppressed electron density at the center of each site. Additional details regarding the structure of this phase are presented below.

The metal--insulator transition at zero magnetic field appears to be first order, with the corresponding critical point denoted by C in Fig.~\ref{key_map}(b). The transition is accompanied by a discontinuous jump in the Fourier components of the
density at the characteristic ordering wavevector.

We also note that simulations with short-range interactions display qualitatively similar phenomenology, although the metal--insulator transition shifts to $V_0\sim 3\,\mathrm{meV}$ for the interaction strengths considered ($V_{\rm PT}=5$ and $\kappa=a^{-1}$). Importantly, however, the density profile of the s-MI in the short-range case is considerably smoother and does not exhibit the pronounced density depletion at the potential minimum.

\subsection{Full Polarization}
Upon applying a magnetic field, the metal--insulator transition shifts toward larger values of the external potential. This behavior can be understood by examining the fully polarized state, $\zeta=1$. At the single-particle level, electrons are forced to occupy the second Bloch band due to the Pauli exclusion principle. Interestingly, the second and third bands touch at symmetry-related momenta, see Appendix \ref{App:DOS}. Therefore, at the single-electron level we expect a semimetal-like behavior. As a consequence, the fully polarized insulating state becomes significantly more susceptible to interaction effects, pushing the metal--insulator transition to approximately $V_0\sim 9\, \mathrm{meV}$.

It is important to emphasize that, for the fully polarized state, a purely $s$-wave contact interaction does not modify the single-particle picture, and one therefore expects a semimetal-like state for arbitrarily small $V_0$. In contrast, both long-range Coulomb interactions and finite-range P\"oschl--Teller interactions yield a finite critical value of $V_0$, demonstrating that finite-range effects qualitatively alter the many-body ground state.
Interestingly, the insulating state that emerges at full polarization reflects the finite-angular-momentum character of the scattering channel. In particular, the localized molecular states acquire a characteristic $p$-wave orbital structure. The orientation of these orbitals spontaneously breaks the sixfold rotational symmetry imposed by the external potential, giving rise to nematic order, as discussed in detail below. We therefore identify the fully polarized insulating phase as an orbital nematic molecular crystal (ON-MC).

\subsection{Response to an external magnetic field}
Having established the limiting cases of zero polarization, $\zeta=0$, and full polarization, $\zeta=1$, we now examine the evolution of the many-body state under an applied magnetic field. The magnetic field creates an interplay between the spin and orbital structure, giving rise to a rich sequence of phase transitions.

For weak external potentials, the polarization initially increases continuously with magnetic field, driving the system into a partially polarized metal phase ($0<\zeta<1$), as shown in Fig.~\ref{key_map}(b). Upon further increasing the field, the system undergoes a first-order metamagnetic transition characterized by a finite jump in polarization into the fully polarized metal state ($\zeta=1$).

\begin{figure}[t]
\includegraphics[width=1.0\columnwidth]{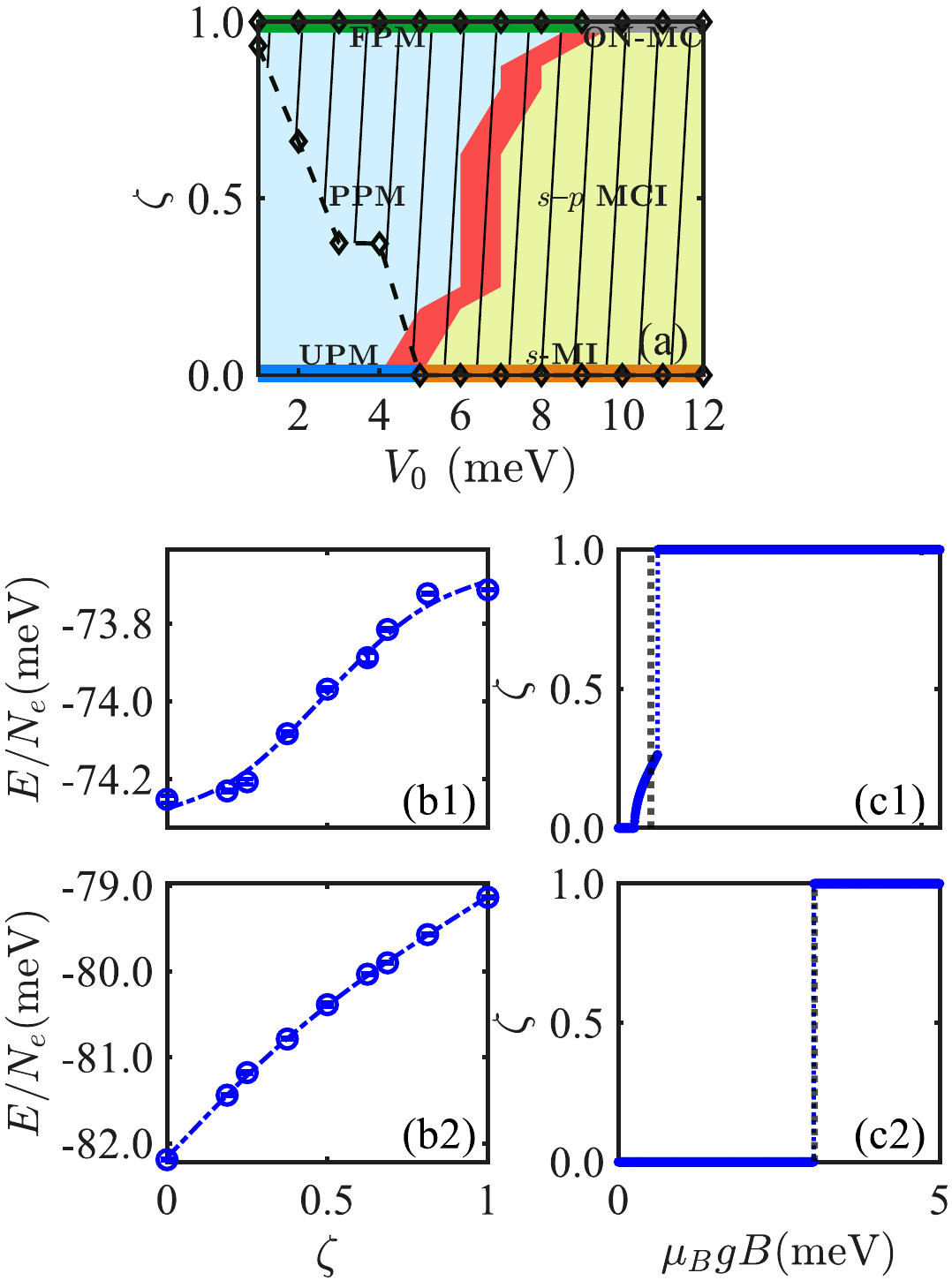}
\caption{(a) Phase diagram in the $\zeta$--$V_0$ plane showing the metal-insulator transition. The hatched area enclosed by black dashed lines denotes the partially polarized region ($0<\zeta<1$), which is inaccessible upon increasing the magnetic field $B$. The red region marks the phase boundary between the metallic and insulating phases. Here, UPM, $s$-MI, PPM, $s$--$p$ MCI, FPM, and ON-MC denote the unpolarized metal (blue), $s$-wave molecular insulator (orange), partially polarized metal (light blue), $s$--$p$ molecular coexistence insulator (light yellow), fully polarized metal (green), and orbital nematic molecular crystal (gray), respectively.
(b1--b2), (c1--c2): The first column shows the energy per particle $E/N_e$ versus $\zeta$, whereas the second column shows $\zeta$ versus the Zeeman energy $\mu_BgB$. The first and second rows correspond to $V_0=3$ and $8\,\mathrm{meV}$, respectively. The vertical dark dotted lines in (c1--c2) denote the energy difference $\left[E(\zeta=1)-E(\zeta=0)\right]/N_e$. The results shown here are obtained for the Coulomb interaction.
}
\label{E_zeta}
\end{figure}

The origin of this discontinuity can be understood by analyzing the energy per particle $E(\zeta)/N_e$ at fixed $V_0$, shown in Fig.~\ref{E_zeta}(b1)--(b2). For weak external potentials, e.g., $V_0=3\, \mathrm{meV}$, the curvature of $E(\zeta)/N_e$ exhibits a nontrivial dependence on polarization. At small $\zeta$, the second derivative $d^2[E/N_e]/d\zeta^2$ is positive, indicating a stable partially polarized state. However, the curvature decreases with increasing $\zeta$ and eventually becomes negative, signaling a crossover from convex to concave behavior.

In the strong-potential regime, above the critical point C, the energy becomes concave over the entire polarization range, i.e., $d^2[E(\zeta)/N_e]/d\zeta^2\le0$. Although $E(\zeta)/N_e$ still increases monotonically with polarization, its slope decreases continuously as the system approaches full polarization.
In this regime, a Maxwell construction yields a lower-energy state corresponding to phase separation between the unpolarized $(\zeta=0)$ and fully polarized $(\zeta=1)$ insulating phases. The relevant energy scale governing this coexistence is the difference between the molecular binding energies per particle of the two phases, $E_B = E_p-E_s$,
where $E_p>E_s$. Since the fully polarized phase consists of localized $p$-wave pairs while the unpolarized phase is formed by localized $s$-wave pairs, an intermediate global polarization $0<\zeta<1$ can be realized by converting only a fraction $\zeta$ of the pairs from the $s$-wave to the $p$-wave configuration. The resulting energy of the phase separated state reads $\left[E(\zeta)-E(0)\right]/N_e\sim E_B\zeta$. Thus, states with intermediate polarization should be understood as phase-separated configurations of the two insulating phases, rather than as homogeneous partially polarized states.

To quantify this behavior, we compute the energy difference between the fully polarized and unpolarized states,
\begin{equation}
\Delta E = \frac{E(\zeta=1)-E(\zeta=0)}{N_e}.
\label{Eq:DeltaE}
\end{equation}
This quantity increases monotonically with $V_0$, reflecting the enhanced relative stability of the unpolarized state at strong periodic potentials.

Above the critical point C, the energy difference per particle $\Delta E$ directly determines the critical magnetic field required to stabilize the fully polarized state, confirming the nearly linear dependence of $E(\zeta)/N_e$ on polarization. In this regime, all intermediate polarized solutions are thermodynamically unstable, as indicated by the hatched region in Fig.~\ref{E_zeta}(a).

By contrast, in the weak-$V_0$ regime, $\Delta E$ does not coincide
with the critical Zeeman energy $\mu_B g B_c$. This indicates that partially polarized states can acquire lower free energy than either endpoint configuration at the same field strength, thereby stabilizing the partially polarized metal phase.

\subsection{Phase Transitions}
The phase boundaries separating the different phases are indicated by the colored lines in Fig.~\ref{key_map}(b). For weak external potentials, increasing the magnetic field drives the sequence $\mathrm{UPM}\rightarrow \mathrm{PPM}\rightarrow \mathrm{FPM}$, corresponding to a continuous evolution from the unpolarized metal to the partially polarized metal, followed by a first-order transition into the fully polarized metal phase.

In the intermediate-$V_0$ regime, the magnetic field instead melts the $s$-wave molecular insulator into the FPM phase (yellow dashed line). This behavior originates from the strong polarization dependence of the metal--insulator transition: the fully polarized state requires a substantially larger external potential to stabilize the insulating phase. As a result, increasing the Zeeman field destabilizes the insulating state and restores metallic behavior.

To characterize the nature of the multicritical points, we examine the derivatives of the free-energy functional $F(B,V_0)$ with respect to $B$ and $V_0$. In particular, we focus on point A in Fig.~\ref{key_map}(b), where the FPM, s-MI, and ON-MC phases coexist. Since both the energy $E$ and the spin polarization $\zeta$ exhibit discontinuities across the phase boundaries, the corresponding derivatives of $F$ can be obtained analytically; the explicit expressions are provided in the Appendix~\ref{dF}.

The derivative with respect to the Zeeman field,
$\partial_{(\mu_B g B)}F(B,V_0)$, evaluated at fixed $V_0$, exhibits a
discontinuous jump from $0$ to $-1$ upon crossing the phase boundary. Since
$\partial_{(\mu_B g B)}F=-\zeta$, this jump directly reflects the abrupt
change in spin polarization from $\zeta=0$ to $\zeta=1$.
By contrast, the derivative with respect to the external potential,
$\partial_{V_0}F(B,V_0)$, evaluated at fixed $B$, displays a richer structure
that depends on the path followed in the phase diagram. On the other hand, $\partial_{V_0}F$ measures the density Fourier components at the
reciprocal lattice vectors imposed by the external periodic potential. It therefore provides a direct probe of how
strongly the real-space charge density is locked to the external potential profile.
This also explains why $|\partial_{V_0}F|$ is larger in the
insulating phases than in the metallic phase. In the insulating regime, the
electrons are more strongly localized near the minima of the external
potential, leading to enhanced density Fourier components at the
reciprocal lattice vectors.
Equivalently, on each
smooth branch of the phase diagram where the spin polarization is fixed,
$\partial_{V_0}F$ coincides with the slope of the optimized energy per particle with
respect to the potential strength. Since $V_0$ enters the Hamiltonian
linearly, the ground-state energy is a concave function of $V_0$; with the
sign convention used here, this implies that the energy slope becomes
increasingly negative as $V_0$ is increased, a more detailed discussion on this behavior is provided in ~\ref{Energy_Sec}.
Along trajectories where the system evolves directly from the FPM to the
s-MI phase, this derivative exhibits a single discontinuity associated with
the metal--insulator transition. However, for paths crossing the sequence
$\mathrm{FPM}\rightarrow \mathrm{ON\text{-}MC}\rightarrow \mathrm{s\text{-}MI}$,
an additional weak discontinuity emerges at the FPM--ON-MC boundary. Although
the FPM and ON-MC phases share the same spin polarization, $\zeta=1$, their
charge ordering and internal molecular structure differ. This difference
changes the Fourier components of the density at the reciprocal
lattice vectors, producing a subtle but finite kink in $\partial_{V_0}F$.
These observations identify point A as a triple point of the phase diagram.

The phase diagram also contains two additional multicritical points, denoted by B and C in Fig.~\ref{key_map}(b). Within the present numerical resolution, point B is located near the termination of the partially polarized metal phase. More precisely, it corresponds to the point at which the lower critical field for the onset of spin polarization and the upper critical field for full polarization merge with the metal--insulator boundary.

The existence of point B further suggests the presence of a small spin gap in the unpolarized metallic phase, as the polarization remains pinned to $\zeta=0$ over a finite range of weak magnetic fields. However, we cannot exclude the possibility that this feature originates from finite-size effects. In particular, it remains possible that points B and C merge in the thermodynamic limit.

\section{Characterization of the Many-Body Phases}

To further clarify the metal--insulator transition and distinguish the competing many-body phases, we employ a set of complementary diagnostics. These include the ground-state energy, real-space density profiles, as well as density and spin correlation functions and the corresponding static structure factors. Taken together, these observables allow us to identify the energetic signatures of the transition and the accompanying evolution of charge and spin correlations.

\subsection{{Energy}}\label{Energy_Sec}

\begin{figure}[h]
\includegraphics[trim=0 0 0 0, clip,width=1\columnwidth]{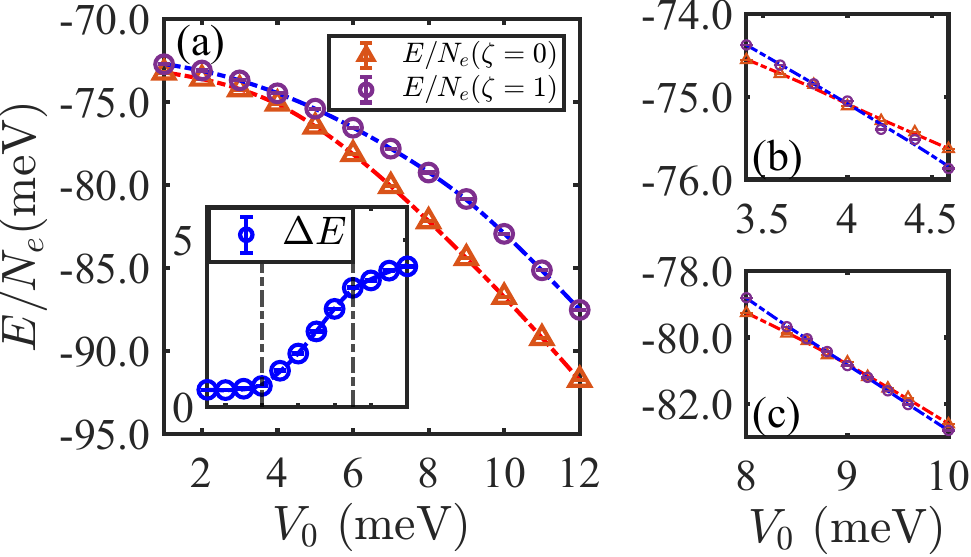}
\caption{
(a) Energy per particle $E/N_e$ in the $\zeta=0$ and $\zeta=1$ sectors as a function of $V_0$. In the inset, we show the energy difference $\Delta E=[E(\zeta=1)-E(\zeta=0)]/N_e$. The dot-dashed lines in the inset represent the metal--insulator transition points in the $\zeta=0$ and $\zeta=1$ sectors, respectively. (b,c) Lowest energies obtained from two-sided annealing for (b) $\zeta=0$ and (c) $\zeta=1$. Red (blue) lines denote the insulating (metallic) variational branches. 
The results shown here are obtained for the Coulomb interaction.
}
\label{E_v0}
\end{figure}

Fig.~\ref{E_v0}(a) shows the energy per particle $E/N_e$ as a function of $V_0$ at fixed spin polarization $\zeta$ for the Coulomb interaction. 
Over the entire range of $V_0$ investigated, the unpolarized state ($\zeta=0$) remains lower in energy than the fully polarized state ($\zeta=1$). The corresponding energy splitting $\Delta E$, defined in Eq.~\eqref{Eq:DeltaE}, increases monotonically with $V_0$ and exhibits three distinct regimes. At small $V_0$, $\Delta E$ depends only weakly on $V_0$ while both the unpolarized and fully polarized states remain metallic. It then increases approximately linearly once the unpolarized state becomes insulating, whereas the fully polarized state remains metallic. Finally, the linear dependence weakens again after the fully polarized state also undergoes a metal--insulator transition.

Each $E/N_e$--$V_0$ curve exhibits a characteristic cusp at a value $V_0^*$, which shifts to larger $V_0$ as the spin polarization increases. This behavior is reflected in the energy splitting $\Delta E$, defined in Eq.~\eqref{Eq:DeltaE}, shown as the blue dot-dashed line in the inset of Fig.~\ref{E_v0}(a). While $\Delta E$ increases monotonically with $V_0$, its slope displays two distinct regimes. For $V_0 \lesssim 4\,\mathrm{meV}$ and $V_0 \gtrsim 9\,\mathrm{meV}$, $\Delta E$ varies only gradually because the $\zeta=0$ and $\zeta=1$ states belong to the same phase, either metallic or insulating. By contrast, in the intermediate region $4\,\mathrm{meV} \lesssim V_0 \lesssim 9\,\mathrm{meV}$, $\Delta E$ grows significantly faster. This enhanced variation arises because the unpolarized state has already undergone the metal--insulator transition while the fully polarized state remains metallic, resulting in markedly different responses to the increasing moir\'e potential. The changes in the slope of $\Delta E(V_0)$ therefore provide a first indication of the distinct transition points in the two spin sectors.

To accurately determine the critical values $V_0^*$ and characterize the nature of the transition, we employ an annealing strategy. Specifically, we approach the transition from small $V_0$ using a variational optimization initialized in a metallic state and from large $V_0$ using an optimization initialized in an insulating state, for both the unpolarized and fully polarized sectors; see Figs.~\ref{E_v0}(b) and \ref{E_v0}(c). This procedure reveals a clear crossing between the energies of the two variational ans\"atze, allowing us to unambiguously determine the metal--insulator transition point within each spin-polarization sector and establishing the transition as first order.

The overall decrease of the energy with increasing $V_0$ follows directly from the Hellmann--Feynman theorem, as discussed in Appendix~\ref{Hellmann_Feynman}. Beyond this monotonic decrease, the energy is lowered progressively more rapidly as the moir\'e potential becomes stronger. This downward curvature can be understood from Jensen's inequality. As proved in Appendix~\ref{Hellmann_Feynman}, for any two moir\'e potential strengths $V_1$ and $V_2$ and $\lambda\in[0,1]$, the fixed-polarization ground-state energy satisfies
\begin{equation}
E[\zeta;\lambda V_1+(1-\lambda)V_2]
\geq
\lambda E(\zeta;V_1)+(1-\lambda)E(\zeta;V_2),
\end{equation}
where $E(\zeta;V_0)$ denotes the ground state energy within the fixed-$\zeta$ sector for a Hamiltonian with moir\'e potential strength $V_0$. This inequality shows that $E(\zeta;V_0)$ is a concave function of $V_0$. Combined with $\partial E/\partial V_0<0$, this concavity implies that the slope becomes increasingly negative with increasing $V_0$, reflecting a stronger sensitivity of the ground-state energy to the moir\'e potential in the large-$V_0$ regime.

\subsection{{Spatial structure across the metal--insulator transition}}
\begin{figure}[h]
\includegraphics[trim=0 0 0 0, clip, width=1\linewidth]
{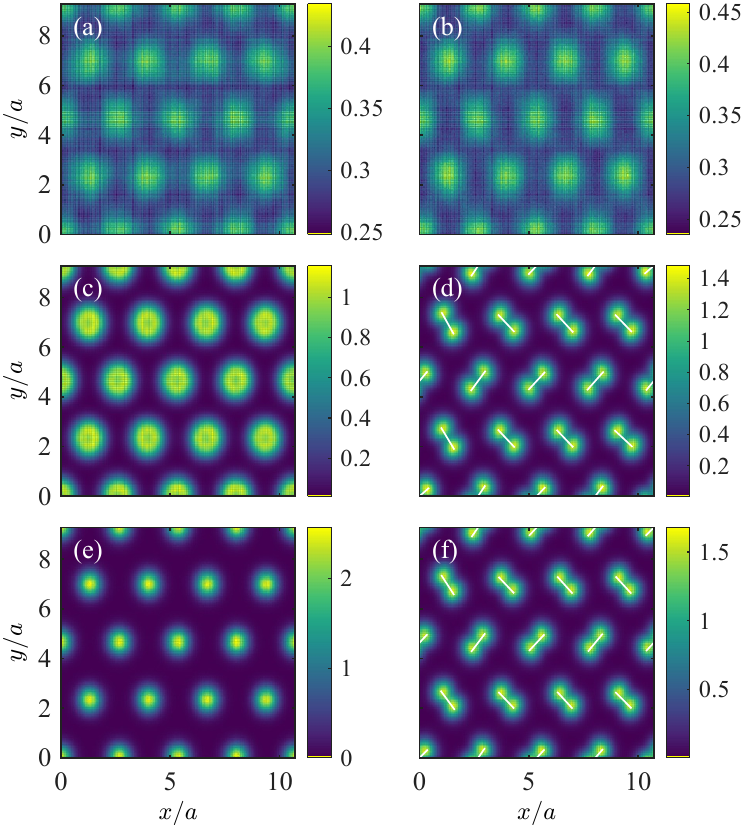}
\caption{
Real-space density profiles in the $\zeta=0$ and $\zeta=1$ sectors.
Panels (a)--(d) show the results for the Coulomb interaction:
(a) $\zeta=0$, $V_0=1\,\mathrm{meV}$;
(b) $\zeta=1$, $V_0=1\,\mathrm{meV}$;
(c) $\zeta=0$, $V_0=11\,\mathrm{meV}$;
(d) $\zeta=1$, $V_0=11\,\mathrm{meV}$.
Panels (e) and (f) show the corresponding results for the P\"oschl--Teller
(PT) potential with $V_{\rm PT}=5$ and $\kappa^{-1}=a$:
(e) $\zeta=0$, $V_0=11\,\mathrm{meV}$;
(f) $\zeta=1$, $V_0=11\,\mathrm{meV}$.
For $V_0=1\,\mathrm{meV}$, the density profiles obtained with the PT
potential are similar to those obtained with the Coulomb interaction and are
therefore not shown. In panels (d) and (f), the quadrupole distribution is also indicated by short white line segments.
}
\label{density_fig}
\end{figure}

\begin{figure}[h]
\includegraphics[trim=0 0 0 0, clip,width=0.98\columnwidth]{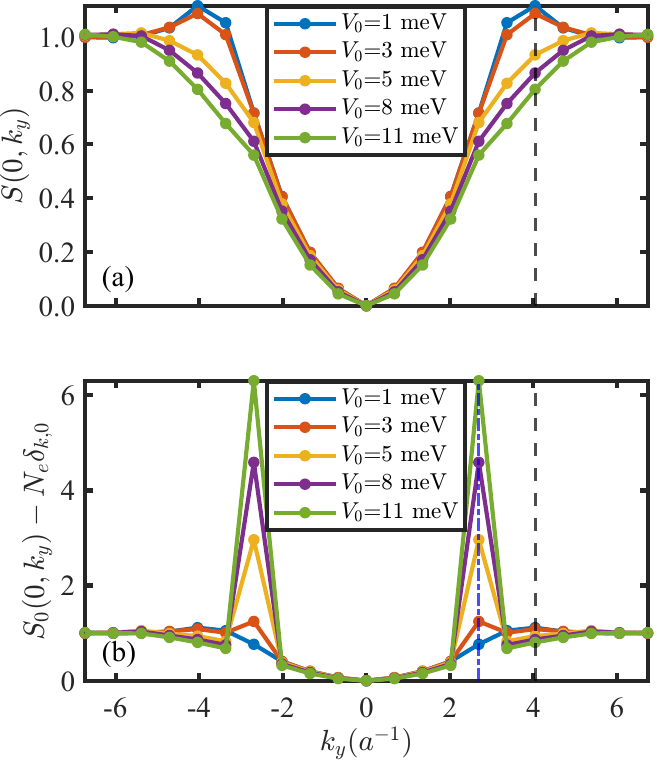}
\caption{
(a) Cut of the connected structure factor $S(0,k_y)$ as a function of $V_0$ in the unpolarized sector. (b) Corresponding cut of the bare structure factor $S_0(0,k_y)- N_e\delta_{k_y,0}$, with the trivial zero-momentum
peak subtracted. The results are obtained for the Coulomb interaction at $\zeta=0$.
}
\label{Sk_Sk0}
\end{figure}

To further elucidate the nature of the metallic and insulating phases, we examine the real-space density profile $n(\mathbf{r})$ together with the static structure factors, see Eqs.~\eqref{density_formula} and~\eqref{Eq:S0k}. The static structure factor $S_0(\bm k)$ contains information about both density fluctuations and the static density modulation. The latter reflects the Fourier components of the average density profile and gives rise to Bragg peaks whenever long-range density order is present. Figures~\ref{density_fig} and~\ref{Sk_Sk0} show typical density profiles $n(\mathbf{r})$ and bare structure factors $S_0(\bm k)$ for the many-body phases encountered in this work.

For weak potentials, the density profile in both the unpolarized and fully polarized sectors exhibits only a modest modulation imposed by the underlying potential. Although the density maxima are aligned with the potential minima, the density remains finite throughout the system, with no significant depletion between neighboring sites. Correspondingly, the bare structure factor displays only weak intensity at the reciprocal lattice vectors $\bm G_M$, indicating that long-range crystalline density order has not yet developed. Additionally, a weak and broad peak appears near $ka\sim 4$ indicating the formation of short-range density correlations, consistent with previous Monte Carlo results for the Fermi-liquid regime~\cite{Ceperley1}. This broad peak appears more prominently in the connected structure factor, see Fig.~\ref{Sk_Sk0}.

\begin{figure}[h]
\includegraphics[width=1.0\columnwidth]{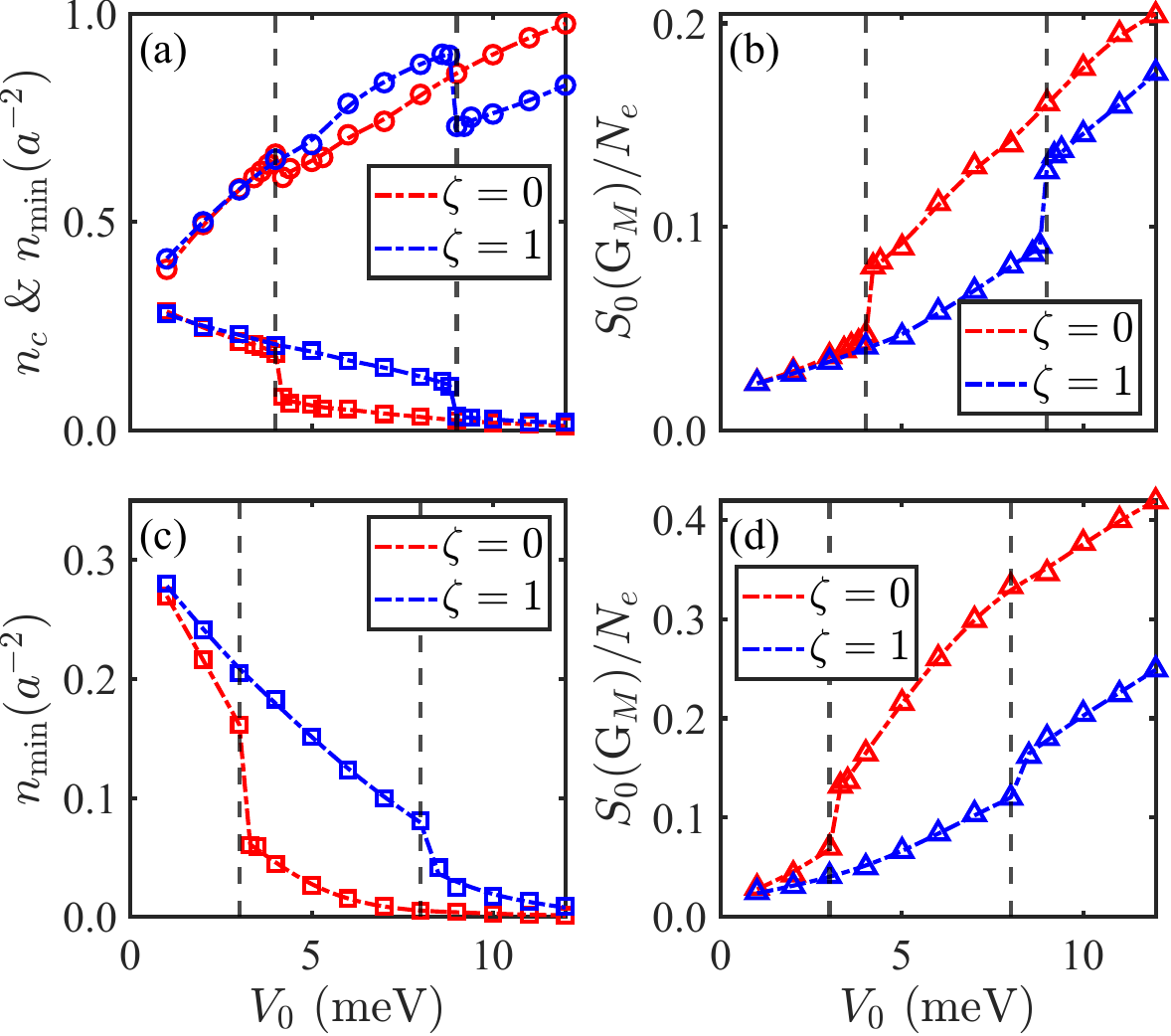}
\caption{
Density modulation as a function of the moir\'e potential strength $V_0$.
(a) Density at the center of a moir\'e-potential minimum and the minimum
density in the $\zeta=0$ and $\zeta=1$ sectors for the Coulomb interaction.
(b) The bare structure factor at the moir\'e reciprocal
lattice vector $\bm{G}_M$ for the Coulomb interaction.
(c) Minimum density $n_{\min}$ in the $\zeta=0$ and $\zeta=1$ sectors for the P\"oschl--Teller interaction with $V_{\rm PT}=5$ and $\kappa^{-1}=a$. In this case, the reduction of the central density $n_c$ across the transition is less pronounced, since the $p$-wave molecular character is not as clearly resolved. Moreover, $n_c$ remains much larger than $n_{\min}$, and is therefore not shown in panel (c).
(d) The bare structure factor at $\bm{G}_M$ for the
P\"oschl--Teller interaction with $V_{\rm PT}=5$ and $\kappa^{-1}=a$.
}
\label{nc_nk}
\end{figure}

The evolution of the density profile provides a useful diagnostic of the metal--insulator transition. To quantify this behavior, we track the electron density at the center of a potential minimum, denoted by $n_c$, together with the minimum density within the simulation unit cell, $n_{\rm min}$, as functions of $V_0$; the results are shown in Fig.~\ref{nc_nk}(a) and~\ref{nc_nk}(c) for the Coulomb and PT interactions, respectively. For the Coulomb interaction, both quantities exhibit a pronounced reduction near the corresponding critical potential $V_0^*$ in both the unpolarized and fully polarized sectors. This behavior signals the progressive depletion of the continuous electronic background and the onset of electron localization within the potential wells. The critical values extracted from the evolution of $n_c$ and $n_{\rm min}$ are in excellent agreement with those obtained from the energy analysis, demonstrating that these real-space observables provide sensitive diagnostics of the metal--insulator transition.
For the PT interaction, $n_{\rm min}$ shows a similar discontinuous decrease across the transition, as shown in Fig.~\ref{nc_nk}(c). 
The evolution of $n_c$ is more complicated than in the Coulomb case, owing to the finite value of the PT potential at $r=0$. In this case, the overlap of two opposite-spin electrons at the molecular center is not strongly penalized, leading to a less straightforward behavior of $n_c$. We thus defer the corresponding results and a more detailed discussion to Fig.~\ref{nc_E_PT_fig} (a) in Appendix~\ref{app:nc_E_PT}.

A complementary momentum-space diagnostic is the intensity of the bare structure factor at the principal reciprocal lattice vector, $S_0(\bm G_M)$, shown in Figs.~\ref{nc_nk}(b) and \ref{nc_nk}(d) for the Coulomb and PT interactions, respectively. In both spin sectors and for both interaction models, $S_0(\bm G_M)$ is strongly enhanced upon crossing the transition. This enhancement reflects the rapid growth of the static density modulation associated with electron localization and the emergence of long-range crystalline order pinned by the periodic potential.
For the PT interaction, $S_0(\bm G_M)$ reaches larger values than in the Coulomb case. This is consistent with the more strongly localized real-space density profile: the finite short-range interaction allows the electrons to concentrate more sharply near the moir\'e minima, producing a larger density contrast between the potential wells and the interstitial regions.

The larger critical potential $V_0^*$ for the metal--insulator transition in the fully polarized sector can be understood from both exchange symmetry and the underlying single-particle band structure. For electrons with identical spins, the Pauli principle suppresses formation of a localized two-electron molecular state. Therefore, electron localization requires a deeper confining potential.
This tendency is further reinforced by the noninteracting band structure. As shown in Appendix~\ref{App:DOS}, the second and third Bloch bands touch at symmetry-related momenta, producing a semimetal-like low-energy spectrum, whereas the unpolarized state is obtained by filling the isolated lowest Bloch band. As a result, the fully polarized metallic phase is more robust against weak periodic potentials. A many-body gap can emerge only when the potential becomes sufficiently strong to localize electrons within the same well.

For the PT interaction parameters used in the main text, the transition occurs at a smaller $V_0^*$ compared to the system with Coulomb interaction. However, we observe that the critical $V_0^*$ for the PT interaction is strongly dependent on the range of the interaction. In the section~\ref{Sec:PT_kappa}, we present additional results for different values of $\kappa$ and discuss how the interaction range affects the insulator transition.

\subsection{Nematic properties of insulating states}

\begin{figure}[h]
\includegraphics[width=1\columnwidth]{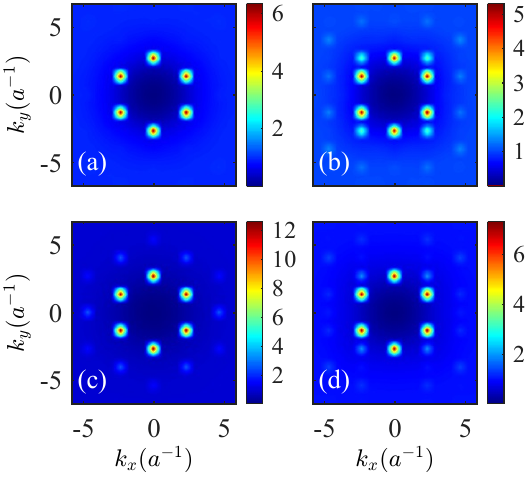}
\caption{Bare structure factors $S_0(\bm{k})-N_e\delta_{\bm{k},\bm{0}}$ at $V_0=11\,\mathrm{meV}$.
(a,b) Results for the Coulomb interaction with (a) $\zeta=0$ and (b) $\zeta=1$.
(c,d) Results for the P\"oschl--Teller interaction with $V_{\rm PT}=5$ and $\kappa^{-1}=a$: (c) $\zeta=0$ and (d) $\zeta=1$.}
\label{Sk0_vs_V0}
\end{figure}

For the unpolarized insulating state we observe two electrons with opposite spin localized near each potential minimum forming a molecular bound state. These molecular states preserve the original $\mathcal{C}_6$ symmetry of the periodic potential for both long-range Coulomb and short-range PT interactions. Consistent with this symmetry-preserving localization, the bare structure factor develops dominant Bragg peaks at the reciprocal lattice vectors $\bm G_M$.

\begin{figure}[!t]
\centering
\includegraphics[width=1.0\columnwidth]{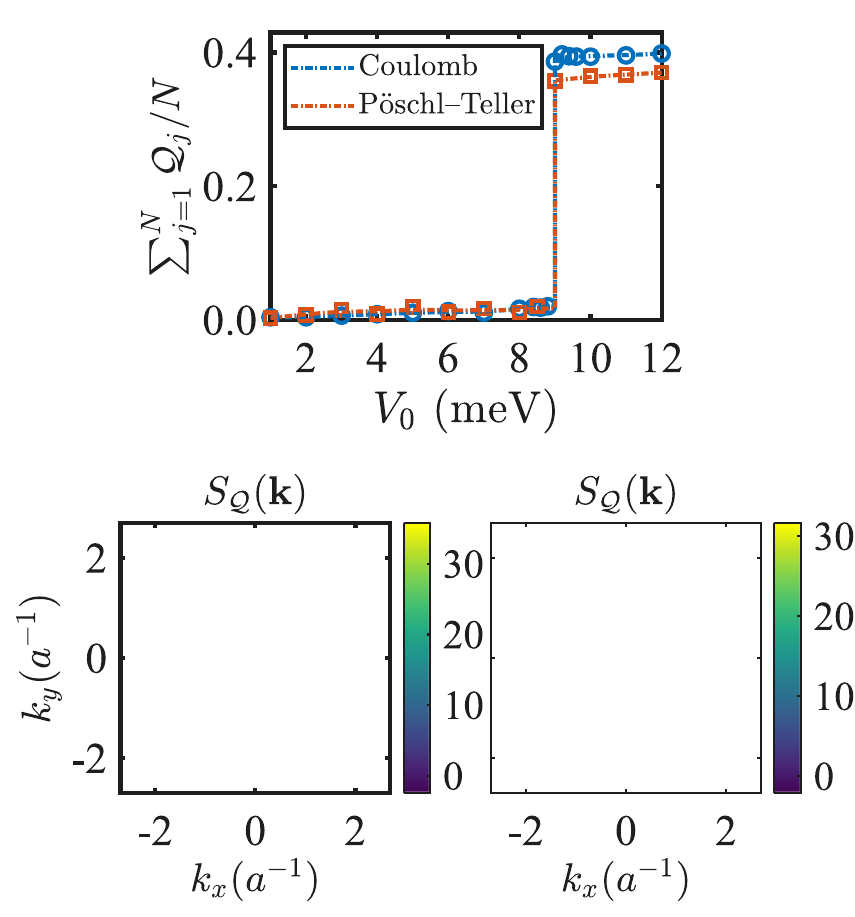}
\caption{
(a) Quadrupole order parameter per moir\'e minimum as a function of the moir\'e potential strength $V_0$ for the Coulomb and P\"oschl--Teller ($V_{\rm PT}=5$, $\kappa^{-1}=a$) interactions at $\zeta=1$.
(b,c) Quadrupole structure factors $S_{\mathcal{Q}}(\bm k)$ for the Coulomb and P\"oschl--Teller ($V_{\rm PT}=5$, $\kappa^{-1}=a$)  interactions at $V_0=11\,\mathrm{meV}$ and $\zeta=1$.
The white dashed lines mark the boundary of the first Brillouin zone, and the open white circles indicate the nearest reciprocal lattice vectors $\bm G_M$.}
\label{quadrupole_fig}
\end{figure}

The fully polarized sector exhibits markedly different behavior. The Pauli exclusion principle prevents two electrons with the same spin from occupying the same local orbital, and the localized states therefore acquire a dumbbell-like shape around each potential minimum. Remarkably, these molecular states spontaneously develop orientational order, with neighboring rows aligned approximately perpendicular to one another. This ordering lowers the rotational symmetry from $\mathcal C_6$ to $\mathcal{C}_2$, producing an orbital-nematic molecular crystal (ON-MC). The symmetry reduction is directly visible in the bare structure factor, see Fig.~\ref{Sk0_vs_V0}(b) and ~\ref{Sk0_vs_V0}(d). In addition to the Bragg peaks at the reciprocal lattice vectors, extra peaks appear near the four corners of a rectangle enclosed within the first Brillouin zone. These additional peaks reflect the enlarged symmetry-broken unit cell associated with the orientational ordering of the molecular states and provide a clear momentum-space signature of the nematic phase.
For the parameters considered in the main text, the fully polarized insulating state with the short-range PT interaction also forms dumbbell-like $p$-wave molecules with an alternating
orientational pattern. In the section~\ref{Sec:PT_kappa}, we provide calculations for a different force range $\kappa^{-1}$ and show that the orientational pattern of the molecules is modified. 
 
The additional peaks in the bare structure factor can be understood from the real-space arrangement of the $p$-wave molecules, as discussed in Appendix~\ref{app:ONMC_S0}. In the ON-MC phase, the peak structure of $S_0(\bm k)$ is controlled by the molecular orientations, or equivalently by the angles between the molecular directors and the $x$-axis. The staggered orientational texture generates additional peaks shifted from the original reciprocal lattice vectors by half of the fundamental reciprocal vector along $k_y$. The intensities of these peaks depend on the polar angle of $\mathbf{k}$ leading to an anisotropic pattern. Similar correlated electronic phases with orbital-nematic character have been discussed in a variety of strongly interacting systems~\cite{Kivelson1998,Fradkin2010,Borzi2007,Matty2022}.

To quantitatively characterize the local orientational order in the ON-MC phase, we introduce the local quadrupole order parameter $\bm{\mathcal{Q}}_j=\mathcal{Q}_j e^{2i\varphi_j}$ at each lattice site $\bm R_j$, see Eq.~\ref{Eq:quadrupole}, which captures the anisotropy of the local density profile. The norm $\mathcal{Q}_j$ measures the strength of the local anisotropy, while the phase $\varphi_j$ denotes the orientation angle of the quadrupole director. 

In Figs.~\ref{density_fig}(d) and \ref{density_fig}(f), we overlay the quadrupole distribution on the density maps using short white line segments. We further plot the average quadrupole norm as a function of $V_0$ for the two interactions in Fig.~\ref{quadrupole_fig}(a). Across the metal-insulator transition, the average quadrupole norm jumps from nearly zero to a finite value, signaling the formation of nematic insulators. Therefore, the quadrupole norm also provides a useful indicator of the transition into the ON-MC phase.

\begin{figure*}[t]
\centering
\includegraphics[
    trim=0 0 0 0,
    clip,
    width=\textwidth
]{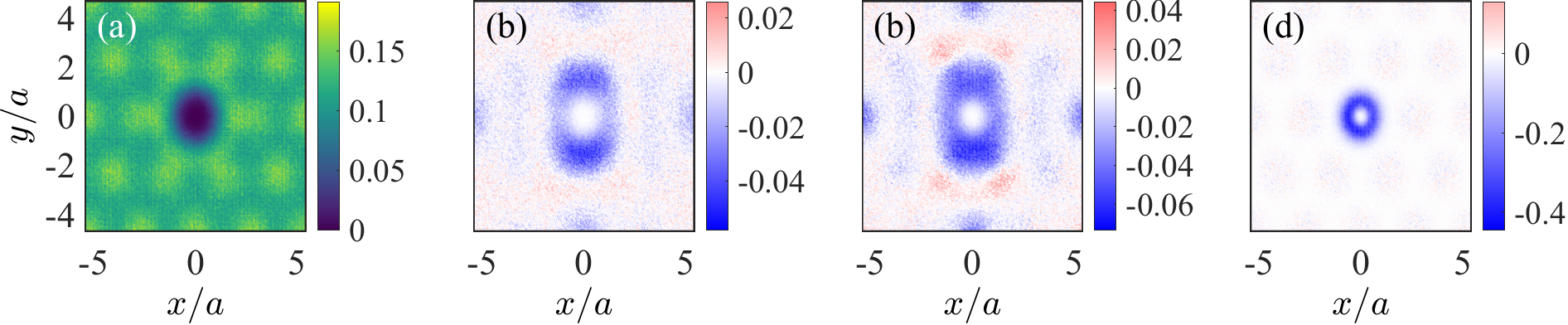}
\caption{
Real-space correlation functions for the Coulomb interaction.
(a) Density-density correlation function $C_n(0,0;x,y)$ for $\zeta=0$
and $V_0=1\,\mathrm{meV}$.
(b--d) Spin-spin correlation function $C_s(0,0;x,y)$ for $\zeta=0$ at
$V_0=1,3,11\,\mathrm{meV}$, respectively.
}
\label{bare_correlation}
\end{figure*}

The spatial coherence of the quadrupole orientations is characterized by the quadrupole structure factor
\begin{equation}
\begin{aligned}
S_{\mathcal{Q}}(\bm k)
&\equiv
\left|
\sum_j \mathcal{Q}_j e^{2i\varphi_j} e^{i\bm k\cdot\bm R_j}
\right|^2
-
\sum_j \mathcal{Q}_j^2
\\
&=
\sum_{j\neq l}
\mathcal{Q}_j\mathcal{Q}_l
\cos\!\left[
2(\varphi_j-\varphi_l)
+\bm k\cdot(\bm R_j-\bm R_l)
\right],
\end{aligned}
\end{equation}
with subtraction removing the on-site self-correlations.

In Figs.~\ref{quadrupole_fig}(b) and \ref{quadrupole_fig}(c), we show $S_{\mathcal{Q}}(\bm k)$ for the two interactions at $V_0=11\,\mathrm{meV}$. 
For both the Coulomb and PT interactions, the staggered orientational pattern
 shifts the dominant peak to $\bm k=\bm b_2/2$, reflecting the doubled unit cell of the orientational order. 

The emergence of the nematic molecular insulator can be understood in terms of an effective classical model, see Appendix~\ref{app:classical_ONMC} for a detailed derivation. In the strong-localization limit, each molecule is described by an orientation angle $\theta_i$ at site $\mathbf{R}_i$. Evaluating the interaction energy between neighboring molecules yields the effective Hamiltonian
\begin{equation}
H_{\rm eff}
=
\sum_{\langle ij\rangle}
\left[
J_1\cos\bigl(2(\theta_i-\theta_j)\bigr)
+
J_2\cos\bigl(2(\theta_i+\theta_j-2\phi_{ij})\bigr)
\right].
\label{E_J1_J2}
\end{equation}
Here, $\phi_{ij}$ denotes the angle of the bond connecting neighboring sites and $J_1$ and $J_2$ are effective interaction strengths. The resulting low-energy theory is a frustrated $XY$ compass model on the triangular lattice. As shown in Appendix~\ref{app:classical_ONMC}, minimizing this Hamiltonian reproduces the orientational pattern obtained in the NQS-VMC simulations.

The two interaction terms encode competing tendencies. The $J_1$ term is minimized by a continuously degenerate manifold of states, whereas the bond-dependent $J_2$ term lifts this degeneracy by coupling the molecular orientations to the underlying lattice directions. The competition between these interactions naturally raises the possibility of multiple finite-temperature phase transitions. In particular, the approximate $U(1)$ symmetry associated with the orientational degree of freedom suggests a Berezinskii--Kosterlitz--Thouless transition at higher temperatures, followed by a lower-temperature transition into the lattice-pinned ordered state. Exploring this rich finite-temperature phase diagram and understanding the role of quantum fluctuations is an interesting direction for future work.

More generally, the effective model indicates that the ratio $J_1/J_2$ is controlled by the range of the microscopic interaction. In two-dimensional materials, this ratio may therefore be tunable by varying the screening environment, for example through the distance to nearby metallic gates, providing an experimental route to engineer different regimes of the frustrated compass model.

\subsection{Magnetic structure}
Although the correlated metallic phase exhibits signatures of short-range density correlations, they respect the symmetry of the periodic potential. The density-density correlation function exhibits a hole in the center and a gear-like structure connecting nearest-neighbor potential minima, see Fig.~\ref{bare_correlation}(a), showing the presence of a correlated metallic background.

More interesting are the spin-$z$ correlation function $C_s$ of the
associated metallic state and its dimensionless counterpart, both of which
display pronounced anisotropic features [see Fig.~\ref{bare_correlation}
for $C_s$ and Appendix~\ref{DimensionlessC} for its dimensionless
counterpart].
This behavior closely resembles the enhanced short-range nematic spin correlations recently reported in variational studies of the two-dimensional electron gas~\cite{ShiweiZhang}. Notably, the interaction strength in our system is $r_s\approx 11.13$, which lies within the nematic spin-correlated liquid (NSCL) regime identified for the interacting electron gas in the absence of a periodic potential~\cite{ShiweiZhang}. These observations suggest that, for weak potentials, the metallic state retains NSCL-like spin correlations.

We next increase the moir\'e potential to $V_0=3\,\mathrm{meV}$, which is closer to the transition point while still lying on the metallic side. At this strength, the peaks in the density--density correlation function $C_n$ become more pronounced, whereas the gear-like structure observed at weaker potential is substantially suppressed. Nevertheless, $C_n$ remains finite in the intermediate regions between neighboring maxima, indicating that the system has not yet developed the fully disconnected density pattern characteristic of a localized insulating state. In the spin sector, the correlation function begins to show a clear magnetic structure, see Fig.~\ref{bare_correlation}(c).

We propose that the observed magnetic structure arises from the interplay between localized electrons and the itinerant metallic background. Near the metal--insulator transition, the electronic density naturally separates into a localized component associated with electrons trapped near potential minima and an extended component that remains delocalized throughout the system. The localized electrons generate strong short-range antiferromagnetic correlations, consistent with the formation of a local singlet. More intriguingly, pronounced magnetic correlations persist at larger distances and exhibit spatial oscillations, which we attribute to an effective interaction mediated by the itinerant electrons, akin to a Ruderman–Kittel–Kasuya–Yosida (RKKY) exchange mechanism.

As the system is driven deeper into the insulating regime, the itinerant component becomes progressively suppressed, reducing the effectiveness of the long-range exchange pathways. Correspondingly, the spin correlation function $C_s$ rapidly decays away from the potential minimum.

\section{Short-range interactions and role of interaction range}\label{Sec:PT_kappa}

\begin{figure}[!t]
\centering
\includegraphics[width=1.0\columnwidth]{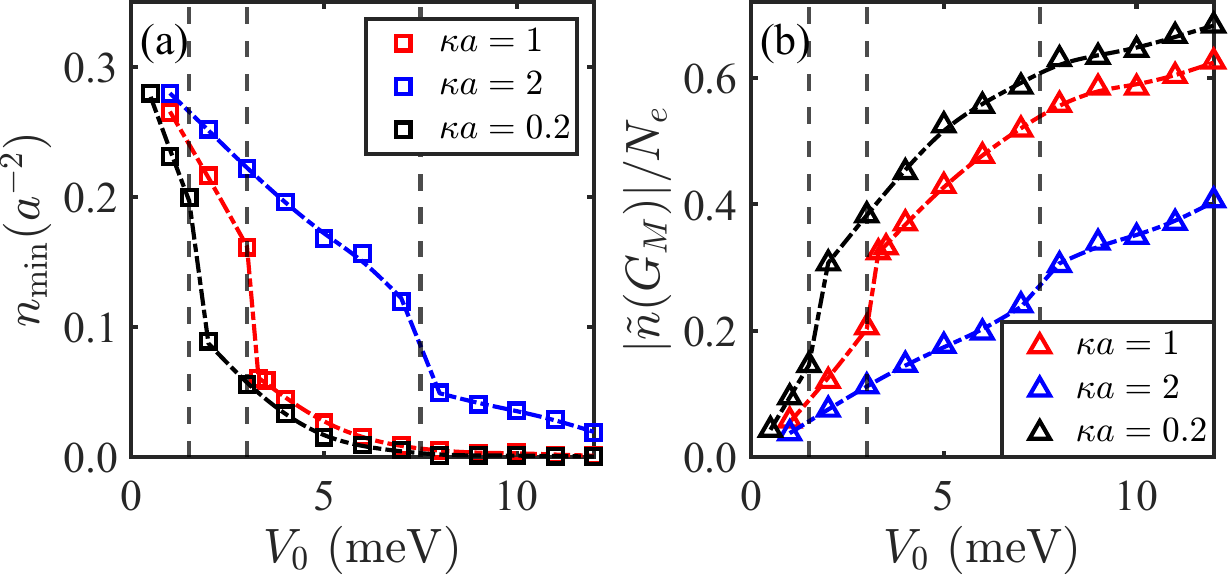}
\caption{
Dependence of the metal--insulator transition on the range parameter
$\kappa$ of the P\"oschl--Teller interaction at fixed $V_{\mathrm{PT}}=5$
and $\zeta=0$.
(a) Minimum density $n_{\rm min}$ as a function of $V_0$.
(b) Density Fourier component $\tilde n(\bm G_M)$ as a function of $V_0$.
The discontinuous decrease of $n_{\rm min}$ and the abrupt enhancement of
$\tilde n(\bm G_M)$ are used to identify the transition point.
The critical potential increases with $\kappa$, indicating that a shorter-range
PT interaction stabilizes the metallic liquid phase over a wider range of
moir\'e potential strengths.
}
\label{PT_kappa_fig}
\end{figure}

The qualitative properties of the many-body phases and their transitions are not specific to long-range Coulomb interactions, but persist for the short-range PT interaction in the parameter regimes studied here. As shown in
Figs.~\ref{nc_nk}, \ref{Sk0_vs_V0}, and \ref{quadrupole_fig}, and discussed in
the previous sections, the PT interaction gives rise to an analogous
metal--insulator transition and molecular-crystal phases characterized by the
same patterns of spatial and rotational symmetry breaking.

To further investigate the role of the interaction range, we study the
dependence of the metal--insulator transition on the parameter $\kappa a$. In
Fig.~\ref{PT_kappa_fig}, we fix $V_{\mathrm{PT}}=5$ and compare
$\kappa a=0.2$, $1$, and $2$ in the unpolarized sector, $\zeta=0$. The
transition is identified from the sharp change in the minimum density
$n_{\rm min}$ and the abrupt enhancement of the density Fourier component
$\tilde n(\bm G_M)$. Across the values considered here, the critical
moir\'e potential increases monotonically with $\kappa$, showing that a
stronger external periodic potential is required to localize the electrons
when the interaction becomes shorter-ranged.

This dependence on $\kappa$ is particularly notable because the integrated
interaction strength is independent of $\kappa$ in the present
parametrization:
\begin{equation}
    \int d^2r\,V(\bm r)
    =
    \widetilde V(\bm q=0)
    =
    \frac{\hbar^2}{m^*}V_{\mathrm{PT}}\,2\pi\ln 2 .
\end{equation}
Nevertheless, the finite-momentum components of the interaction remain
strongly dependent on $\kappa$. In particular, increasing $\kappa$ makes the
interaction more localized in real space and correspondingly broadens its
momentum-space profile, enhancing its weight at momenta of the order of the
moir\'e reciprocal lattice vectors. The results in
Fig.~\ref{PT_kappa_fig} therefore show that the transition is controlled not
only by the integrated interaction strength, but also by the spatial range
and finite-momentum structure of the interaction.

A simple interpretation follows from the competition between the interaction
and the moir\'e-induced density modulation. For small $\kappa$, the
interaction varies only weakly over distances comparable to the moir\'e
lattice spacing, and its Fourier weight is concentrated near zero momentum.
Consequently, it only weakly penalizes density modulations at $\bm G_M$,
allowing the electrons to localize at relatively small $V_0$. For larger
$\kappa$, the broader momentum-space interaction has greater weight at
$\bm G_M$ and at higher reciprocal-lattice harmonics. It therefore imposes a
larger interaction-energy cost on the strongly modulated and spatially
localized density profile of the insulating state. A deeper moir\'e
potential is then required to overcome this cost, shifting the
metal--insulator transition to larger $V_0$, in agreement with the numerical
results shown in Fig.~\ref{PT_kappa_fig}.

\begin{figure}[!t]
\includegraphics[trim=0 0 0 0, clip,width=1.0\columnwidth]{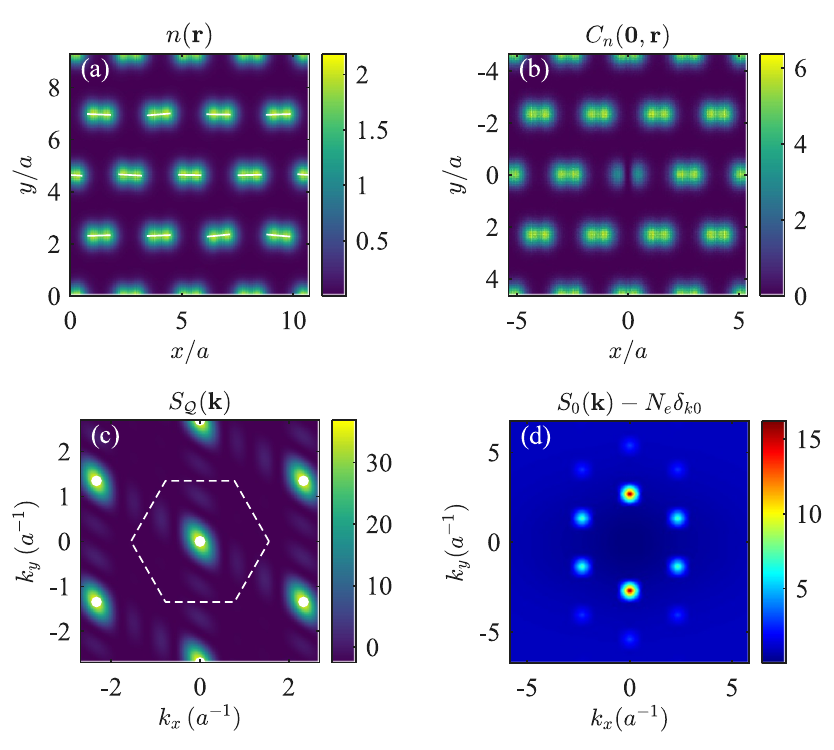}
\caption{For the PT interaction with $\kappa a=0.35$ and $V_{\mathrm{PT}}=40$, together with the moir\'e potential $V_0=12\,\mathrm{meV}$, the lowest-energy configuration in the fully polarized sector exhibits a uniform orientational pattern. Panels (a)--(d) show representative observables associated with this pattern: (a) the real-space density and local quadrupole distribution, (b) the density correlation function $C_n(\bm 0,\bm r)$, (c) the quadrupole structure factor $S_{\mathcal{Q}}(\bm k)$, and (d) the bare density structure factor $S_0(\bm k)-N_e\delta_{\bm k,0}$.
}
\label{quadrupole_parallel_fig}
\end{figure}

Having established the dependence of the metal--insulator transition on the
interaction range in the unpolarized sector, we next examine whether tuning
the interaction parameters can also stabilize a different orientational
ordering pattern within the ON-MC phase. The effective frustrated $XY$
compass model introduced in Eq.~\eqref{E_J1_J2} predicts that different
orientational configurations can be favored as the effective couplings are
varied. To test this prediction, we consider a different PT parameter set,
$\kappa a=0.35$ and $V_{\mathrm{PT}}=40$, together with a moir\'e potential
strength of $V_0=12\,\mathrm{meV}$. For a $4\times4$ system in the fully
polarized sector, the corresponding effective couplings place the system in
the uniformly aligned regime. This prediction is consistent with our NQS
many-body results, as illustrated by the real-space density distribution in
Fig.~\ref{quadrupole_parallel_fig}(a).
Figure~\ref{quadrupole_parallel_fig}(b) shows the corresponding density
correlation function. Although a few local molecular density profiles in
Fig.~\ref{quadrupole_parallel_fig}(a) are slightly tilted away from the
$x$ direction, the spatially averaged correlation function reveals a clear
global alignment along the $x$ axis. The averaging over different reference
positions suppresses local distortions and finite-size fluctuations while
retaining the dominant orientational order.
Figures~\ref{quadrupole_parallel_fig}(c) and
\ref{quadrupole_parallel_fig}(d) display the quadrupole structure factor and
the bare density structure factor, respectively. The nearly uniform
alignment of the molecular orbitals produces a dominant quadrupole peak at
$\bm k=0$, as expected for uniform orientational order. Meanwhile, the bare
density structure factor,
$S_0(\bm{k})-N_e\delta_{\bm{k},0}$, develops enhanced peaks along the $k_y$ direction, consistent with the phenomenological analysis of Appendix~\ref{app:ONMC_S0}, showing that the dominant peaks appear perpendicular to the real-space orientation of the molecular orbitals.

\section{Exciton absorption spectrum}
\label{sec:exciton-spectrum}
To identify possible experimental signatures of the ON--MC phase, we
calculate the exciton absorption spectrum within a static-density
approximation. Specifically, we replace the electronic density operator by
the ground-state density \(n(\bm r)\) obtained from the NQS calculation. The exciton then experiences the effective periodic potential
\begin{equation}
U_X(\bm R)
=
\int d^2r'\,
V_{e\text{-}X}(\bm R-\bm r')n(\bm r').
\end{equation}

Following Ref.~\cite{PhysRevX.15.011049}, we model the exciton--electron interaction as
\begin{equation}
V_{e\text{-}X}(r)
=
-\frac{W_0}{\left(a_X^2+r^2\right)^2}
+
P_0 e^{-r/a_X}.
\label{eq:exciton-electron-interaction}
\end{equation}
Here, \(W_0>0\) controls the attractive component of the interaction,
\(a_X>0\) sets its characteristic range, and \(P_0>0\) determines the strength
of the short-range effective repulsion. For the same-valley
exciton--electron channel, this repulsive contribution should be understood
as a phenomenological representation of the Pauli exclusion principle, which is not
captured explicitly when the exciton is treated as a structureless
quasiparticle. Importantly,
Eq.~\eqref{eq:exciton-electron-interaction} is rotationally symmetric and
therefore introduces no additional angular anisotropy. Consequently, all
symmetry-breaking features of the effective exciton potential originate
from the electronic density of the ON--MC phase.
Using the Fourier-transform convention $\widetilde V_{e\text{-}X}(\bm q)=\int d^2r\,e^{-i\bm q\cdot\bm r}V_{e\text{-}X}(r),$ and writing $n(\bm r)=\sum_{\bm G}\tilde n_{\bm G}e^{i\bm G\cdot\bm r}$, the Fourier components of the periodic mean-field potential are $U_{\bm G}=\widetilde V_{e\text{-}X}(G)\tilde n_{\bm G}$, where \(\bm G\) is a reciprocal vector of the enlarged ON--MC unit cell and \(G=|\bm G|\). 

We employ a phenomenological density ansatz that captures the essential
features of the ON--MC order, $n(\bm r)=\sum_j n_0(\bm r-\bm R_j;\bm R_j).$
Here,
\(\bm R_j=m_{jx}\bm a_1+m_{jy}\bm a_2\) denotes the position of the
\(j\)th moir\'e minimum, with
\(\bm a_1=a_M(1,0)\) and
\(\bm a_2=a_M\left(\frac{1}{2},\frac{\sqrt{3}}{2}\right)\).
The local density profile around each moir\'e minimum is modeled as a
superposition of an isotropic \(s\)-wave component and an orientational
\(p\)-wave component,
\begin{equation}
\begin{split}
n_0(\bm r;\bm R_j)
&={}
\eta_s\frac{2\alpha_s}{\pi}e^{-2\alpha_s r^2}\\
&+
\eta_p\frac{8\alpha_p^2}{\pi}
r^2e^{-2\alpha_p r^2}
\cos^2\left[
\varphi+\gamma(\bm R_j)
\right].
\end{split}
\label{eq:sp-density-ansatz}
\end{equation}
\(\varphi\) is the polar angle of \(\bm r\), while
\(\alpha_s^{-1/2}\) and \(\alpha_p^{-1/2}\) characterize the spatial
extents of the two density components. The normalization factors are chosen
such that the two terms integrate to \(\eta_s\) and \(\eta_p\),
respectively, with \(\eta_s+\eta_p=2\) at filling \(\nu=2\). The parameters
\(\eta_s\), \(\eta_p\), \(\alpha_s\), and \(\alpha_p\) are determined by
fitting Eq.~\eqref{eq:sp-density-ansatz} to the real-space electronic
density obtained from the NQS calculation. In the calculations presented
here, we use
\(\eta_s=0.6883\), \(\eta_p=1.3117\),
\(\alpha_s a_M^2=13.754\), and
\(\alpha_p a_M^2=12.118\).
For \(a_M=10.2\,\mathrm{nm}\), these values correspond to
\(\alpha_s\simeq0.1322\,\mathrm{nm}^{-2}\) and
\(\alpha_p\simeq0.1165\,\mathrm{nm}^{-2}\).
The quantities \(\eta_s\) and \(\eta_p\) should be understood as effective
weights of the two density components rather than occupations of individual
single-particle orbitals.
The local orientation is parameterized as
\begin{equation}
\gamma(\bm R_j)
=
\gamma_0+(-1)^{m_{jy}}\Delta.
\end{equation}
Within the convention adopted in Eq.~\eqref{eq:sp-density-ansatz}, the
geometric axis of the \(p\)-wave component is oriented along
\(-\gamma(\bm R_j)\) modulo \(\pi\). Because the orientation alternates
between neighboring moir\'e minima along the \(\bm a_2\) direction, the
ON--MC order doubles the unit cell in this direction.
The isotropic \(s\)-wave component is identical on the two sublattices and
therefore contributes only at reciprocal vectors of the original moir\'e
superlattice. By contrast, the staggered \(p\)-wave component also generates
folded Fourier components shifted by half of the corresponding moir\'e
reciprocal vector. These additional Fourier components provide scattering
channels that are absent in the undoubled density pattern and can therefore
produce additional Umklapp resonances associated with the formation of the
ON--MC phase.

We diagonalize the effective exciton Hamiltonian in the reciprocal-vector
basis,
\begin{equation}
H_{\bm G\bm G'}(\bm k)
=
\frac{\hbar^2|\bm k+\bm G|^2}{2M_X}
\delta_{\bm G\bm G'}
+
U_{\bm G-\bm G'},
\end{equation}
to obtain the exciton eigenenergies and eigenstates, where \(M_X\) is the
exciton center-of-mass mass. In the following calculations, we take
\(M_X=1.2m_e\). At normal incidence, the photon carries negligible in-plane
momentum and therefore couples directly only to the \(\bm G=\bm 0\)
component of the exciton wave function. The optical weight of the \(s\)th
eigenstate at \(\bm k=\bm 0\) is consequently $w_s=\left|c_{s,\bm 0}(\bm G=\bm 0)\right|^2,$ where \(c_{s,\bm 0}(\bm G)\) denotes the expansion coefficient of the \(|\bm G\rangle\) plane-wave component in the $s$th eigenstate of the effective exciton Hamiltonian at Bloch momentum \(\bm k=\bm 0\).
Introducing a phenomenological Lorentzian broadening, we calculate the
exciton absorption spectrum as
\begin{equation}
I(E)
=
\sum_s
w_s
\frac{\Gamma/\pi}
{\left[E-E_s(\bm 0)\right]^2+\Gamma^2},
\end{equation}
where \(E_s(\bm 0)\) and \(w_s\) denote the energy and optical weight of
the \(s\)th exciton eigenstate at \(\bm k=\bm 0\), respectively, and
\(\Gamma\) is the phenomenological half-width at half-maximum. Further
details of the mean-field construction and the derivation of the optical
matrix elements are provided in Appendix~\ref{app:exciton-absorption}.

The periodic mean-field potential mixes the optically bright bare
\(\bm G=\bm 0\) exciton with finite-momentum exciton states, thereby
redistributing its optical weight among the resulting eigenstates.
Consequently, states dominated by finite-\(\bm G\) components can acquire
a nonzero \(\bm G=\bm 0\) component and appear as Umklapp resonances in
the absorption spectrum. In the ON--MC phase, the anisotropic molecular
density and the enlarged unit cell generate additional Fourier components
that couple the original moir\'e reciprocal-vector sector to the
half-shifted sector. This coupling produces additional Umklapp resonances
associated with the ON--MC order.

\begin{figure}[h]
\centering
\includegraphics[
    trim=0 0 0 0,
    clip,
    width=\columnwidth
]{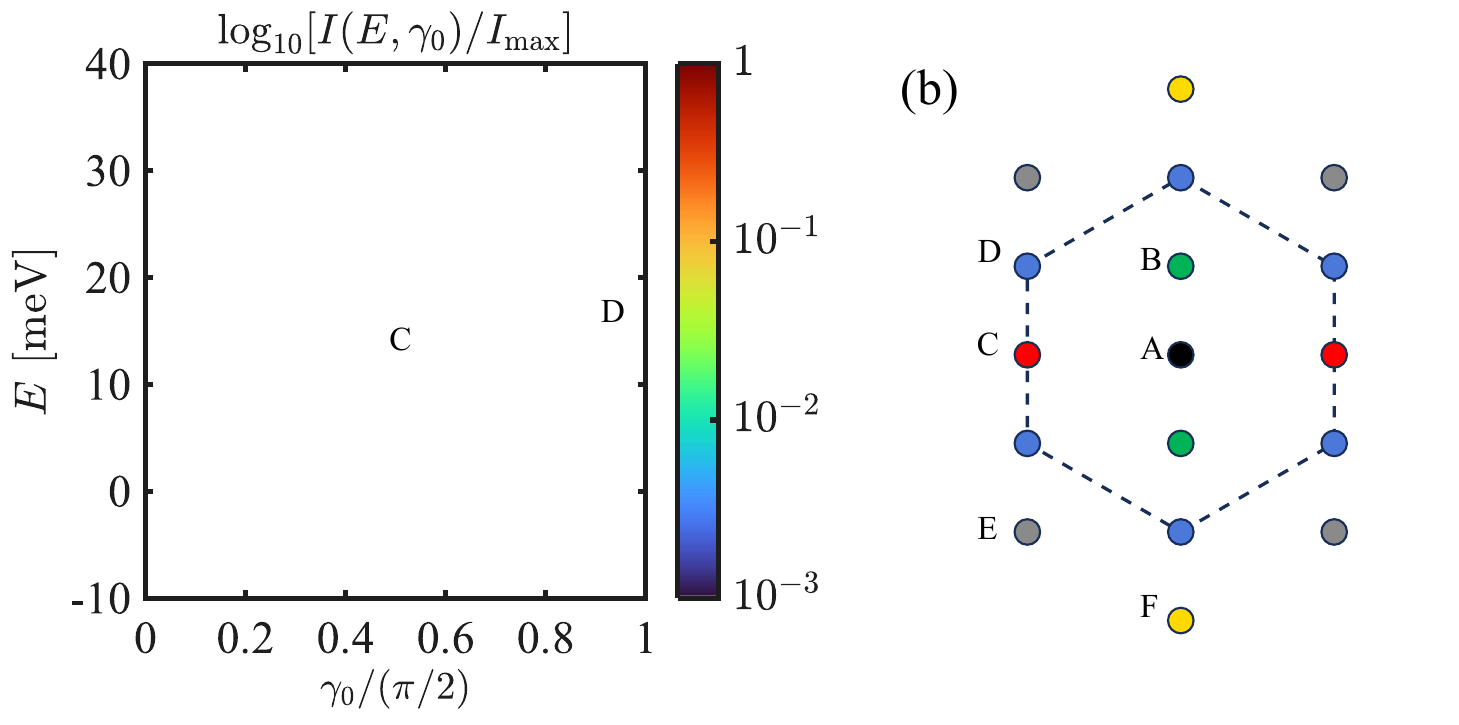}
\caption{
Exciton absorption spectrum generated by the static mean-field potential of the ON--MC electronic density.
We use the rotationally symmetric exciton--electron interaction
$V_{e-X}(r)=-W_0/(a_X^2+r^2)^2+P_0e^{-r/a_X}$,
with $W_0=296.7~{\rm meV\,nm^4}$, $P_0=114~{\rm meV}$, and
$a_X=1.1~{\rm nm}$, and include a Lorentzian broadening with half width at
half maximum $\Gamma=1.0~{\rm meV}$.
(a) Exciton absorption as a function of energy $E$ and the mean orientation
angle $\gamma_0$, with the staggered canting angle fixed at
$\Delta=\pi/4$. The intensity is shown on a logarithmic scale as
$\log_{10}[I(E,\gamma_0)/I_{\rm max}]$, where $I_{\rm max}$ is the global
maximum over the entire two-dimensional spectrum. The spectrum is symmetric
about $\gamma_0=\pi/4$, since the density configurations at $\gamma_0$ and
$\pi/2-\gamma_0$ are related by a mirror transformation.
(b) Reciprocal-space momenta associated with the main spectral features.
The labels A--F denote the momentum shells
$|\bm G|=0$, $G_M/2$, $\sqrt{3}G_M/2$, $G_M$,
$\sqrt{7}G_M/2$, and $3G_M/2$, respectively.
The dashed hexagon indicates the first moir\'e Brillouin zone.
The corresponding labels in panel (a) identify the visible absorption
resonances. The B feature associated with $|\bm G|=G_M/2$ is strongly
suppressed for the present interaction parameters.
}
\label{exciton_scan}
\end{figure}

We consider a parameter regime close to the exciton--electron binding
threshold, taking \(W_0=296.7~{\rm meV\,nm^4}\),
\(P_0=114~{\rm meV}\), and \(a_X=1.1~{\rm nm}\). For fixed \(W_0\)
and \(a_X\), this value of \(P_0\) lies slightly above the critical
repulsive strength at which the isolated exciton--electron bound state
disappears. Figure~\ref{exciton_scan}(a) shows the absorption spectrum
as a function of the mean orientation angle \(\gamma_0\) at fixed
staggered canting angle \(\Delta=\pi/4\). Because the Umklapp
resonances are much weaker than the dominant exciton peak, the spectrum
is displayed on a logarithmic scale,
\(\log_{10}[I(E,\gamma_0)/I_{\rm max}]\). Here we define a single global
maximum, $I_{\rm max}=\max_{E,\gamma_0} I(E,\gamma_0)$,
over the entire two-dimensional spectrum, so that all values of
\(\gamma_0\) are normalized by the same reference intensity.
Figure~\ref{exciton_scan}(b) shows the reciprocal-space momenta
associated with the main spectral features. The labeled momentum shells
allow the higher-energy resonances in panel (a) to be identified with
specific reciprocal vectors of the ON--MC density.

The dominant low-energy resonance, labeled \(A\), is continuously connected
to the bare \(\bm G=\bm 0\) exciton and retains most of its oscillator
strength. The net positive spatial average of the exciton--electron
interaction produces an overall blueshift of approximately
\(1.58~{\rm meV}\) relative to the bare exciton energy. The higher-energy
features \(B\)--\(F\) are associated predominantly with finite-\(\bm G\)
plane-wave sectors belonging to different reciprocal-momentum shells.
For the present interaction parameters, the \(B\) resonance associated
with \(|\bm G|=G_M/2\) is strongly suppressed and is not clearly resolved
in the absorption spectrum, whereas the higher-momentum resonances
\(C\)--\(F\) remain visible.

Varying \(\gamma_0\) modifies the Fourier components of the ON--MC electronic density and mainly affects the oscillator strengths of the Umklapp resonances. The corresponding energy shifts are much less pronounced because the exciton--electron interaction is weak in the present parameter regime. The spectrum is symmetric about
\(\gamma_0=\pi/4\), because the molecular configurations at
\(\gamma_0\) and \(\pi/2-\gamma_0\) are related by a mirror
transformation. The systematic redistribution of spectral weight among
the weak Umklapp resonances as \(\gamma_0\) is varied therefore provides
a potential optical signature of the orientational order in the ON--MC
phase.

To identify the momentum-space origin of the Umklapp resonances, we
decompose each exciton eigenstate into reciprocal-vector shells. The
primitive reciprocal vectors of the triangular moir\'e lattice are
\begin{equation}
\bm b_1
=
G_M\left(\frac{\sqrt{3}}{2},-\frac{1}{2}\right),
\quad
\bm b_2=G_M(0,1),
\quad
G_M=\frac{4\pi}{\sqrt{3}a_M}.
\end{equation}
Because the ON--MC order doubles the unit cell along \(\bm a_2\), the
reciprocal vectors of the enlarged cell are $\bm G_{mn}=m\bm b_1+\frac{n}{2}\bm b_2.$
Even \(n\) corresponds to the original moir\'e reciprocal lattice,
whereas odd \(n\) defines the half-shifted sector generated by the
unit-cell doubling.

For the \(\gamma_0\) scan in Fig.~\ref{exciton_scan}(a), the higher-energy
resonances can be associated with the reciprocal-momentum shells shown in
Fig.~\ref{exciton_scan}(b). At \(\gamma_0=\pi/2\), the visible resonances
\(D\) and \(E\) are dominated by the
\(|\bm G|=G_M\) and \(|\bm G|=\sqrt{7}G_M/2\) shells, respectively.
Their energy separations from the dominant resonance, approximately
\(15.1\) and \(28.5~{\rm meV}\), are close to the corresponding
free-exciton kinetic energies.
As \(\gamma_0\) approaches \(\pi/4\), the additional resonances \(C\) and
\(F\) become visible. They originate predominantly from the
\(|\bm G|=\sqrt{3}G_M/2\) and \(|\bm G|=3G_M/2\) half-shifted shells,
with energy separations of approximately \(12.7\) and
\(37.0~{\rm meV}\), respectively. These half-shifted Fourier components
are suppressed at \(\gamma_0=\pi/2\) by the angular structure of the
ON--MC density. The \(B\) resonance associated with
\(|\bm G|=G_M/2\) remains too weak to be clearly resolved for the
present interaction parameters.

The relative intensities of these Umklapp resonances range from
approximately \(0.5\%\) to \(4\%\) of the dominant exciton peak,
suggesting that they may be experimentally resolvable. Different ON--MC
ordering patterns produce distinct sets of Umklapp resonances, with
characteristic energies and spectral weights. Exciton absorption
spectroscopy can therefore provide a direct probe of the underlying
ON--MC orientational order.

\section{ Discussion and Outlook}
\label{sec:discussion}

We have studied interacting continuum electrons in a triangular moir\'e
potential at filling $\nu=2$ using variational Monte Carlo with a
message-passing neural-network quantum state. The main result is a
spin-selective molecular crystallization transition. The unpolarized system
forms an $s$-wave molecular crystal at relatively weak moir\'e potential,
whereas the fully polarized system remains metallic over a wider range and
crystallizes only at stronger confinement into an orbital-nematic molecular
crystal with reduced spatial symmetry.

This result places the moir\'e metal--insulator transition between two
familiar limits. It is not the homogeneous Wigner-crystallization transition,
because the external moir\'e potential explicitly pins the charge density and
drives a first-order transition at a density where the corresponding uniform
electron gas remains metallic. It is also not a conventional single-orbital
Mott transition, because the relevant local degree of freedom at $\nu=2$ is a
two-electron molecular state whose internal orbital symmetry depends on spin.
Opposite-spin electrons can occupy a spatially symmetric $s$-wave molecular
configuration, while Fermi statistics require fully polarized electrons to
have an antisymmetric spatial wave function, making the lowest local molecular
configuration $p$-wave-like.
 This molecular constraint shifts the crystallization
threshold, produces discontinuous magnetization curves, and gives the
$B$--$V_0$ phase diagram its multicritical structure.

The two-dimensional fermionic system with repulsive interactions considered here can also be realized with ultracold polar molecules, a rapidly developing platform for strongly correlated quantum matter. In these systems, microwave dressing provides a repulsive shielding core that suppresses collisional losses~\cite{Karman2018,Doyle2021}, and, more importantly, offers a flexible route to engineering the effective intermolecular interaction~\cite{Deng2023,Deng2025}. The substantially enhanced lifetime enabled by microwave shielding~\cite{Doyle2021} has recently allowed the realization of degenerate gases of fermionic molecules~\cite{Luo2022a,Luo2022b,Chen2024} and Bose--Einstein condensates of bosonic molecules~\cite{Will2024,Will2026,Wang2026}. These advances have also stimulated theoretical proposals for a variety of many-body phases, including $p$-wave superfluidity~\cite{Deng2023}, self-bound bosonic droplets~\cite{Jin2025,Langen2024,Deng2025,Zhang2025,Pohl2025}, and supersolid phases~\cite{zhang2025supersolid}.

For the present purpose, fermionic polar molecules such as NaK or LiRb can be confined by a transverse trap and an in-plane optical lattice, thereby realizing an effective two-dimensional triangular geometry.  A dual-microwave shielding scheme can then be implemented: a linearly polarized microwave field propagating in the \(xy\) plane and a circularly polarized microwave field propagating along the \(z\) direction couple the rotational ground state to the first excited rotational manifold \(\ket{J=1,M=1,0,-1}\).  The two microwave fields are blue detuned, with independently tunable frequencies and Rabi couplings.  In the \(xy\) plane, this dressing scheme generates an effective intermolecular interaction of the form $V_{\rm{eff}}(r)=-C_3/r^3+C_6/r^6$~\cite{Deng2025}, where \(C_6>0\) provides the short-range shielding barrier that suppresses collisional losses. The long-range dipolar coefficient \(C_3\) can be tuned, and can even change sign, by varying the Rabi frequency of the linearly polarized microwave field. This provides a direct experimental knob for controlling both the strength and the character of the effective interaction. Since the recently realized single-component gases of fermionic molecules~\cite{Luo2022a,Luo2022b} naturally correspond to a fully spin-polarized regime, this platform is particularly well suited for probing the polarized molecular crystal predicted here, including its associated orbital-nematic ordering tendency.

\section{Acknowledgments }
We thank Leonard Bleiziffer for helpful discussions. W. Z., I. M., and E. A. D. acknowledge support from the SNSF project 200021$_{ }$212899; the SNSF Sinergia grant CRSII--222792; the Swiss State Secretariat for Education, Research and Innovation (contract number UeM019-1); and NCCR SPIN, a National Centre of Competence in Research (SNSF grant number 225153). 
 T. S. acknowledges support from the National Key Research and Development Program of China (Grant No. 2021YFA0718304) and the National Natural Science Foundation of China (NSFC) (Grants No. 12525413 and No. 12135018).

\bibliography{ref}

@article{MPNN1,
  title = {Message-passing neural quantum states for the homogeneous electron gas},
  author = {Pescia, Gabriel and Nys, Jannes and Kim, Jane and Lovato, Alessandro and Carleo, Giuseppe},
  journal = {Phys. Rev. B},
  volume = {110},
  issue = {3},
  pages = {035108},
  numpages = {11},
  year = {2024},
  month = {Jul},
  publisher = {American Physical Society},
  doi = {10.1103/PhysRevB.110.035108},
  url = {https://link.aps.org/doi/10.1103/PhysRevB.110.035108}
}

@article{MPNN2,
  author    = {Jane Kim and Gabriel Pescia and Bryce Fore and Jannes Nys and Giuseppe Carleo and Stefano Gandolfi and Morten Hjorth-Jensen and Alessandro Lovato},
  title     = {Neural-network quantum states for ultra-cold Fermi gases},
  journal   = {Communications Physics},
  year      = {2024},
  volume    = {7},
  number    = {1},
  pages      = {148},
  doi       = {10.1038/s42005-024-01613-w},
  url       = {https://doi.org/10.1038/s42005-024-01613-w},
  issn      = {2399-3650}
}

@article{PBC_varible,
  title = {Discovering Quantum Phase Transitions with Fermionic Neural Networks},
  author = {Cassella, Gino and Sutterud, Halvard and Azadi, Sam and Drummond, N. D. and Pfau, David and Spencer, James S. and Foulkes, W. M. C.},
  journal = {Phys. Rev. Lett.},
  volume = {130},
  issue = {3},
  pages = {036401},
  numpages = {6},
  year = {2023},
  month = {Jan},
  publisher = {American Physical Society},
  doi = {10.1103/PhysRevLett.130.036401},
  url = {https://link.aps.org/doi/10.1103/PhysRevLett.130.036401}
}

@article{ShiweiZhang,
  title = {Unified Variational Approach Description of Ground-State Phases of the Two-Dimensional Electron Gas},
  author = {Smith, Conor and Chen, Yixiao and Levy, Ryan and Yang, Yubo and Morales, Miguel A. and Zhang, Shiwei},
  journal = {Phys. Rev. Lett.},
  volume = {133},
  issue = {26},
  pages = {266504},
  numpages = {6},
  year = {2024},
  month = {Dec},
  publisher = {American Physical Society},
  doi = {10.1103/PhysRevLett.133.266504},
  url = {https://link.aps.org/doi/10.1103/PhysRevLett.133.266504}
}

@article{
CarleoTroyer2017,
author = {Giuseppe Carleo  and Matthias Troyer },
title = {Solving the quantum many-body problem with artificial neural networks},
journal = {Science},
volume = {355},
number = {6325},
pages = {602-606},
year = {2017},
doi = {10.1126/science.aag2302},
URL = {https://www.science.org/doi/abs/10.1126/science.aag2302},
eprint = {https://www.science.org/doi/pdf/10.1126/science.aag2302}}

@article{FermiNet2020,
  title = {Ab initio solution of the many-electron Schr\"odinger equation with deep neural networks},
  author = {Pfau, David and Spencer, James S. and Matthews, Alexander G. D. G. and Foulkes, W. M. C.},
  journal = {Phys. Rev. Res.},
  volume = {2},
  issue = {3},
  pages = {033429},
  numpages = {20},
  year = {2020},
  month = {Sep},
  publisher = {American Physical Society},
  doi = {10.1103/PhysRevResearch.2.033429},
  url = {https://link.aps.org/doi/10.1103/PhysRevResearch.2.033429}
}

@article{PauliNet2020,
  author = {Hermann, Jan and Sch{\"a}tzle, Zeno and No{\'e}, Frank},
  title = {Deep-neural-network solution of the electronic Schr{\"o}dinger equation},
  journal = {Nature Chemistry},
  volume = {12},
  number = {10},
  pages = {891--897},
  year = {2020},
  doi = {10.1038/s41557-020-0544-y},
  url = {https://doi.org/10.1038/s41557-020-0544-y}
}

@article{Ceperley1,
  title = {Ground state of the two-dimensional electron gas},
  author = {Tanatar, B. and Ceperley, D. M.},
  journal = {Phys. Rev. B},
  volume = {39},
  issue = {8},
  pages = {5005--5016},
  numpages = {0},
  year = {1989},
  month = {Mar},
  publisher = {American Physical Society},
  doi = {10.1103/PhysRevB.39.5005},
  url = {https://link.aps.org/doi/10.1103/PhysRevB.39.5005}
}

@article{Ceperley2,
  title = {Ground State of the Electron Gas by a Stochastic Method},
  author = {Ceperley, D. M. and Alder, B. J.},
  journal = {Phys. Rev. Lett.},
  volume = {45},
  issue = {7},
  pages = {566--569},
  numpages = {0},
  year = {1980},
  month = {Aug},
  publisher = {American Physical Society},
  doi = {10.1103/PhysRevLett.45.566},
  url = {https://link.aps.org/doi/10.1103/PhysRevLett.45.566}
}

@misc{LiangFu2024,
      title={Simulating moir\'e quantum matter with neural network}, 
      author={Di Luo and David D. Dai and Liang Fu},
      year={2024},
      eprint={2406.17645},
      archivePrefix={arXiv},
      primaryClass={cond-mat.str-el},
      url={https://arxiv.org/abs/2406.17645}, 
}

@article{LiangFu2025,
  title = {Self-attention neural network for solving correlated electron problems in solids},
  author = {Geier, Max and Nazaryan, Khachatur and Zaklama, Timothy and Fu, Liang},
  journal = {Phys. Rev. B},
  volume = {112},
  issue = {4},
  pages = {045119},
  numpages = {16},
  year = {2025},
  month = {Jul},
  publisher = {American Physical Society},
  doi = {10.1103/qxc3-bkc7},
  url = {https://link.aps.org/doi/10.1103/qxc3-bkc7}
}

@article{LiangFu2020,
  title = {Moir\'e quantum chemistry: Charge transfer in transition metal dichalcogenide superlattices},
  author = {Zhang, Yang and Yuan, Noah F. Q. and Fu, Liang},
  journal = {Phys. Rev. B},
  volume = {102},
  issue = {20},
  pages = {201115(R)},
  numpages = {6},
  year = {2020},
  month = {Nov},
  publisher = {American Physical Society},
  doi = {10.1103/PhysRevB.102.201115},
  url = {https://link.aps.org/doi/10.1103/PhysRevB.102.201115}
}

@Article{XuOLtriangular,
author={Xu, Muqing
and Kendrick, Lev Haldar
and Kale, Anant
and Gang, Youqi
and Ji, Geoffrey
and Scalettar, Richard T.
and Lebrat, Martin
and Greiner, Markus},
title={{Frustration- and doping-induced magnetism in a Fermi--Hubbard simulator}},
journal={Nature},
year={2023},
month={Aug},
day={01},
volume={620},
number={7976},
pages={971-976},
issn={1476-4687},
doi={10.1038/s41586-023-06280-5},
url={https://doi.org/10.1038/s41586-023-06280-5}
}

@Article{Lebrat2024,
  author    = {Lebrat, Martin and Xu, Muqing and Kendrick, Lev Haldar and Kale, Anant and Gang, Youqi and Seetharaman, Pranav and Morera, Ivan and Khatami, Ehsan and Demler, Eugene and Greiner, Markus},
  journal   = {Nature},
  title     = {{Observation of Nagaoka polarons in a Fermi–Hubbard quantum simulator}},
  year      = {2024},
  issn      = {1476-4687},
  month     = may,
  number    = {8011},
  pages     = {317--322},
  volume    = {629},
  publisher = {Springer Science and Business Media LLC},
  url       = {https://doi.org/10.1038/s41586-024-07272-9},
}

@Article{Prichard2024,
  author    = {Prichard, Max L. and Spar, Benjamin M. and Morera, Ivan and Demler, Eugene and Yan, Zoe Z. and Bakr, Waseem S.},
  journal   = {Nature},
  title     = {{Directly imaging spin polarons in a kinetically frustrated Hubbard system}},
  year      = {2024},
  issn      = {1476-4687},
  month     = may,
  number    = {8011},
  pages     = {323--328},
  volume    = {629},
  publisher = {Springer Science and Business Media LLC},
  url       = {https://doi.org/10.1038/s41586-024-07356-6},
}

@article{ImadaMI_REV,
  title = {Metal-insulator transitions},
  author = {Imada, Masatoshi and Fujimori, Atsushi and Tokura, Yoshinori},
  journal = {Rev. Mod. Phys.},
  volume = {70},
  issue = {4},
  pages = {1039--1263},
  numpages = {0},
  year = {1998},
  month = {Oct},
  publisher = {American Physical Society},
  doi = {10.1103/RevModPhys.70.1039},
  url = {https://link.aps.org/doi/10.1103/RevModPhys.70.1039}
}

@article{GargMI_2006,
  title = {Can Correlations Drive a Band Insulator Metallic?},
  author = {Garg, Arti and Krishnamurthy, H. R. and Randeria, Mohit},
  journal = {Phys. Rev. Lett.},
  volume = {97},
  issue = {4},
  pages = {046403},
  numpages = {4},
  year = {2006},
  month = {Jul},
  publisher = {American Physical Society},
  doi = {10.1103/PhysRevLett.97.046403},
  url = {https://link.aps.org/doi/10.1103/PhysRevLett.97.046403}
}

@article{KancharlaMI_2007,
  title = {Band insulator to Mott insulator transition in a bilayer Hubbard model},
  author = {Kancharla, S. S. and Okamoto, S.},
  journal = {Phys. Rev. B},
  volume = {75},
  issue = {19},
  pages = {193103},
  numpages = {4},
  year = {2007},
  month = {May},
  publisher = {American Physical Society},
  doi = {10.1103/PhysRevB.75.193103},
  url = {https://link.aps.org/doi/10.1103/PhysRevB.75.193103}
}

@article{ScalettarMI_2007,
  title = {Quantum Monte Carlo Study of an Interaction-Driven Band-Insulator--to--Metal Transition},
  author = {Paris, N. and Bouadim, K. and Hebert, F. and Batrouni, G. G. and Scalettar, R. T.},
  journal = {Phys. Rev. Lett.},
  volume = {98},
  issue = {4},
  pages = {046403},
  numpages = {4},
  year = {2007},
  month = {Jan},
  publisher = {American Physical Society},
  doi = {10.1103/PhysRevLett.98.046403},
  url = {https://link.aps.org/doi/10.1103/PhysRevLett.98.046403}
}

@article{Li_REV_2016,
doi = {10.1088/0034-4885/79/11/116401},
url = {https://doi.org/10.1088/0034-4885/79/11/116401},
year = {2016},
month = {sep},
publisher = {IOP Publishing},
volume = {79},
number = {11},
pages = {116401},
author = {Li, Xiaopeng and Liu, W Vincent},
title = {Physics of higher orbital bands in optical lattices: a review},
journal = {Reports on Progress in Physics}
}

@article{Luhmann_2012,
doi = {10.1088/1367-2630/14/3/033021},
url = {https://doi.org/10.1088/1367-2630/14/3/033021},
year = {2012},
month = {mar},
publisher = {IOP Publishing},
volume = {14},
number = {3},
pages = {033021},
author = {Lühmann, Dirk-Sören and Jürgensen, Ole and Sengstock, Klaus},
title = {Multi-orbital and density-induced tunneling of bosons in optical lattices},
journal = {New Journal of Physics}
}

@Article{Mazurenko2017,
author={Mazurenko, Anton
and Chiu, Christie S.
and Ji, Geoffrey
and Parsons, Maxwell F.
and Kan{\'a}sz-Nagy, M{\'a}rton
and Schmidt, Richard
and Grusdt, Fabian
and Demler, Eugene
and Greif, Daniel
and Greiner, Markus},
title={A cold-atom Fermi--Hubbard antiferromagnet},
journal={Nature},
year={2017},
month={May},
day={01},
volume={545},
number={7655},
pages={462-466},
issn={1476-4687},
doi={10.1038/nature22362},
url={https://doi.org/10.1038/nature22362}
}

@article{Bloch2012_review,
  title     = "Quantum simulations with ultracold quantum gases",
  author    = "Bloch, Immanuel and Dalibard, Jean and Nascimb{\`e}ne, Sylvain",
  journal   = "Nat. Phys.",
  publisher = "Springer Science and Business Media LLC",
  volume    =  8,
  number    =  4,
  pages     = "267--276",
  month     =  apr,
  year      =  2012
}

@Article{Koepsell2019,
author={Koepsell, Joannis
and Vijayan, Jayadev
and Sompet, Pimonpan
and Grusdt, Fabian
and Hilker, Timon A.
and Demler, Eugene
and Salomon, Guillaume
and Bloch, Immanuel
and Gross, Christian},
title={Imaging magnetic polarons in the doped Fermi--Hubbard model},
journal={Nature},
year={2019},
month={Aug},
day={01},
volume={572},
number={7769},
pages={358-362},
issn={1476-4687},
doi={10.1038/s41586-019-1463-1},
url={https://doi.org/10.1038/s41586-019-1463-1}
}

@article{Koepsell_doping_2021,
author = {Joannis Koepsell  and Dominik Bourgund  and Pimonpan Sompet  and Sarah Hirthe  and Annabelle Bohrdt  and Yao Wang  and Fabian Grusdt  and Eugene Demler  and Guillaume Salomon  and Christian Gross  and Immanuel Bloch },
title = {Microscopic evolution of doped Mott insulators from polaronic metal to Fermi liquid},
journal = {Science},
volume = {374},
number = {6563},
pages = {82-86},
year = {2021},
doi = {10.1126/science.abe7165},
URL = {https://www.science.org/doi/abs/10.1126/science.abe7165},
eprint = {https://www.science.org/doi/pdf/10.1126/science.abe7165}}

@article{chalopin_observation_2026,
	title = {Observation of emergent scaling of spin–charge correlations at the onset of the pseudogap},
	volume = {123},
	url = {https://www.pnas.org/doi/10.1073/pnas.2525539123},
	doi = {10.1073/pnas.2525539123},
	number = {4},
	urldate = {2026-02-27},
	journal = {Proceedings of the National Academy of Sciences},
	publisher = {Proceedings of the National Academy of Sciences},
	author = {Chalopin, Thomas and Bojović, Petar and Wang, Si and Franz, Titus and Sinha, Aritra and Wang, Zhenjiu and Bourgund, Dominik and Obermeyer, Johannes and Grusdt, Fabian and Bohrdt, Annabelle and Pollet, Lode and Wietek, Alexander and Georges, Antoine and Hilker, Timon and Bloch, Immanuel},
	month = jan,
	year = {2026},
	pages = {e2525539123},
}

@misc{kendrick2025pseudogapfermihubbardquantumsimulator,
      title={Pseudogap in a Fermi-Hubbard quantum simulator}, 
      author={Lev Haldar Kendrick and Anant Kale and Youqi Gang and Alexander Dennisovich Deters and Martin Lebrat and Aaron W. Young and Markus Greiner},
      year={2025},
      eprint={2509.18075},
      archivePrefix={arXiv},
      primaryClass={cond-mat.quant-gas},
      url={https://arxiv.org/abs/2509.18075}, 
}

@Article{MakRev2022,
author={Mak, Kin Fai
and Shan, Jie},
title={Semiconductor moir{\'e} materials},
journal={Nature Nanotechnology},
year={2022},
month={Jul},
day={01},
volume={17},
number={7},
pages={686-695},
issn={1748-3395},
doi={10.1038/s41565-022-01165-6},
url={https://doi.org/10.1038/s41565-022-01165-6}
}

@Article{AndreiRev2021,
author={Andrei, Eva Y.
and Efetov, Dmitri K.
and Jarillo-Herrero, Pablo
and MacDonald, Allan H.
and Mak, Kin Fai
and Senthil, T.
and Tutuc, Emanuel
and Yazdani, Ali
and Young, Andrea F.},
title={The marvels of moir{\'e} materials},
journal={Nature Reviews Materials},
year={2021},
month={Mar},
day={01},
volume={6},
number={3},
pages={201-206},
issn={2058-8437},
doi={10.1038/s41578-021-00284-1},
url={https://doi.org/10.1038/s41578-021-00284-1}
}

@article{Xu2020a,
  title = {Correlated Insulating States at Fractional Fillings of Moir\'e Superlattices},
  author = {Xu, Yang and Liu, Song and Rhodes, Daniel A. and Watanabe, Kenji and Taniguchi, Takashi and Hone, James and Elser, Veit and Mak, Kin Fai and Shan, Jie},
  year = {2020},
  month = nov,
  journal = {Nature},
  volume = {587},
  number = {7833},
  pages = {214--218},
  issn = {1476-4687},
  doi = {10.1038/s41586-020-2868-6}
  }

@article{Tang2020a,
  author = {Tang, Yanhao and Li, Lizhong and Li, Tingxin and Xu, Yang and Liu, Song and Barmak, Katayun and Watanabe, Kenji and Taniguchi, Takashi and MacDonald, Allan H. and Shan, Jie and Mak, Kin Fai},
  year = {2020},
  month = mar,
  journal = {Nature},
  volume = {579},
  number = {7799},
  pages = {353--358},
  issn = {1476-4687},
  doi = {10.1038/s41586-020-2085-3},
  title = {Simulation of {{Hubbard}} Model Physics in {{WSe}}{$_2$}/{{WS}}{$_2$} Moir\'e Superlattices}
}

@article{Tang2023,
  title = {Evidence of Frustrated Magnetic Interactions in a {{Wigner}}\textendash{{Mott}} Insulator},
  author = {Tang, Yanhao and Su, Kaixiang and Li, Lizhong and Xu, Yang and Liu, Song and Watanabe, Kenji and Taniguchi, Takashi and Hone, James and Jian, Chao-Ming and Xu, Cenke and Mak, Kin Fai and Shan, Jie},
  year = {2023},
  month = mar,
  journal = {Nature Nanotechnology},
  volume = {18},
  number = {3},
  pages = {233--237},
  issn = {1748-3395},
  doi = {10.1038/s41565-022-01309-8}
}

@article{Regan2020,
  author = {Regan, Emma C. and Wang, Danqing and Jin, Chenhao and Bakti Utama, M. Iqbal and Gao, Beini and Wei, Xin and Zhao, Sihan and Zhao, Wenyu and Zhang, Zuocheng and Yumigeta, Kentaro and Blei, Mark and Carlstr{\"o}m, Johan D. and Watanabe, Kenji and Taniguchi, Takashi and Tongay, Sefaattin and Crommie, Michael and Zettl, Alex and Wang, Feng},
  year = {2020},
  month = mar,
  journal = {Nature},
  volume = {579},
  number = {7799},
  pages = {359--363},
  issn = {1476-4687},
  doi = {10.1038/s41586-020-2092-4},
  title = {Mott and Generalized {{Wigner}} Crystal States in {{WSe}}{$_2$}/{{WS}}{$_2$} Moir\'e Superlattices}
}

@article{Wang2020,
  title = {Correlated Electronic Phases in Twisted Bilayer Transition Metal Dichalcogenides},
  author = {Wang, Lei and Shih, En-Min and Ghiotto, Augusto and Xian, Lede and Rhodes, Daniel A. and Tan, Cheng and Claassen, Martin and Kennes, Dante M. and Bai, Yusong and Kim, Bumho and Watanabe, Kenji and Taniguchi, Takashi and Zhu, Xiaoyang and Hone, James and Rubio, Angel and Pasupathy, Abhay N. and Dean, Cory R.},
  year = {2020},
  month = aug,
  journal = {Nature Materials},
  volume = {19},
  number = {8},
  pages = {861--866},
  issn = {1476-4660},
  doi = {10.1038/s41563-020-0708-6}
}

@article{Shimazaki2020,
  title = {Strongly Correlated Electrons and Hybrid Excitons in a Moir\'e Heterostructure},
  author = {Shimazaki, Yuya and Schwartz, Ido and Watanabe, Kenji and Taniguchi, Takashi and Kroner, Martin and Imamo{\u g}lu, Ata{\c c}},
  year = {2020},
  month = apr,
  journal = {Nature},
  volume = {580},
  number = {7804},
  pages = {472--477},
  issn = {1476-4687},
  doi = {10.1038/s41586-020-2191-2}
}

@Article{Huang2021,
author={Huang, Xiong
and Wang, Tianmeng
and Miao, Shengnan
and Wang, Chong
and Li, Zhipeng
and Lian, Zhen
and Taniguchi, Takashi
and Watanabe, Kenji
and Okamoto, Satoshi
and Xiao, Di
and Shi, Su-Fei
and Cui, Yong-Tao},
title={Correlated insulating states at fractional fillings of the WS2/WSe2 moir{\'e} lattice},
journal={Nature Physics},
year={2021},
month={Jun},
day={01},
volume={17},
number={6},
pages={715-719},
issn={1745-2481},
doi={10.1038/s41567-021-01171-w},
url={https://doi.org/10.1038/s41567-021-01171-w}
}

@Article{Zhao2023,
author={Zhao, Wenjin
and Shen, Bowen
and Tao, Zui
and Han, Zhongdong
and Kang, Kaifei
and Watanabe, Kenji
and Taniguchi, Takashi
and Mak, Kin Fai
and Shan, Jie},
title={Gate-tunable heavy fermions in a moir{\'e} Kondo lattice},
journal={Nature},
year={2023},
month={Apr},
day={01},
volume={616},
number={7955},
pages={61-65},
issn={1476-4687},
doi={10.1038/s41586-023-05800-7},
url={https://doi.org/10.1038/s41586-023-05800-7}
}

@Article{Li2021,
author={Li, Tingxin
and Jiang, Shengwei
and Shen, Bowen
and Zhang, Yang
and Li, Lizhong
and Tao, Zui
and Devakul, Trithep
and Watanabe, Kenji
and Taniguchi, Takashi
and Fu, Liang
and Shan, Jie
and Mak, Kin Fai},
title={Quantum anomalous Hall effect from intertwined moir{\'e} bands},
journal={Nature},
year={2021},
month={Dec},
day={01},
volume={600},
number={7890},
pages={641-646},
issn={1476-4687},
doi={10.1038/s41586-021-04171-1},
url={https://doi.org/10.1038/s41586-021-04171-1}
}

@Article{Ciorciaro2023,
author={Ciorciaro, L.
and Smole{\'{n}}ski, T.
and Morera, I.
and Kiper, N.
and Hiestand, S.
and Kroner, M.
and Zhang, Y.
and Watanabe, K.
and Taniguchi, T.
and Demler, E.
and {\.{I}}mamo{\u{g}}lu, A.},
title={Kinetic magnetism in triangular moir{\'e} materials},
journal={Nature},
year={2023},
month={Nov},
day={01},
volume={623},
number={7987},
pages={509-513},
issn={1476-4687},
doi={10.1038/s41586-023-06633-0},
url={https://doi.org/10.1038/s41586-023-06633-0}
}

@Article{Tao2024,
author={Tao, Zui
and Zhao, Wenjin
and Shen, Bowen
and Li, Tingxin
and Kn{\"u}ppel, Patrick
and Watanabe, Kenji
and Taniguchi, Takashi
and Shan, Jie
and Mak, Kin Fai},
title={Observation of spin polarons in a frustrated moir{\'e} Hubbard system},
journal={Nature Physics},
year={2024},
month={May},
day={01},
volume={20},
number={5},
pages={783-787},
issn={1745-2481},
doi={10.1038/s41567-024-02434-y},
url={https://doi.org/10.1038/s41567-024-02434-y}
}

@article{FengWang2024,
author = {Hongyuan Li  and Ziyu Xiang  and Aidan P. Reddy  and Trithep Devakul  and Renee Sailus  and Rounak Banerjee  and Takashi Taniguchi  and Kenji Watanabe  and Sefaattin Tongay  and Alex Zettl  and Liang Fu  and Michael F. Crommie  and Feng Wang },
title = {Wigner molecular crystals from multielectron moiré artificial atoms},
journal = {Science},
volume = {385},
number = {6704},
pages = {86-91},
year = {2024},
doi = {10.1126/science.adk1348},
URL = {https://www.science.org/doi/abs/10.1126/science.adk1348},
eprint = {https://www.science.org/doi/pdf/10.1126/science.adk1348}}

@article{Jichen2025,
  author  = {Li, Xiang and Qian, Yubing and Ren, Weiluo and Xu, Yang and Chen, Ji},
  title   = {Emergent Wigner phases in moir{\'e} superlattice from deep learning},
  journal = {Communications Physics},
  year    = {2025},
  volume  = {8},
  number  = {1},
  pages   = {364},
  doi     = {10.1038/s42005-025-02282-z},
  url     = {https://doi.org/10.1038/s42005-025-02282-z},
  issn    = {2399-3650}
}

@article{NQSDynamics-2026,
  title = {Time-Dependent Neural Galerkin Method for Quantum Dynamics},
  author = {Sinibaldi, Alessandro and Hendry, Douglas and Vicentini, Filippo and Carleo, Giuseppe},
  journal = {Phys. Rev. Lett.},
  volume = {136},
  issue = {12},
  pages = {120402},
  numpages = {10},
  year = {2026},
  month = {Mar},
  publisher = {American Physical Society},
  doi = {10.1103/kqvx-dl54},
  url = {https://link.aps.org/doi/10.1103/kqvx-dl54}
}

@article{VandeWalle_2025,
doi = {10.1088/2632-2153/ae0f39},
url = {https://doi.org/10.1088/2632-2153/ae0f39},
year = {2025},
month = {oct},
publisher = {IOP Publishing},
volume = {6},
number = {4},
pages = {045011},
author = {Van de Walle, Anka and Schmitt, Markus and Bohrdt, Annabelle},
title = {Many-body dynamics with explicitly time-dependent neural quantum states},
journal = {Machine Learning: Science and Technology}
}

@article{Nys2024,
  author  = {Nys, Jannes and Pescia, Gabriel and Sinibaldi, Alessandro and Carleo, Giuseppe},
  title   = {Ab-initio variational wave functions for the time-dependent many-electron Schr{\"o}dinger equation},
  journal = {Nature Communications},
  year    = {2024},
  volume  = {15},
  number  = {1},
  pages   = {9404},
  doi     = {10.1038/s41467-024-53672-w},
  url     = {https://doi.org/10.1038/s41467-024-53672-w},
  issn    = {2041-1723}
}

@article{PhysRevResearch.2.013284,
  title = {Neural-network quantum states at finite temperature},
  author = {Irikura, Naoki and Saito, Hiroki},
  journal = {Phys. Rev. Res.},
  volume = {2},
  issue = {1},
  pages = {013284},
  numpages = {6},
  year = {2020},
  month = {Mar},
  publisher = {American Physical Society},
  doi = {10.1103/PhysRevResearch.2.013284},
  url = {https://link.aps.org/doi/10.1103/PhysRevResearch.2.013284}
}

@article{PhysRevLett.120.240503,
  title = {Latent Space Purification via Neural Density Operators},
  author = {Torlai, Giacomo and Melko, Roger G.},
  journal = {Phys. Rev. Lett.},
  volume = {120},
  issue = {24},
  pages = {240503},
  numpages = {5},
  year = {2018},
  month = {Jun},
  publisher = {American Physical Society},
  doi = {10.1103/PhysRevLett.120.240503},
  url = {https://link.aps.org/doi/10.1103/PhysRevLett.120.240503}
}

@misc{nys2025fermionicneuralgibbsstates,
      title={Fermionic neural Gibbs states}, 
      author={Jannes Nys and Juan Carrasquilla},
      year={2025},
      eprint={2512.04663},
      archivePrefix={arXiv},
      primaryClass={quant-ph},
      url={https://arxiv.org/abs/2512.04663}, 
}

@article{NQS_Hubbard,
  title = {Backflow Transformations via Neural Networks for Quantum Many-Body Wave Functions},
  author = {Luo, Di and Clark, Bryan K.},
  journal = {Phys. Rev. Lett.},
  volume = {122},
  issue = {22},
  pages = {226401},
  numpages = {6},
  year = {2019},
  month = {Jun},
  publisher = {American Physical Society},
  doi = {10.1103/PhysRevLett.122.226401},
  url = {https://link.aps.org/doi/10.1103/PhysRevLett.122.226401}
}

@article{
NQS_Hubbard2,
author = {Javier Robledo Moreno  and Giuseppe Carleo  and Antoine Georges  and James Stokes },
title = {Fermionic wave functions from neural-network constrained hidden states},
journal = {Proceedings of the National Academy of Sciences},
volume = {119},
number = {32},
pages = {e2122059119},
year = {2022},
doi = {10.1073/pnas.2122059119},
URL = {https://www.pnas.org/doi/abs/10.1073/pnas.2122059119},
eprint = {https://www.pnas.org/doi/pdf/10.1073/pnas.2122059119}}

@article{NQS_Hubbard3,
  title = {Autoregressive neural quantum states of Fermi Hubbard models},
  author = {Ibarra-Garc\'{\i}a-Padilla, Eduardo and Lange, Hannah and Melko, Roger G. and Scalettar, Richard T. and Carrasquilla, Juan and Bohrdt, Annabelle and Khatami, Ehsan},
  journal = {Phys. Rev. Res.},
  volume = {7},
  issue = {1},
  pages = {013122},
  numpages = {14},
  year = {2025},
  month = {Feb},
  publisher = {American Physical Society},
  doi = {10.1103/PhysRevResearch.7.013122},
  url = {https://link.aps.org/doi/10.1103/PhysRevResearch.7.013122}
}

@article{NQS_nuclear,
  title = {Variational Monte Carlo Calculations of $A\ensuremath{\le}4$ Nuclei with an Artificial Neural-Network Correlator Ansatz},
  author = {Adams, Corey and Carleo, Giuseppe and Lovato, Alessandro and Rocco, Noemi},
  journal = {Phys. Rev. Lett.},
  volume = {127},
  issue = {2},
  pages = {022502},
  numpages = {6},
  year = {2021},
  month = {Jul},
  publisher = {American Physical Society},
  doi = {10.1103/PhysRevLett.127.022502},
  url = {https://link.aps.org/doi/10.1103/PhysRevLett.127.022502}
}

@article{NQS_nuclear2,
  title = {Hidden-nucleons neural-network quantum states for the nuclear many-body problem},
  author = {Lovato, Alessandro and Adams, Corey and Carleo, Giuseppe and Rocco, Noemi},
  journal = {Phys. Rev. Res.},
  volume = {4},
  issue = {4},
  pages = {043178},
  numpages = {8},
  year = {2022},
  month = {Dec},
  publisher = {American Physical Society},
  doi = {10.1103/PhysRevResearch.4.043178},
  url = {https://link.aps.org/doi/10.1103/PhysRevResearch.4.043178}
}

@article{NQS_nuclear3,
  title = {Dilute neutron star matter from neural-network quantum states},
  author = {Fore, Bryce and Kim, Jane M. and Carleo, Giuseppe and Hjorth-Jensen, Morten and Lovato, Alessandro and Piarulli, Maria},
  journal = {Phys. Rev. Res.},
  volume = {5},
  issue = {3},
  pages = {033062},
  numpages = {6},
  year = {2023},
  month = {Jul},
  publisher = {American Physical Society},
  doi = {10.1103/PhysRevResearch.5.033062},
  url = {https://link.aps.org/doi/10.1103/PhysRevResearch.5.033062}
}

@article{NQS_chemistry,
  title = {Neural network backflow for ab initio quantum chemistry},
  author = {Liu, An-Jun and Clark, Bryan K.},
  journal = {Phys. Rev. B},
  volume = {110},
  issue = {11},
  pages = {115137},
  numpages = {10},
  year = {2024},
  month = {Sep},
  publisher = {American Physical Society},
  doi = {10.1103/PhysRevB.110.115137},
  url = {https://link.aps.org/doi/10.1103/PhysRevB.110.115137}
}

@article{PhysRevA.79.063412,
  title = {Accessing the N\'eel phase of ultracold fermionic atoms in a simple-cubic optical lattice},
  author = {Mathy, C. J. M. and Huse, David A.},
  journal = {Phys. Rev. A},
  volume = {79},
  issue = {6},
  pages = {063412},
  numpages = {7},
  year = {2009},
  month = {Jun},
  publisher = {American Physical Society},
  doi = {10.1103/PhysRevA.79.063412},
  url = {https://link.aps.org/doi/10.1103/PhysRevA.79.063412}
}

@article{PhysRevA.79.043626,
  title = {Multiband superfluidity and superfluid--to--band-insulator transition of strongly interacting fermionic atoms in an optical lattice},
  author = {Burkov, A. A. and Paramekanti, Arun},
  journal = {Phys. Rev. A},
  volume = {79},
  issue = {4},
  pages = {043626},
  numpages = {5},
  year = {2009},
  month = {Apr},
  publisher = {American Physical Society},
  doi = {10.1103/PhysRevA.79.043626},
  url = {https://link.aps.org/doi/10.1103/PhysRevA.79.043626}
}

@article{Kivelson1998,
  author  = {Kivelson, S. A. and Fradkin, E. and Emery, V. J.},
  title   = {Electronic liquid-crystal phases of a doped Mott insulator},
  journal = {Nature},
  year    = {1998},
  volume  = {393},
  number  = {6685},
  pages   = {550--553},
  doi     = {10.1038/31177},
  url     = {https://doi.org/10.1038/31177},
  issn    = {1476-4687}
}

@article{Fradkin2010,
   author = "Fradkin, Eduardo and Kivelson, Steven A. and Lawler, Michael J. and Eisenstein, James P. and Mackenzie, Andrew P.",
   title = "Nematic Fermi Fluids in Condensed Matter Physics", 
   journal= "Annual Review of Condensed Matter Physics",
   year = "2010",
   volume = "1",
   number = "Volume 1, 2010",
   pages = "153-178",
   doi = "https://doi.org/10.1146/annurev-conmatphys-070909-103925",
   url = "https://www.annualreviews.org/content/journals/10.1146/annurev-conmatphys-070909-103925",
   publisher = "Annual Reviews",
   issn = "1947-5462",
   type = "Journal Article",
  }

@article{
Borzi2007,
author = {R. A. Borzi  and S. A. Grigera  and J. Farrell  and R. S. Perry  and S. J. S. Lister  and S. L. Lee  and D. A. Tennant  and Y. Maeno  and A. P. Mackenzie },
title = {Formation of a Nematic Fluid at High Fields in Sr<sub>3</sub>Ru<sub>2</sub>O<sub>7</sub>},
journal = {Science},
volume = {315},
number = {5809},
pages = {214-217},
year = {2007},
doi = {10.1126/science.1134796},
URL = {https://www.science.org/doi/abs/10.1126/science.1134796},
eprint = {https://www.science.org/doi/pdf/10.1126/science.1134796}}

@article{Matty2022,
  author  = {Matty, Michael and Kim, Eun-Ah},
  title   = {Melting of generalized Wigner crystals in transition metal dichalcogenide heterobilayer Moir{\'e} systems},
  journal = {Nature Communications},
  year    = {2022},
  volume  = {13},
  number  = {1},
  pages   = {7098},
  doi     = {10.1038/s41467-022-34683-x},
  url     = {https://doi.org/10.1038/s41467-022-34683-x},
  issn    = {2041-1723}
}

@Article{Luo2022a,
author={Schindewolf, Andreas
and Bause, Roman
and Chen, Xing-Yan
and Duda, Marcel
and Karman, Tijs
and Bloch, Immanuel
and Luo, Xin-Yu},
title={Evaporation of microwave-shielded polar molecules to quantum degeneracy},
journal={Nature},
year={2022},
month={Jul},
day={01},
volume={607},
number={7920},
pages={677-681},
issn={1476-4687},
doi={10.1038/s41586-022-04900-0},
url={https://doi.org/10.1038/s41586-022-04900-0}
}

@article{Luo2022b,  
  title={Field-linked resonances of polar molecules},     
      author={Chen, Xing-Yan and Schindewolf, Andreas and Eppelt, Sebastian and Bause, Roman and Duda, Marcel and Biswas, Shrestha and Karman, Tijs and Hilker, Timon and Bloch, Immanuel and Luo, Xin-Yu},    
      journal = {Nature},
	 volume = {614},
	 number = {},
	 pages = {59--63},
	 year = {2022},
	 doi = {10.1038/s41586-022-05651-8}
}

@misc{Will2024,
      title={Three-Body Recombination of Ultracold Microwave-Shielded Polar Molecules}, 
      author={Ian Stevenson and Shayamal Singh and Ahmed Elkamshishy and Niccoló Bigagli and Weijun Yuan and Siwei Zhang and Chris H. Greene and Sebastian Will},
      year={2024},
      eprint={2407.04901},
      archivePrefix={arXiv},
      primaryClass={cond-mat.quant-gas},
      url={https://arxiv.org/abs/2407.04901}, 
}

@article{Deng2023,
  title = {Effective Potential and Superfluidity of Microwave-Shielded Polar Molecules},
  author = {Deng, Fulin and Chen, Xing-Yan and Luo, Xin-Yu and Zhang, Wenxian and Yi, Su and Shi, Tao},
  journal = {Phys. Rev. Lett.},
  volume = {130},
  issue = {18},
  pages = {183001},
  numpages = {6},
  year = {2023},
  doi = {10.1103/PhysRevLett.130.183001},
}

@misc{Deng2025,
      title={Two- and many-body physics of ultracold molecules dressed by dual microwave fields}, 
      author={Fulin Deng and Xinyuan Hu and Wei-Jian Jin and Su Yi and Tao Shi},
      year={2025},
      eprint={2501.05210},
      archivePrefix={arXiv},
      primaryClass={cond-mat.quant-gas},
      url={https://arxiv.org/abs/2501.05210}, 
}

@article{Jin2025,
  title = {Bose-Einstein Condensates of Microwave-Shielded Polar Molecules},
  author = {Jin, Wei-Jian and Deng, Fulin and Yi, Su and Shi, Tao},
  journal = {Phys. Rev. Lett.},
  volume = {134},
  issue = {23},
  pages = {233003},
  numpages = {7},
  year = {2025},
  month = {Jun},
  publisher = {American Physical Society},
  doi = {10.1103/b8y9-yvz9},
  url = {https://link.aps.org/doi/10.1103/b8y9-yvz9}
}

@article{Wigner,
  title = {On the Interaction of Electrons in Metals},
  author = {Wigner, E.},
  journal = {Phys. Rev.},
  volume = {46},
  issue = {11},
  pages = {1002--1011},
  numpages = {0},
  year = {1934},
  month = {Dec},
  publisher = {American Physical Society},
  doi = {10.1103/PhysRev.46.1002},
  url = {https://link.aps.org/doi/10.1103/PhysRev.46.1002}
}

@article{Feynman1956,
  title = {Energy Spectrum of the Excitations in Liquid Helium},
  author = {Feynman, R. P. and Cohen, Michael},
  journal = {Phys. Rev.},
  volume = {102},
  issue = {5},
  pages = {1189--1204},
  numpages = {0},
  year = {1956},
  month = {Jun},
  publisher = {American Physical Society},
  doi = {10.1103/PhysRev.102.1189},
  url = {https://link.aps.org/doi/10.1103/PhysRev.102.1189}
}

@article{Backflow1,
  title = {Effects of backflow correlation in the three-dimensional electron gas: Quantum Monte Carlo study},
  author = {Kwon, Yongkyung and Ceperley, D. M. and Martin, Richard M.},
  journal = {Phys. Rev. B},
  volume = {58},
  issue = {11},
  pages = {6800--6806},
  numpages = {0},
  year = {1998},
  month = {Sep},
  publisher = {American Physical Society},
  doi = {10.1103/PhysRevB.58.6800},
  url = {https://link.aps.org/doi/10.1103/PhysRevB.58.6800}
}

@article{Backflow2,
  title = {Effects of three-body and backflow correlations in the two-dimensional electron gas},
  author = {Kwon, Yongkyung and Ceperley, D. M. and Martin, Richard M.},
  journal = {Phys. Rev. B},
  volume = {48},
  issue = {16},
  pages = {12037--12046},
  numpages = {0},
  year = {1993},
  month = {Oct},
  publisher = {American Physical Society},
  doi = {10.1103/PhysRevB.48.12037},
  url = {https://link.aps.org/doi/10.1103/PhysRevB.48.12037}
}

@article{Attaccalite2002,
  title = {Correlation Energy and Spin Polarization in the 2D Electron Gas},
  author = {Attaccalite, Claudio and Moroni, Saverio and Gori-Giorgi, Paola and Bachelet, Giovanni B.},
  journal = {Phys. Rev. Lett.},
  volume = {88},
  issue = {25},
  pages = {256601},
  numpages = {4},
  year = {2002},
  month = {Jun},
  publisher = {American Physical Society},
  doi = {10.1103/PhysRevLett.88.256601},
  url = {https://link.aps.org/doi/10.1103/PhysRevLett.88.256601}
}

@article{PhysRevB.79.085414,
  title = {Quantum Monte Carlo study of the ground state of the two-dimensional Fermi fluid},
  author = {Drummond, N. D. and Needs, R. J.},
  journal = {Phys. Rev. B},
  volume = {79},
  issue = {8},
  pages = {085414},
  numpages = {8},
  year = {2009},
  month = {Feb},
  publisher = {American Physical Society},
  doi = {10.1103/PhysRevB.79.085414},
  url = {https://link.aps.org/doi/10.1103/PhysRevB.79.085414}
}

@article{Karman2018,
  title = {Microwave Shielding of Ultracold Polar Molecules},
  author = {Karman, Tijs and Hutson, Jeremy M.},
  journal = {Phys. Rev. Lett.},
  volume = {121},
  issue = {16},
  pages = {163401},
  numpages = {5},
  year = {2018},
  month = {Oct},
  publisher = {American Physical Society},
  doi = {10.1103/PhysRevLett.121.163401},
  url = {https://link.aps.org/doi/10.1103/PhysRevLett.121.163401}
}

@article{Doyle2021,
author = {Loïc Anderegg  and Sean Burchesky  and Yicheng Bao  and Scarlett S. Yu  and Tijs Karman  and Eunmi Chae  and Kang-Kuen Ni  and Wolfgang Ketterle  and John M. Doyle },
title = {Observation of microwave shielding of ultracold molecules},
journal = {Science},
volume = {373},
number = {6556},
pages = {779-782},
year = {2021},
doi = {10.1126/science.abg9502},
URL = {https://www.science.org/doi/abs/10.1126/science.abg9502}}

@Article{Chen2024,
author={Chen, Xing-Yan
and Biswas, Shrestha
and Eppelt, Sebastian
and Schindewolf, Andreas
and Deng, Fulin
and Shi, Tao
and Yi, Su
and Hilker, Timon A.
and Bloch, Immanuel
and Luo, Xin-Yu},
title={Ultracold field-linked tetratomic molecules},
journal={Nature},
year={2024},
month={Feb},
day={01},
volume={626},
number={7998},
pages={283-287},
issn={1476-4687},
doi={10.1038/s41586-023-06986-6},
url={https://doi.org/10.1038/s41586-023-06986-6}
}

@Article{Will2026,
author={Zhang, Siwei
and Yuan, Weijun
and Bigagli, Niccolò
and Kwak, Haneul
and Karman, Tijs
and Stevenson, Ian
and Will, Sebastian},
title={Observation of self-bound droplets of ultracold dipolar molecules},
journal={Nature},
year={2026},
month={Mar},
day={01},
volume={651},
number={8106},
pages={601-606},
issn={1476-4687},
doi={10.1038/s41586-026-10245-9},
url={https://doi.org/10.1038/s41586-026-10245-9}
}

@Article{Wang2026,
author={Shi, Zhaopeng
and Huang, Zerong
and Deng, Fulin
and Jin, Wei-Jian
and Yi, Su
and Shi, Tao
and Wang, Dajun},
title={Bose–Einstein condensate of ultracold sodium–rubidium molecules with tunable dipolar interactions},
journal={Nature Physics},
year={2026},
month={Jul},
day={09},
volume={},
number={},
pages={},
issn={1745-2481},
doi={10.1038/s41567-026-03362-9},
url={https://doi.org/10.1038/s41567-026-03362-9}
}

@article{Langen2024,
  title = {Dipolar Droplets of Strongly Interacting Molecules},
  author = {Langen, Tim and Boronat, Jordi and S\'anchez-Baena, Juan and Bomb\'{\i}n, Ra\'ul and Karman, Tijs and Mazzanti, Ferran},
  journal = {Phys. Rev. Lett.},
  volume = {134},
  issue = {5},
  pages = {053001},
  numpages = {7},
  year = {2025},
  month = {Feb},
  publisher = {American Physical Society},
  doi = {10.1103/PhysRevLett.134.053001},
  url = {https://link.aps.org/doi/10.1103/PhysRevLett.134.053001}
}

@article{Zhang2025,
  title = {Quantum Phases for Finite-Temperature Gases of Bosonic Polar Molecules Shielded by Dual Microwaves},
  author = {Zhang, Wei and Chen, Kun and Yi, Su and Shi, Tao},
  journal = {PRX Quantum},
  volume = {6},
  issue = {4},
  pages = {040307},
  numpages = {11},
  year = {2025},
  month = {Oct},
  publisher = {American Physical Society},
  doi = {10.1103/9cxl-d9zg},
  url = {https://link.aps.org/doi/10.1103/9cxl-d9zg}
}

@article{Pohl2025,
  title = {Self-Bound Superfluid Membranes and Monolayer Crystals of Ultracold Polar Molecules},
  author = {Ciardi, Matteo and Pedersen, Kasper R\o{}nning and Langen, Tim and Pohl, Thomas},
  journal = {Phys. Rev. Lett.},
  volume = {135},
  issue = {15},
  pages = {153401},
  numpages = {8},
  year = {2025},
  month = {Oct},
  publisher = {American Physical Society},
  doi = {10.1103/v7gw-xy36},
  url = {https://link.aps.org/doi/10.1103/v7gw-xy36}
}

@misc{zhang2025supersolid,
      title={Supersolid Phases in Ultracold Gases of Microwave Shielded Polar Molecules}, 
      author={Wei Zhang and Hongye Liu and Fulin Deng and Kun Chen and Su Yi and Tao Shi},
      year={2025},
      eprint={2506.23820},
      archivePrefix={arXiv},
      primaryClass={cond-mat.quant-gas},
      url={https://arxiv.org/abs/2506.23820}, 
}

@Article{Schafer2020,
author={Sch{\"a}fer, Florian
and Fukuhara, Takeshi
and Sugawa, Seiji
and Takasu, Yosuke
and Takahashi, Yoshiro},
title={Tools for quantum simulation with ultracold atoms in optical lattices},
journal={Nature Reviews Physics},
year={2020},
month={Aug},
day={01},
volume={2},
number={8},
pages={411-425},
issn={2522-5820},
doi={10.1038/s42254-020-0195-3},
url={https://doi.org/10.1038/s42254-020-0195-3}
}

@Article{Brown2020,
author={Brown, Peter T.
and Guardado-Sanchez, Elmer
and Spar, Benjamin M.
and Huang, Edwin W.
and Devereaux, Thomas P.
and Bakr, Waseem S.},
title={Angle-resolved photoemission spectroscopy of a Fermi--Hubbard system},
journal={Nature Physics},
year={2020},
month={Jan},
day={01},
volume={16},
number={1},
pages={26-31},
issn={1745-2481},
doi={10.1038/s41567-019-0696-0},
url={https://doi.org/10.1038/s41567-019-0696-0}
}

@Article{Prichard2025,
author={Prichard, Max L.
and Ba, Zengli
and Morera, Ivan
and Spar, Benjamin M.
and Huse, David A.
and Demler, Eugene
and Bakr, Waseem S.},
title={Magnon-polarons in the Fermi--Hubbard model},
journal={Nature Physics},
year={2025},
month={Oct},
day={01},
volume={21},
number={10},
pages={1548-1554},
issn={1745-2481},
doi={10.1038/s41567-025-03004-6},
url={https://doi.org/10.1038/s41567-025-03004-6}
}

@article{PhysRevX.15.011049,
  title = {Confined Trions and Mott-Wigner States in a Purely Electrostatic Moir\'e Potential},
  author = {Kiper, Natasha and Adlong, Haydn S. and Christianen, Arthur and Kroner, Martin and Watanabe, Kenji and Taniguchi, Takashi and \ifmmode \dot{I}\else \.{I}\fi{}mamo\ifmmode \breve{g}\else \u{g}\fi{}lu, Atac},
  journal = {Phys. Rev. X},
  volume = {15},
  issue = {1},
  pages = {011049},
  numpages = {18},
  year = {2025},
  month = {Mar},
  publisher = {American Physical Society},
  doi = {10.1103/PhysRevX.15.011049},
  url = {https://link.aps.org/doi/10.1103/PhysRevX.15.011049}
}

\clearpage

\appendix

\onecolumngrid

\section{Details of the Message-Passing Neural-Network Ansatz}
\label{app_MPNN}

We first encode the electronic coordinates in a form that explicitly respects the periodic boundary conditions of the simulation cell. For each electron $i=1,\ldots,N_e$, we define the one-body input feature
\begin{equation}
\begin{aligned}
\bm x_i^{(0)}=
\big(
&\sin(\bm G_1\!\cdot\!\bm r_i),
\sin(\bm G_2\!\cdot\!\bm r_i),
\cos(\bm G_1\!\cdot\!\bm r_i),
\cos(\bm G_2\!\cdot\!\bm r_i), \|\bm r_i\|_{\rm PBC},
s_i
\big),
\end{aligned}
\end{equation}
where $s_i=+1$ and $s_i=-1$ denote spin-up and spin-down electrons, respectively. Here $\bm G_{1,2}$ are the primitive reciprocal lattice vectors of the simulation cell, and $\|\bm r_i\|_{\rm PBC}$ denotes the periodic distance defined in Ref.~\cite{PBC_varible}. Since the external moir\'e potential explicitly breaks continuous translational invariance, the absolute positions of the electrons are retained as physically relevant input variables.

For each pair of electrons, we introduce the corresponding two-body feature
\begin{equation}
\begin{aligned}
\bm x_{ij}^{(0)}=
\big(
&\sin(\bm G_1\!\cdot\!\bm r_{ij}),
\sin(\bm G_2\!\cdot\!\bm r_{ij}),
\cos(\bm G_1\!\cdot\!\bm r_{ij}),
\cos(\bm G_2\!\cdot\!\bm r_{ij}), \|\bm r_{ij}\|_{\rm PBC},
s_{ij}
\big),
\end{aligned}
\end{equation}
where $\bm r_{ij}=\bm r_i-\bm r_j$ and $s_{ij}=s_i s_j$. The one-body and two-body input features are then lifted to higher-dimensional representations as
\begin{equation}
\bm g_i^{(0)}=[\bm x_i^{(0)},\bm h_i^{(0)}],
\qquad
\bm g_{ij}^{(0)}=[\bm x_{ij}^{(0)},\bm h_{ij}^{(0)}],
\end{equation}
where $[\cdot,\cdot]$ denotes vector concatenation. The initial hidden embeddings are generated by multilayer perceptrons (MLPs),
\begin{equation}
\bm h_i^{(0)}=f^{(0)}(\bm x_i^{(0)}),
\qquad
\bm h_{ij}^{(0)}=\tilde f^{(0)}(\bm x_{ij}^{(0)}).
\end{equation}

The central component of the ansatz is a message-passing neural network, which iteratively propagates pairwise information among electrons. At layer $l$, we construct the message from electron $j$ to electron $i$ as
\begin{equation}
\bm m_{ij}^{(l)}
=
\mathrm{GELU}
\!\left(
\sum_{n}
\bm Q_{in}^{(l)}
\odot
\bm K_{nj}^{(l)}
\right)
\odot
\eta^{(l)}\!\left(\bm g_{ij}^{(l)}\right),
\end{equation}
where $\odot$ denotes element-wise multiplication. The query and key vectors are obtained from the two-body features through linear projections,
\begin{equation}
\bm Q_{ij}^{(l)}=W_Q^{(l)} \bm g_{ij}^{(l)},
\qquad
\bm K_{ij}^{(l)}=W_K^{(l)} \bm g_{ij}^{(l)}.
\end{equation}

The messages are used to update both the one-body and two-body hidden features. Specifically, for $l=0,\ldots,L-1$, we define
\begin{equation}
\bm h_i^{(l+1)}
=
f^{(l+1)}
\left(
\left[
\bm g_i^{(l)},
\sum_{j\neq i}\bm m_{ij}^{(l)}
\right]
\right),\qquad \bm h_{ij}^{(l+1)}
=
\tilde f^{(l+1)}
\left(
\left[
\bm g_{ij}^{(l)},
\bm m_{ij}^{(l)}
\right]
\right).
\end{equation}
The updated one-body and two-body features are then formed by reattaching the original input variables,
\begin{equation}
\bm g_i^{(l+1)}
=
[\bm x_i^{(0)},\bm h_i^{(l+1)}],
\qquad
\bm g_{ij}^{(l+1)}
=
[\bm x_{ij}^{(0)},\bm h_{ij}^{(l+1)}].
\end{equation}
Thus, the bare geometric and spin information is reintroduced at every layer through skip connections. This improves numerical stability, facilitates optimization, and helps the network retain direct access to the original physical coordinates while building many-body correlations. In general, the functions $\eta^{(l)}$, $f^{(l)}$, and $\tilde f^{(l)}$ are allowed to be layer dependent.

After $L$ message-passing layers, the backflow displacement of each electron is obtained from the final one-body feature by a linear projection,
\begin{equation}
\delta \bm r_i
=
W\,\bm g_i^{(L)}
\equiv
\delta \bm r_i(\bm r_i,\{\bm r_{j\neq i}\}).
\end{equation}
Because $\bm g_i^{(L)}$ is constructed from periodic input features, the resulting backflow coordinate is compatible with the periodic boundary conditions. Moreover, $\delta\bm r_i$ is an explicit many-body function of the full electronic configuration. The construction also preserves the required permutation symmetry: under any exchange of two electrons $j$ and $k$ with $j,k\neq i$, the displacement $\delta\bm r_i$ remains invariant.

The variational wave function is written as a neural Jastrow--backflow state.
For each spin sector $\sigma\in\{\uparrow,\downarrow\}$, we define the
spin-resolved Slater matrix
\begin{equation}
    D_{ia}^{\sigma}
    =
    \Phi_{a}^{\sigma}
    \left(
        \bm r_i;\delta\bm r_i
    \right),
    \qquad
    i,a=1,\ldots,N_{\sigma},
\end{equation}
where the electron index $i$ is defined separately within each spin sector.
The full variational wave function is then given by
\begin{equation}
    \Psi(\bm R)
    =
    J(\bm R)\,
    \det D^{\uparrow}\,
    \det D^{\downarrow}.
\end{equation}
Here $\bm R=(\bm r_1,\ldots,\bm r_{N_e})^{T}$ denotes the full electronic
configuration. The factor $J(\bm R)$ is a global Jastrow factor parameterized
by neural networks, while the orbitals
$\Phi_a^{\sigma}(\bm r_i;\delta\bm r_i)$ depend separately on the bare
electronic coordinate $\bm r_i$ and the message-passing backflow variable
$\delta\bm r_i$. In the fully polarized sector, the wave function contains
only a single Slater determinant.

\begin{figure*}[t]
\centering
\includegraphics[width=1.0\textwidth]{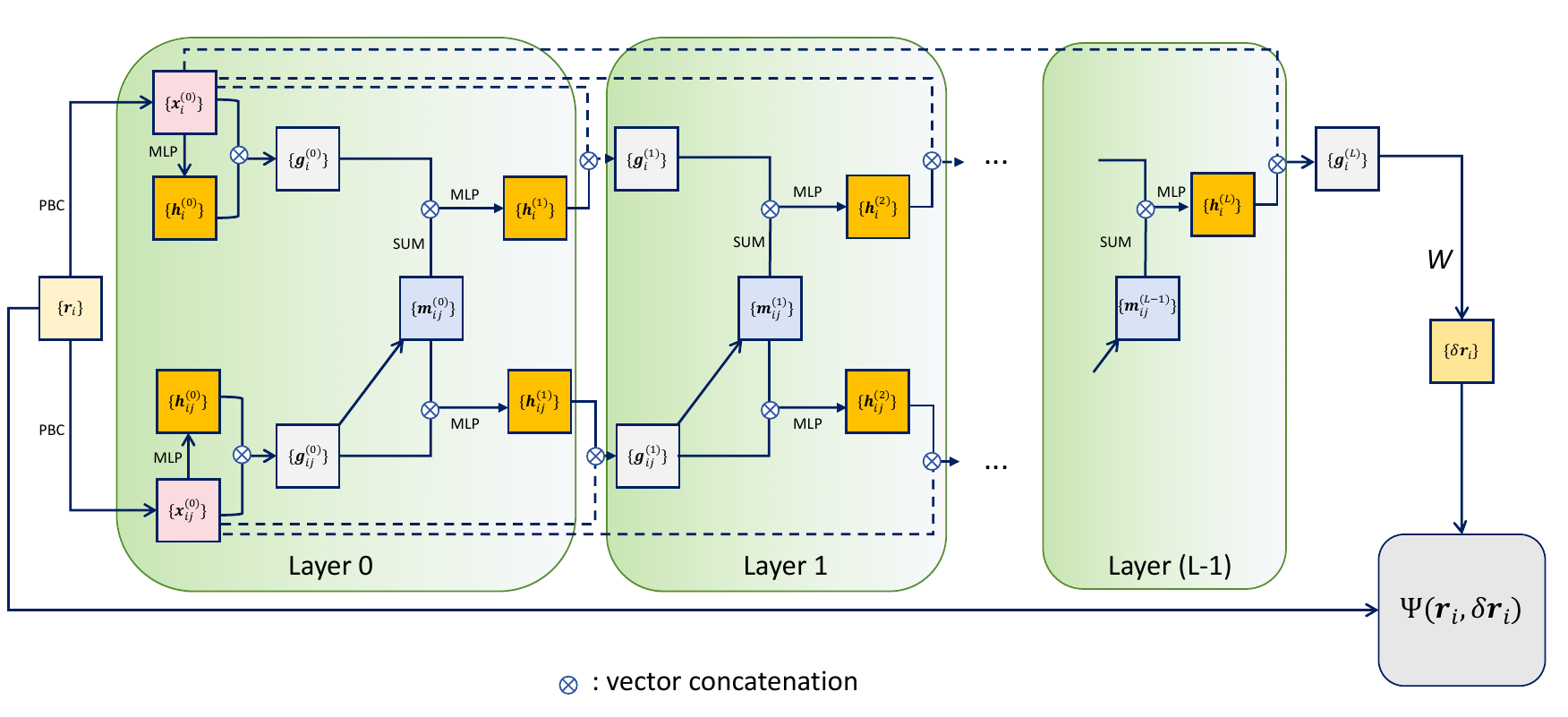}
\caption{Sketch map of MPNN.}
\label{Sketch_map_MPNN}
\end{figure*}

We provide a schematic illustration of the algorithm in Fig.~\ref{Sketch_map_MPNN}. Unless otherwise specified, we use $\tanh(x)$ as the nonlinear activation function in all multilayer perceptrons (MLPs), and include bias vectors in all affine layers. The main neural-network and simulation parameters are summarized in Table~\ref{table_NQS}.

\begin{table}[h]
    \centering
    \begin{tabular}{|c|c|c|c|}
        \hline
        System size & $4\times4$ & Filling fraction $\nu$ & 2 \\
        \hline
        Dimension of $\bm h_i^{(l)}$ & 16 & Dimension of $\bm h_{ij}^{(l)}$ & 16 \\ 
        \hline
        Hidden dimension of MLPs & 64 & Bias included & Yes \\
        \hline
        Number of layers $L$ & 3 & Total parameters & 36605 \\
        \hline
    \end{tabular}
    \caption{Main simulation and neural-network parameters used in the message-passing neural-network ansatz.}
    \label{table_NQS}
\end{table}

The variational parameters are optimized using the Kronecker-factored approximate curvature (KFAC) optimizer. The optimization parameters are summarized in Table~\ref{table_opt}.

\begin{table}[h]
    \centering
    \begin{tabular}{|c|c|c|c|}
        \hline
        Learning rate & $10^{-3}$--$10^{-2}$ & Decay & 1.0 \\
        \hline
        L2 regularization & 0.0 & Momentum & 0.0 \\
        \hline
        EMA decay rate & 0.95 & Damping & $10^{-4}$--$10^{-3}$ \\
        \hline
        Batch size & 1024 & Norm constraint & $10^{-3}$ \\
        \hline
    \end{tabular}
    \caption{Optimization parameters used in the KFAC optimization.}
    \label{table_opt}
\end{table}

All simulations are performed on NVIDIA A100 GPUs.

\section{Derivatives of $F(B,V_0)$ near the triple point}
\label{dF}

We analyze the nonanalytic behavior of the zero-temperature free energy per
particle,
\begin{equation}
F(B,V_0)
=
\frac{E(B,V_0)}{N_e}
-
\mu_B g B\,\zeta(B,V_0),
\end{equation}
near the triple point, where different spin-polarization and crystalline
phases become nearly degenerate. Here
$E(B,V_0)\equiv E[\zeta(B,V_0),V_0]$ denotes the variational total
energy excluding the Zeeman contribution, while
\begin{equation}
    \zeta
    =
    \frac{N_\uparrow-N_\downarrow}{N_e}
\end{equation}
is the dimensionless spin polarization. Thus, $F(B,V_0)$ is also understood
as the free energy per particle.

We first consider the derivative with respect to the magnetic field at fixed
$V_0$. Close to the first-order spin-polarization boundary, the relevant
competing states are the unpolarized state with energy $E_0(V_0)$ and the
fully polarized state with energy $E_1(V_0)$. The free energy can therefore
be written as
\begin{equation}
\begin{aligned}
F(B,V_0)
=&\,
\frac{E_0(V_0)}{N_e}\,\Theta[B_c(V_0)-B]
+
\left[
\frac{E_1(V_0)}{N_e}-\mu_B g B
\right]\Theta[B-B_c(V_0)].
\end{aligned}
\end{equation}
where $\Theta(x)$ is the Heaviside step function. The transition field
$B_c(V_0)$ is determined by the level-crossing condition
\begin{equation}
\frac{E_1(V_0)-E_0(V_0)}{N_e}
=
\mu_B g B_c(V_0).
\end{equation}
Taking the derivative at fixed $V_0$ gives
\begin{equation}
\begin{aligned}
\left.
\frac{\partial F}{\partial B}
\right|_{V_0}
=&
-\mu_B g\,\Theta[B-B_c(V_0)]
+
\left[
\frac{E_1(V_0)-E_0(V_0)}{N_e}
-\mu_B g B_c(V_0)
\right]
\delta[B-B_c(V_0)].
\end{aligned}
\end{equation}
The coefficient of the delta-function term vanishes by the level-crossing
condition. Hence,
\begin{equation}
\left.
\frac{\partial F}{\partial B}
\right|_{V_0}
=
-\mu_B g\,\Theta[B-B_c(V_0)]
=
-\mu_B g\,\zeta(B,V_0).
\end{equation}
Thus, $\left.\partial F/\partial B\right|_{V_0}$ jumps discontinuously across
the spin-polarization transition, as expected for a first-order transition
with a discontinuous magnetization.

We next consider the derivative with respect to the moir\'e potential
strength at fixed magnetic field. Near the transition boundary, the free energy per particle takes the form
\begin{equation}
\begin{aligned}
F(B,V_0)
=&\,
\frac{E_0(V_0)}{N_e}\,\Theta[V_0-V_c(B)]
+
\left[
\frac{E_1(V_0)}{N_e}-\mu_B g B
\right]\Theta[V_c(B)-V_0],
\end{aligned}
\end{equation}
where $V_c(B)$ denotes the transition moir\'e potential at fixed $B$. The
transition condition is
\begin{equation}
\frac{E_1[V_c(B)]-E_0[V_c(B)]}{N_e}
=
\mu_B g B.
\end{equation}
Differentiating with respect to $V_0$ at fixed $B$, we obtain
\begin{equation}
\begin{aligned}
\left.
\frac{\partial F}{\partial V_0}
\right|_{B}
=&
\frac{\partial (E_0/N_e)}{\partial V_0}\,
\Theta[V_0-V_c(B)]
+
\frac{\partial (E_1/N_e)}{\partial V_0}\,
\Theta[V_c(B)-V_0]
+
\left[
\frac{E_0(V_0)-E_1(V_0)}{N_e}+\mu_B g B
\right]\delta[V_0-V_c(B)].
\end{aligned}
\end{equation}
Again, the delta-function term vanishes on the transition boundary. Therefore,
\begin{equation}
\left.
\frac{\partial F}{\partial V_0}
\right|_{B}
=
\frac{\partial (E_0/N_e)}{\partial V_0}\,
\Theta[V_0-V_c(B)]
+
\frac{\partial (E_1/N_e)}{\partial V_0}\,
\Theta[V_c(B)-V_0].
\end{equation}
This derivative is generally discontinuous across the transition boundary
whenever the two competing phases have different slopes with respect to
$V_0$. Along the path FPM $\rightarrow$ ON-MC $\rightarrow$ s-MI,
the structure is further enriched because $\partial(E_1/N_e)/\partial V_0$ itself
is discontinuous at the FPM--ON-MC transition. The free energy surface
therefore contains multiple first-derivative discontinuities emanating from
the vicinity of the triple point.

\begin{figure}[h]
\includegraphics[trim=0 0 0 0, clip,width=0.5\columnwidth]{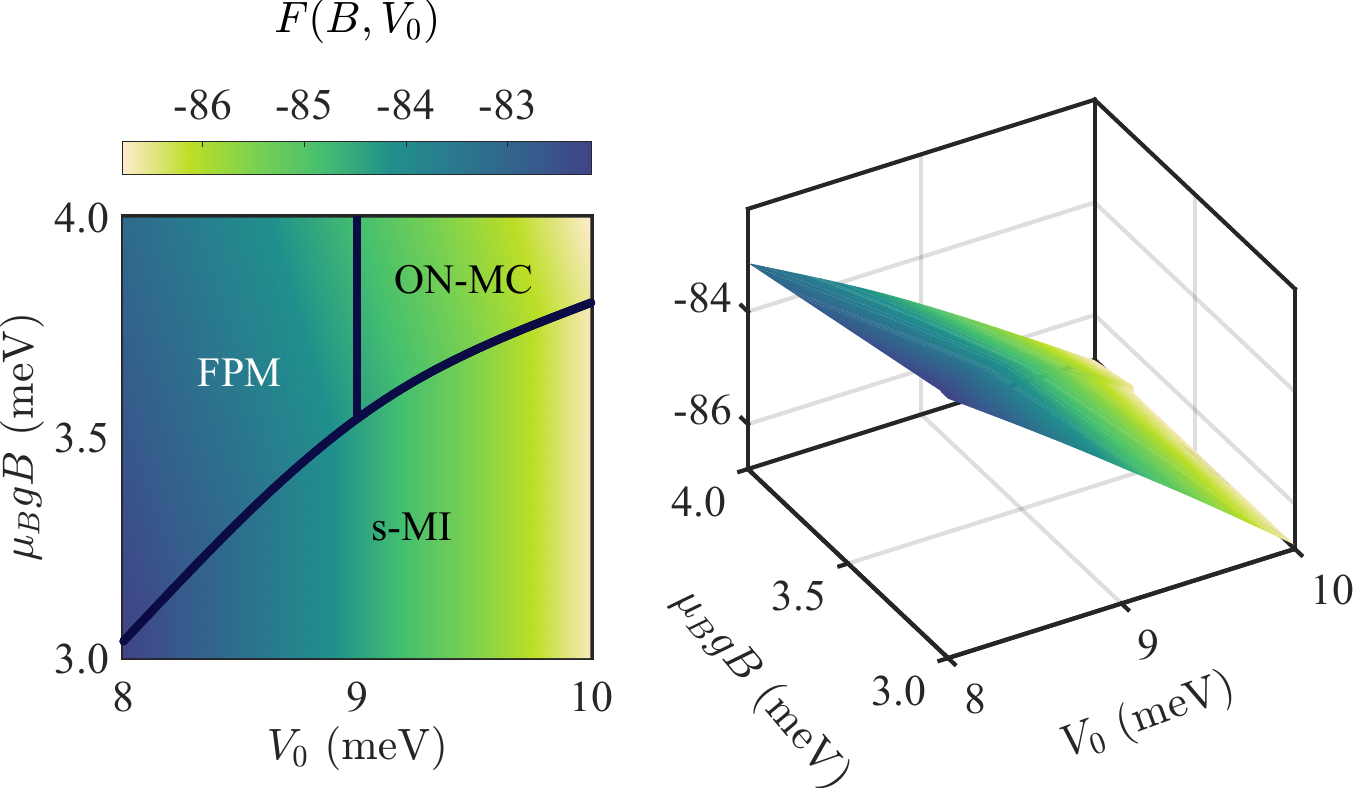}
\caption{Free energy $F(B,V_0)$ near the triple point A shown in
Fig.~\ref{key_map}(b). The piecewise structure of the free-energy surface
reflects the crossings between competing spin and crystalline sectors. The
resulting first-derivative discontinuities mark the first-order transition
boundaries in the $B$--$V_0$ plane. The results shown here are obtained for the Coulomb interaction.}
\label{F_B_V0}
\end{figure}

\section{Hellmann--Feynman theorem and concavity in $V_0$}
\label{Hellmann_Feynman}

In this appendix, we discuss two general properties of the lowest total energy $E_\zeta(V_0)$ in a fixed spin-polarization sector $\zeta$. First,
$E_\zeta(V_0)$ is a concave function of the moir\'e potential strength
$V_0$. Second, for positive $V_0$, its derivative is nonpositive,
consistent with the fact that the moir\'e potential lowers the energy by
favoring density accumulation near its minima.

We write the many-body Hamiltonian as $\hat H(V_0)=\hat H_0+V_0\hat{\mathcal V}_M$,
where $\hat H_0$ contains the kinetic and interaction terms, and
$\hat{\mathcal V}_M$ is the moir\'e potential operator with unit amplitude.
Explicitly, $\hat{\mathcal V}_M=\sum_{i=1}^{N_e}v_M(\bm r_i)$, and $v_M(\bm r)=-\sum_{n=1}^{6}e^{i\bm G_M^{(n)}\cdot\bm r}$.
Equivalently, pairing opposite reciprocal-lattice vectors gives $    v_M(\bm r)=-2\sum_{\alpha=1}^{3}\cos\left(\bm G_M^{(\alpha)}\cdot\bm r\right)$.
The spatial average of $v_M(\bm r)$ over the moir\'e unit cell vanishes.
Importantly, the following analysis does not rely on the specific form of
the interaction potential.

Since $\hat H(V_0)$ conserves the total spin-$z$ component, the Hilbert space
can be decomposed into sectors labeled by $\zeta$. We define the lowest
energy in a fixed-$\zeta$ sector as
\begin{equation}
    E_\zeta(V_0)
    =
    \min_{\substack{
        |\psi\rangle\in\mathcal H_\zeta\\
        \langle\psi|\psi\rangle=1
    }}
    \langle\psi|\hat H(V_0)|\psi\rangle.
\end{equation}

We first show that $E_\zeta(V_0)$ is concave in $V_0$. For
$0<\lambda<1$, the linearity of the Hamiltonian in $V_0$ gives
\begin{equation}
    \hat H\!\left[
        \lambda V_1+(1-\lambda)V_2
    \right]
    =
    \lambda\hat H(V_1)
    +
    (1-\lambda)\hat H(V_2).
\end{equation}
For any normalized state $|\psi\rangle\in\mathcal H_\zeta$, the variational
principle implies
\begin{equation}
    E_\zeta(V_1)
    \leq
    \langle\psi|\hat H(V_1)|\psi\rangle,
    \qquad
    E_\zeta(V_2)
    \leq
    \langle\psi|\hat H(V_2)|\psi\rangle.
\end{equation}
Therefore,
\begin{equation}
\begin{aligned}
    &\lambda E_\zeta(V_1)
    +(1-\lambda)E_\zeta(V_2)\leq\left\langle\psi\left|
        \hat H\!\left[
            \lambda V_1+(1-\lambda)V_2
        \right]
    \right|\psi\right\rangle.
\end{aligned}
\end{equation}
Choosing $|\psi\rangle$ to be the ground state of
$\hat H[\lambda V_1+(1-\lambda)V_2]$ in the same $\zeta$ sector yields
\begin{equation}
    E_\zeta\!\left[
        \lambda V_1+(1-\lambda)V_2
    \right]
    \geq
    \lambda E_\zeta(V_1)
    +
    (1-\lambda)E_\zeta(V_2).
\end{equation}
Thus, $E_\zeta(V_0)$ is a concave function of $V_0$.

At points where $E_\zeta(V_0)$ is differentiable, the Hellmann--Feynman
theorem gives
\begin{equation}
    \frac{\partial E_\zeta(V_0)}{\partial V_0}
    =
    \left\langle
        \psi_{\zeta,0}(V_0)
        \left|
            \hat{\mathcal V}_M
        \right|
        \psi_{\zeta,0}(V_0)
    \right\rangle,
    \label{eq:HF_V0}
\end{equation}
where $|\psi_{\zeta,0}(V_0)\rangle$ is the ground state in the fixed
$\zeta$ sector with moir\'e potential strength $V_0$. At a first-order
transition or level crossing, the same statement applies separately to the
appropriate one-sided derivatives.

We define the density Fourier component as $\widetilde n(\bm q)=\left\langle\sum_{i=1}^{N_e}e^{i\bm q\cdot\bm r_i}\right\rangle.$
Substituting the definition of $\hat{\mathcal V}_M$ into
Eq.~\eqref{eq:HF_V0}, we find
\begin{equation}
    \frac{\partial E_\zeta(V_0)}{\partial V_0}
    =
    -\sum_{n=1}^{6}
    \widetilde n\!\left(
        \bm G_M^{(n)}
    \right).
\end{equation}

For $V_0=0$, the Hamiltonian is translationally invariant. A translationally
invariant state has a uniform density, and hence $\left\langle\hat{\mathcal V}_M\right\rangle=0,$
because the spatial average of $v_M(\bm r)$ vanishes. Using such a ground
state of $\hat H(0)$ as a variational state for $V_0>0$ gives $E_\zeta(V_0)\leq E_\zeta(0)$.
Consequently, the right derivative of $E_\zeta(V_0)$ at $V_0=0$ is
nonpositive. Since the slope of a concave function cannot increase with
$V_0$, all one-sided derivatives at positive $V_0$ satisfy
\begin{equation}
    \frac{\partial E_\zeta(V_0)}{\partial V_0}
    \leq
    0.
\end{equation}
This establishes the monotonic decrease of the lowest energy
in each fixed spin-polarization sector as the moir\'e potential strength is
increased.

The same conclusion can also be obtained from perturbation theory away from
degeneracies. Let $ E_{\zeta,n}$ denote the total eigenenergy of the
$n$th eigenstate $|\psi_{\zeta,n}\rangle$ in the fixed-$\zeta$ sector, where
$n=0$ labels the ground state. 
Differentiating Eq.~\eqref{eq:HF_V0} with respect to $V_0$ gives
\begin{equation}
\begin{aligned}
    \frac{\partial^2E_{\zeta,0}}{\partial V_0^2}
    =
    \Big[
    &
    \left(
        \partial_{V_0}
        \langle\psi_{\zeta,0}|
    \right)
    \hat{\mathcal V}_M
    |\psi_{\zeta,0}\rangle+
    \langle\psi_{\zeta,0}|
    \hat{\mathcal V}_M
    \left(
        \partial_{V_0}
        |\psi_{\zeta,0}\rangle
    \right)
    \Big].
\end{aligned}
\end{equation}
For a nondegenerate ground state, first-order perturbation theory yields
\begin{equation}
    \partial_{V_0}|\psi_{\zeta,0}\rangle
    =
    \sum_{n\neq0}
    \frac{
        \langle\psi_{\zeta,n}|
        \hat{\mathcal V}_M
        |\psi_{\zeta,0}\rangle
    }{
         E_{\zeta,0}
        -
         E_{\zeta,n}
    }
    |\psi_{\zeta,n}\rangle.
\end{equation}
Here $|\psi_{\zeta,n}\rangle$ denotes the $n$th eigenstate in the
fixed-$\zeta$ sector, with corresponding total eigenenergy
$E_{\zeta,n}$, and $n=0$ labels the ground state. Substituting this
expression into the second derivative gives
\begin{equation}
    \frac{\partial^2E_{\zeta,0}}{\partial V_0^2}
    =
   2
    \sum_{n\neq0}
    \frac{
        \left|
            \langle\psi_{\zeta,n}|
            \hat{\mathcal V}_M
            |\psi_{\zeta,0}\rangle
        \right|^2
    }{
         E_{\zeta,0}
        -
        E_{\zeta,n}
    }
    \leq
    0,
\end{equation}
since $E_{\zeta,n}>E_{\zeta,0}$
for all excited states above a nondegenerate ground state. This perturbative
expression is the local version of the general concavity proved above. It is
valid away from degeneracies, while the variational argument remains valid
globally and also covers possible nonanalytic points.

\section{Fourier transform of density}

\begin{figure}[h]
\includegraphics[width=0.5\columnwidth]{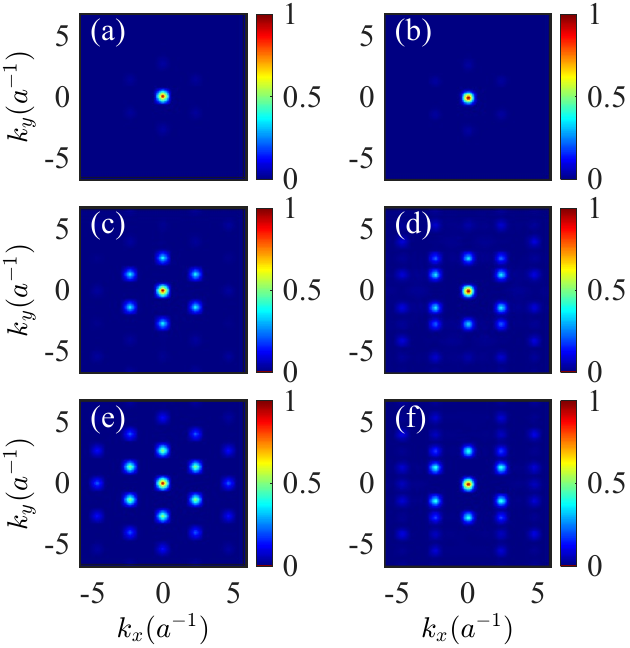}
\caption{
Fourier amplitude of the real-space density in the $\zeta=0$ and $\zeta=1$ sectors.
Panels (a)--(d) show the results for the Coulomb interaction:
(a) $\zeta=0$, $V_0=1\,\mathrm{meV}$;
(b) $\zeta=0$, $V_0=11\,\mathrm{meV}$;
(c) $\zeta=1$, $V_0=1\,\mathrm{meV}$;
(d) $\zeta=1$, $V_0=11\,\mathrm{meV}$.
Panels (e) and (f) show the corresponding insulating states for the P\"oschl--Teller
(PT) interaction with $V_{\rm PT}=5$ and $\kappa^{-1}=a$:
(e) $\zeta=0$, $V_0=11\,\mathrm{meV}$;
(f) $\zeta=1$, $V_0=11\,\mathrm{meV}$.
For $V_0=1\,\mathrm{meV}$, the PT results are similar to those obtained with the Coulomb interaction and are therefore not shown.
The normalization is chosen such that the peak at $\bm k=0$ equals unity.
}
\label{nk_fig}
\end{figure}

To further elucidate the nature of the metallic and insulating phases, we examine the Fourier amplitude of the density profile, $\tilde n(\bm k)$. Fig.~\ref{nk_fig} shows $\tilde n(\bm k)$ for representative moir\'e potential strengths. For the Coulomb interaction, we compare $V_0=1\,\mathrm{meV}$ and $V_0=11\,\mathrm{meV}$, corresponding to the metallic and insulating regimes, respectively. For the P\"oschl--Teller (PT) interaction, we show the insulating states at $V_0=11\,\mathrm{meV}$.

For weak moir\'e potentials, $\tilde n(\bm k)$ displays only weak features at the moir\'e reciprocal lattice vectors $\bm G_M$, indicating that a pronounced static density modulation has not yet developed. By contrast, for strong moir\'e potentials, sharp Bragg peaks emerge, signaling the formation of a molecular crystalline state pinned by the moir\'e potential.
For the Coulomb interaction in the unpolarized sector, the dominant peaks appear at the reciprocal lattice vectors $\bm G_M$, consistent with a symmetry-preserving molecular crystal. In the fully polarized sector, however, $\tilde n(\bm k)$ exhibits a richer peak structure. In addition to the peaks at $\bm G_M$, extra peaks appear away from the reciprocal lattice vectors, reflecting the enlarged unit cell and the reduced rotational symmetry of the orbital-nematic molecular crystal. Similar correlated electronic phases with orbital-nematic character have been discussed in a variety of strongly interacting systems~\cite{Kivelson1998,Fradkin2010,Borzi2007,Matty2022}.

The PT interaction leads to a distinct momentum-space pattern in the insulating regime. For $\zeta=0$, the peaks at $\bm G_M$ are stronger than in the Coulomb case, and additional higher-order peaks appear at more distant reciprocal lattice vectors. This reflects the stronger density modulation produced by the more localized molecular states. For $\zeta=1$, we do not observe the extra peaks away from the reciprocal lattice vectors that appear in the Coulomb case. Instead, the Bragg peaks along the $k_y$ direction are strongly enhanced relative to the other reciprocal-lattice peaks, consistent with a different orientational pattern of the $p$-wave molecules.

\begin{figure}[h]
\includegraphics[width=0.5\columnwidth]{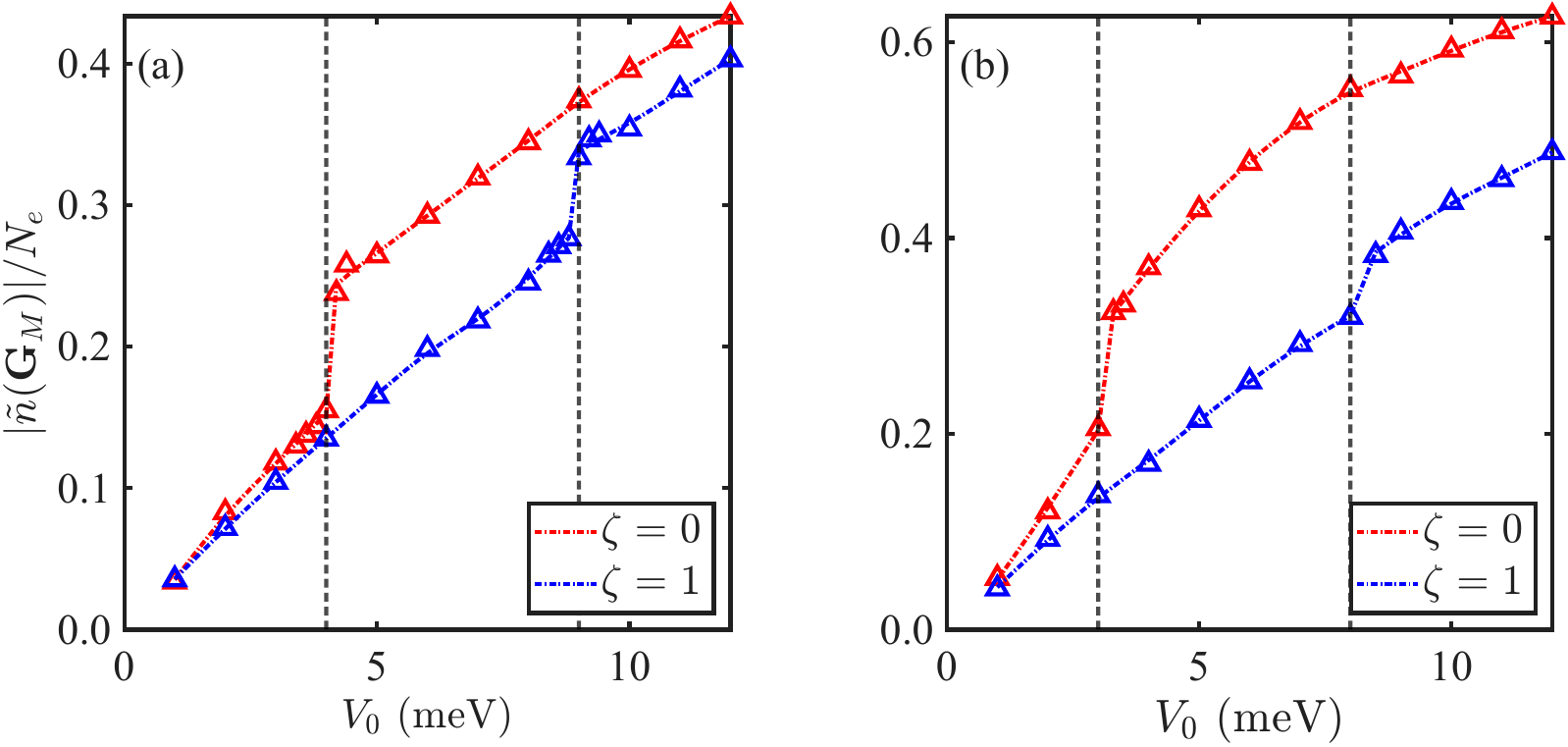}
\caption{
Fourier component of the density at the principal reciprocal lattice vector $\bm G_M$ as a function of $V_0$ for
(a) the Coulomb interaction and
(b) the P\"oschl--Teller interaction ($V_{\rm PT}=5$ and $\kappa^{-1}=a$).
}
\label{nk_interaction}
\end{figure}

Figure~\ref{nk_interaction} summarizes this behavior through the Fourier component $\tilde n(\bm G_M)$ as a function of $V_0$. Across the metal--insulator transition, $\tilde n(\bm G_M)$ exhibits an abrupt enhancement, demonstrating that the density modulation at the principal moir\'e reciprocal lattice vector provides a direct momentum-space diagnostic of electron localization. The larger value of $\tilde n(\bm G_M)$ for the PT interaction is consistent with the stronger localization and larger density contrast in the insulating phase.

\section{Dimensionless correlation functions}
\label{DimensionlessC}

\begin{figure}[h]
\includegraphics[trim=0 0 0 0, clip,width=0.5\columnwidth]{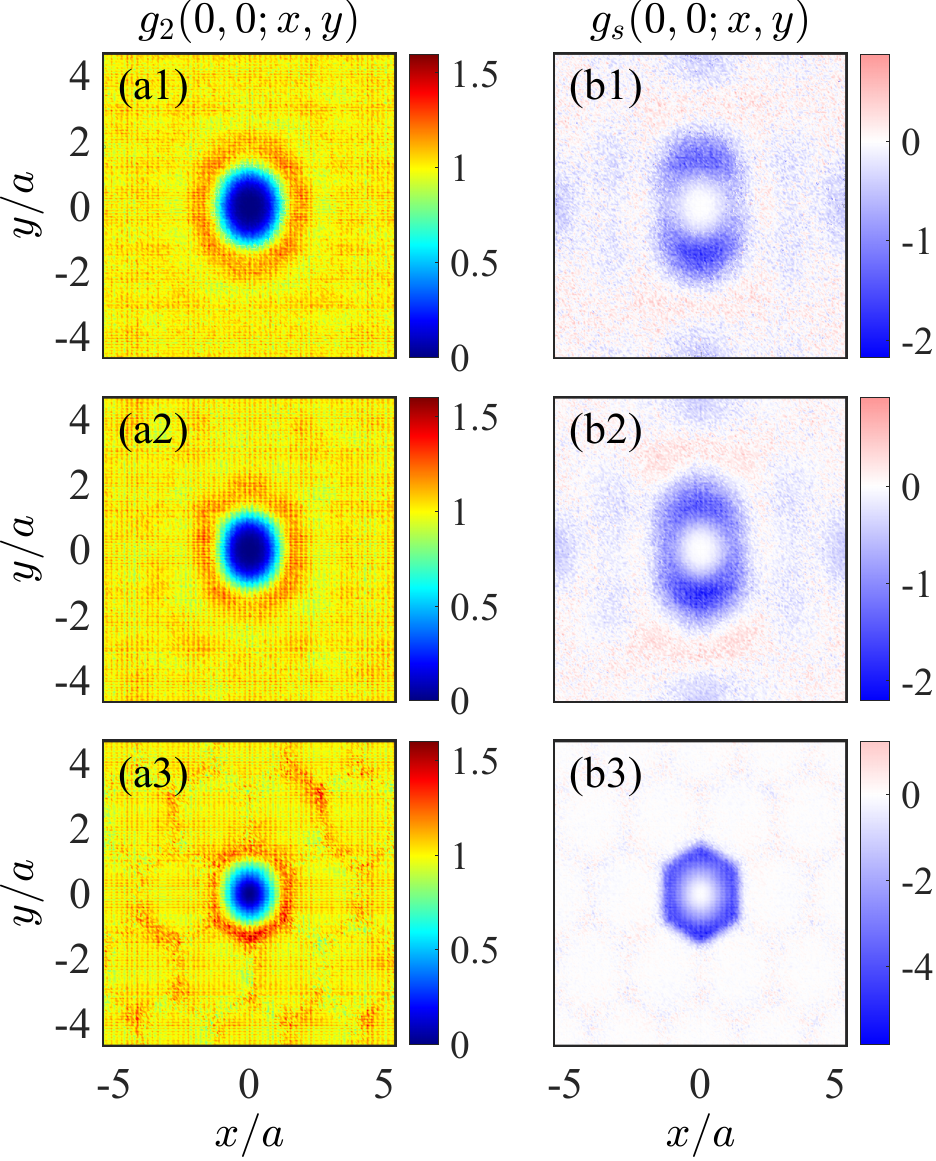}
\caption{Dimensionless real-space correlation functions in the unpolarized sector. Panels (a1)--(a3) show the dimensionless density--density correlation function $g_2(0,0;x,y)$, while panels (b1)--(b3) show the dimensionless spin correlation function $g_s(0,0;x,y)$. The moir\'e potential strengths are $V_0=1\,\mathrm{meV}$ for (a1,b1), $V_0=3\,\mathrm{meV}$ for (a2,b2), and $V_0=8\,\mathrm{meV}$ for (a3,b3). With increasing $V_0$, the dimensionless correlations become increasingly shaped by the moir\'e unit cell and are enhanced near the Wigner--Seitz-cell boundary, revealing short-range density and spin correlations beyond the static density distribution. The results shown here are obtained for the Coulomb interaction.}
\label{correlation}
\end{figure}

We further analyze the dimensionless correlation functions in the unpolarized sector, as shown in Fig.~\ref{correlation}. The dimensionless pair correlation function is defined as
\begin{equation}
g_2(\bm 0;\bm r)
=
\frac{C_n(\bm 0,\bm r)}
{n(\bm 0)n(\bm r)},
\end{equation}
while the dimensionless spin correlation function is defined as
\begin{equation}
g_s(\bm 0;\bm r)
=
\sum_{\sigma,\sigma'}
\left(2\delta_{\sigma\sigma'}-1\right)
g_{\sigma\sigma'}(\bm 0;\bm r).
\end{equation}
Here, the spin-resolved dimensionless pair correlation function is given by
\begin{equation}
g_{\sigma\sigma'}(\bm 0;\bm r)
=
\frac{
\langle \hat{n}_{\sigma}(\bm 0)\hat{n}_{\sigma'}(\bm r)\rangle
}{
n_{\sigma}(\bm 0)n_{\sigma'}(\bm r)
}.
\end{equation}

In the weak moir\'e-potential regime, exemplified by $V_0=1\,\mathrm{meV}$, the dimensionless density--density correlation function $g_2(\bm 0;\bm r)$ remains approximately isotropic, consistent with a Fermi-liquid-like state. By contrast, the dimensionless spin correlation function $g_s(\bm 0;\bm r)$ already exhibits pronounced anisotropic features. This separation between nearly isotropic charge correlations and strongly anisotropic spin correlations closely resembles the nematic spin-correlated liquid (NSCL) proposed by Zhang and collaborators~\cite{ShiweiZhang}. Since the interaction parameter in our system is $r_s\simeq 11.13$, which lies in the regime where such spin-nematic correlations are expected to be important, we interpret the weak-$V_0$ state as an NSCL-like liquid weakly modulated by the moir\'e potential.

As the moir\'e potential is increased to $V_0=3\,\mathrm{meV}$, the outer resonance peaks in $g_2(0,0;x,y)$ begin to lose their isotropic character, indicating that the charge correlations are increasingly shaped by the moir\'e unit cell. At the same time, the short-range antiferromagnetic structure in $g_s(0,0;x,y)$ becomes more clearly resolved. Thus, increasing $V_0$ enhances the coupling between the preexisting anisotropic spin correlations and the spatial modulation imposed by the moir\'e potential.

For a larger moir\'e potential, such as $V_0=8\,\mathrm{meV}$, the region where $g_2$ exhibits significant nontrivial correlations shrinks substantially. The dominant correlation features become concentrated near the boundary of the Wigner--Seitz cell, signaling a clear departure from the nearly isotropic structure observed at weak potential. A similar spatial organization appears in $g_s(0,0;x,y)$, whose strongest response is also located near the Wigner--Seitz-cell boundary.

This behavior reflects the combined effects of Coulomb correlations, itinerancy, and moir\'e-induced density modulation. The strong short-range Coulomb repulsion suppresses configurations in which two electrons approach one another, causing both $g_2(0,0;x,y)$ and $g_s(0,0;x,y)$ to vanish near $(x,y)=(0,0)$ and giving rise to a ring-like correlation structure around the reference electron. As $V_0$ increases, this structure is further reshaped by the moir\'e unit cell, leading to enhanced dimensionless correlations near the cell boundary. Importantly, regions with relatively low density do not necessarily exhibit weak dimensionless spin correlations. This indicates that the spin correlations are not determined solely by the static density distribution, but also reflect genuine many-body spin fluctuations encoded in the wave function.

\section{Connected structure factor}
\label{app_connected_Sk}

\begin{figure}[h]
\includegraphics[trim=0 0 0 0, clip,width=0.5\columnwidth]{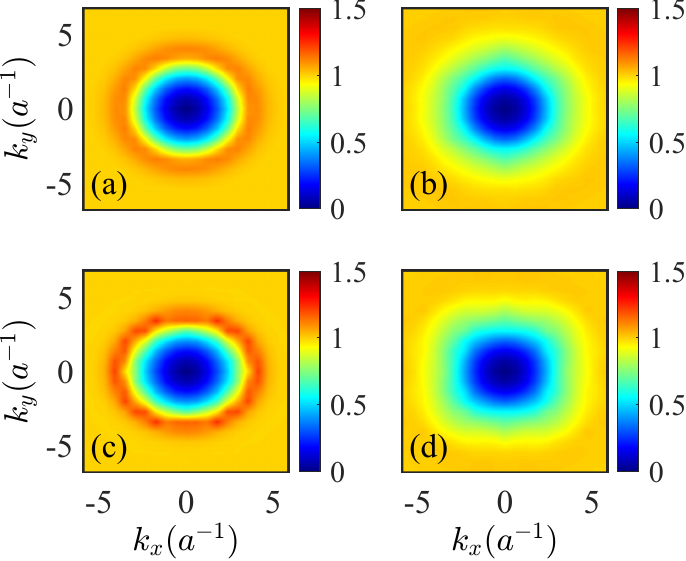}
\caption{Connected structure factor $S(\bm k)$ in representative weak- and strong-moir\'e regimes. Panels (a)--(d) correspond to $(V_0,\zeta)=(1\,\mathrm{meV},0)$, $(11\,\mathrm{meV},0)$, $(1\,\mathrm{meV},1)$, and $(11\,\mathrm{meV},1)$, respectively. In the weak-potential regime, $S(\bm k)$ is nearly isotropic and displays a broad finite-$k$ peak around $k\simeq 4\,a^{-1}$, characteristic of Fermi-liquid-like correlations. In the localized crystalline regime, the subtraction of the inhomogeneous density profile removes the Bragg-like peaks associated with the static density modulation. For the fully polarized state, the additional symmetry breaking of the density pattern produces a strongly anisotropic, nearly rectangular structure in momentum space. The results shown here are obtained for the Coulomb interaction.}
\label{Sk_vs_V0_connected}
\end{figure}

We further compute the connected structure factor $S(\bm k)$ for the unpolarized ($\zeta=0$) and fully polarized ($\zeta=1$) states at representative moir\'e potential strengths, as shown in Fig.~\ref{Sk_vs_V0_connected}. In the weak-potential regime, exemplified by $V_0=1\,\mathrm{meV}$, $S(\bm k)$ is nearly isotropic for both spin polarizations. In each case, it exhibits a broad finite-$k$ peak around $k\simeq 4\,a^{-1}$, consistent with the short-range correlations of a Fermi-liquid-like state.

The behavior changes qualitatively in the strong-potential regime. At $V_0=11\,\mathrm{meV}$, where the system enters the localized crystalline regime, the connected structure factor of the unpolarized state shows no pronounced Bragg-like peaks. This behavior should be contrasted with the conventional structure-factor signature of Wigner crystallization in a translationally invariant system. In the present case, the inhomogeneous density profile induced by the moir\'e potential has been explicitly subtracted in $S(\bm k)$; consequently, the peaks associated with the static density modulation are removed by construction. The remaining signal therefore probes correlations beyond the one-body density profile. For the fully polarized state, the crystalline density pattern further lowers the lattice symmetry, and this symmetry breaking appears in the connected structure factor as a strongly anisotropic, nearly rectangular momentum-space structure.

\section{Band structures and density of states of the moir\'e system}
\label{App:DOS}

We present the noninteracting band structures and density of states (DOS) of the moir\'e system. To obtain a smooth DOS, we approximate the Dirac delta function by a Gaussian kernel and compute
\begin{equation}
\begin{aligned}
D(E)
&=
\frac{1}{N_{\bm k}}
\sum_{n}
\sum_{\bm k\in \mathrm{1BZ}}
\delta\left(E-E_n(\bm k)\right) \simeq
\frac{1}{N_{\bm k}}
\sum_{n}
\sum_{\bm k\in \mathrm{1BZ}}
\frac{1}{\sqrt{2\pi}\sigma}
\exp\left[
-\frac{\left(E-E_n(\bm k)\right)^2}{2\sigma^2}
\right],
\end{aligned}
\end{equation}
where $E_n(\bm k)$ is the energy eigenvalue with Bloch momentum $\bm k$ and band index $n$, and $N_{\bm k}$ denotes the number of sampled Bloch momenta in the first Brillouin zone (1BZ). In our calculations, we use a Gaussian broadening $\sigma=0.05E_0$, with the energy unit defined as $E_0=\hbar^2/(m^*a^2)$. 

\begin{figure}[h]
\includegraphics[width=0.6\columnwidth]{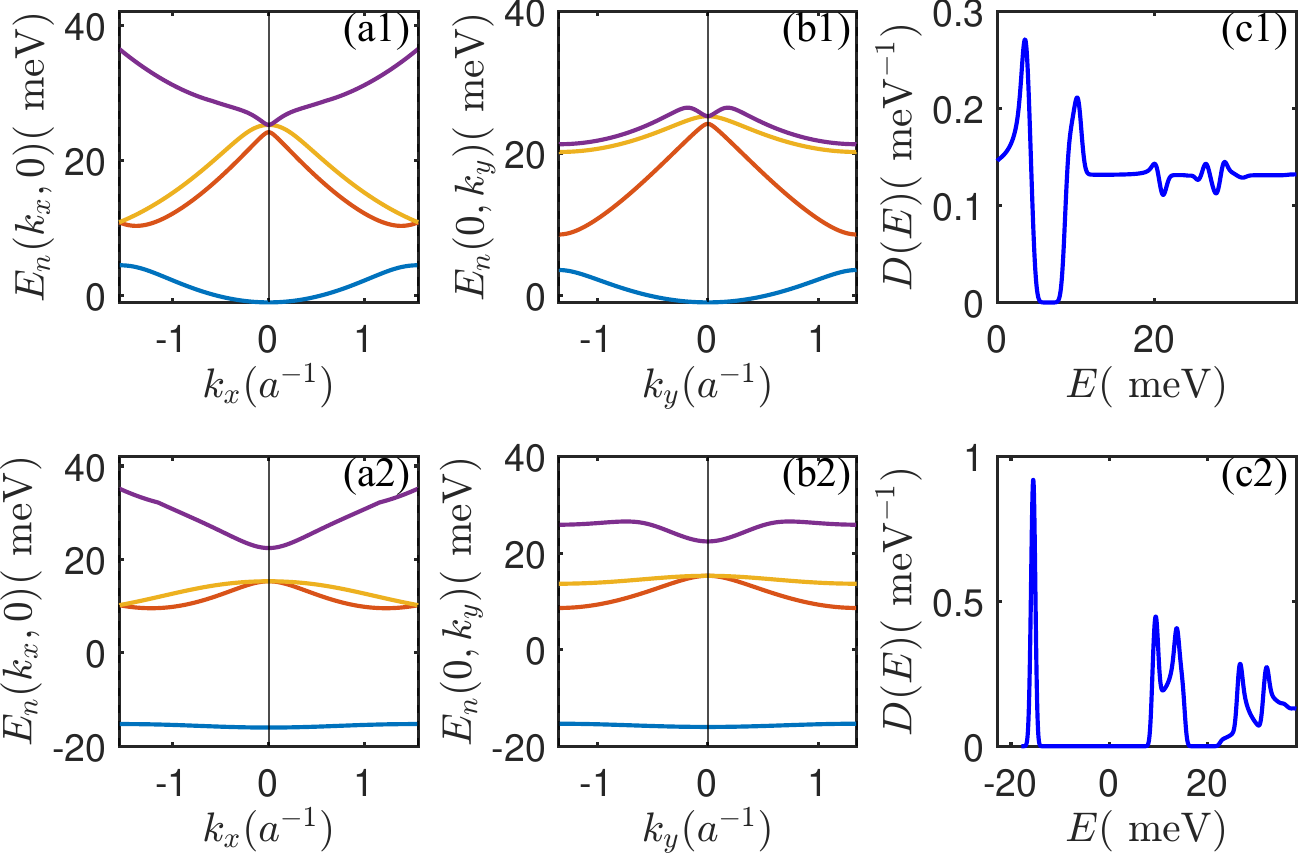}
\caption{Band cuts and density of states of the noninteracting moir\'e system. Panels (a1) and (a2) show cuts of the first four bands along the $k_x$ direction, while panels (b1) and (b2) show the corresponding cuts along the $k_y$ direction. Panels (c1) and (c2) show the density of states. The first row, (a1)--(c1), corresponds to $V_0=2\,\mathrm{meV}$, and the second row, (a2)--(c2), corresponds to $V_0=8\,\mathrm{meV}$.}
\label{fig:dos_band_structure}
\end{figure}

As shown in Fig.~\ref{fig:dos_band_structure}, the noninteracting moir\'e
spectrum already exhibits a finite global gap between the first and second
bands even for a relatively weak moir\'e potential, $V_0=2\,\mathrm{meV}$.
This gap is directly reflected in the DOS by the absence of states between
the first band and the higher-energy bands. Increasing the moir\'e potential
strength further enhances this band separation, as seen by comparing the
results for $V_0=2\,\mathrm{meV}$ and $V_0=8\,\mathrm{meV}$. In contrast, the
second and third bands are not separated by a global gap. The band cuts show
that these two bands touch, or become nearly degenerate, at certain momenta
in the Brillouin zone. Correspondingly, the DOS does not display a clear gap
between the second and third bands. These single-particle results provide a
useful reference for the interacting calculations: at filling $\nu=2$, the
noninteracting system is expected to form a band insulator once the lowest
band is isolated, while the higher bands remain connected in energy.

\section{$n_c$ and $E/N_e$ for P\"oschl--Teller interaction}\label{app:nc_E_PT}

\begin{figure}[h]
\centering
\includegraphics[width=0.5\columnwidth]{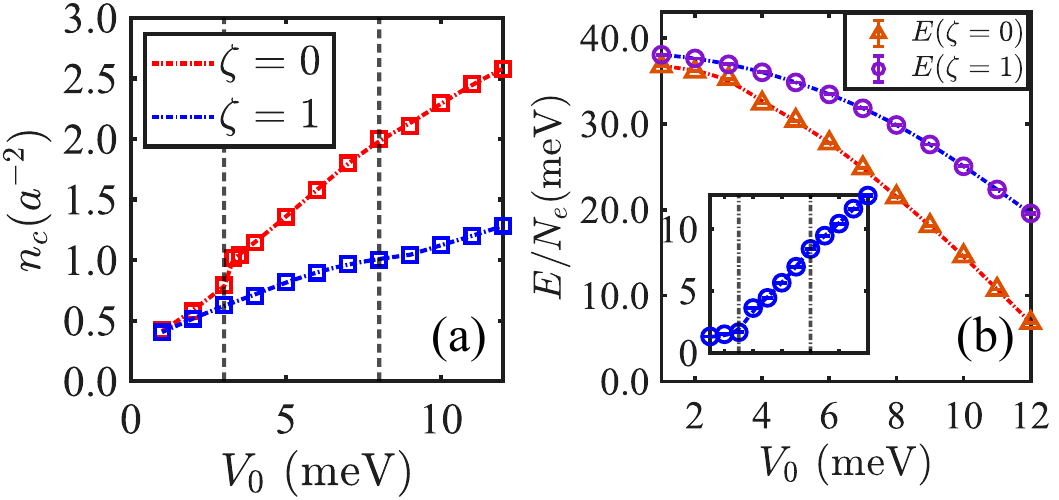}
\caption{
(a) Density at the center of a moir\'e-potential minimum $n_c$ in the $\zeta=0$ and $\zeta=1$ sectors for the PT interaction ($V_{\rm PT}=5$, $\kappa^{-1}=a$). (b) Energy per electron of $\zeta=0$ and $\zeta=1$ sectors with various $V_0$ for the PT interaction ($V_{\rm PT}=5$, $\kappa^{-1}=a$).  In the inset, we show the energy difference $\Delta E(\zeta=1)\equiv \left[E(\zeta=1)-E(\zeta=0)\right]/N_e$. The dot-dashed lines in the inset represent the metal--insulator transition points of $\zeta=0$ and $\zeta=1$ sectors, respectively.  
}
\label{nc_E_PT_fig}
\end{figure}

In Fig.~\ref{nc_E_PT_fig}, we show the evolution of $n_c$ and the average energy per particle $E/N_e$ as a function of $V_0$ for the PT interaction ($V_{\rm PT}=5$, $\kappa^{-1}=a$). For $\zeta=0$, $n_c$ still exhibits a discontinuous jump across the metal-insulator transition. However, in contrast to the Coulomb case, this jump corresponds to an increase rather than a decrease of $n_c$. This can be understood from the fact that the PT potential remains finite at $r=0$, so that the molecule does not need to suppress the density at its center in order to reduce the interaction energy. This is also consistent with the density profile discussed in the main text. For $\zeta=1$, the jump of $n_c$ across the transition is much less pronounced than in the Coulomb case.
The average energy per particle $E/N_e$ decreases with increasing $V_0$ for both spin polarizations, and the decrease becomes progressively faster at larger $V_0$. This behavior is consistent with the analysis presented before. The energy difference between the two spin-polarized states, shown by the blue curve in the inset, displays a more pronounced cusp at the metal-insulator transition of the $\zeta=0$ branch, whereas its variation near the transition of the $\zeta=1$ branch is less evident than in the Coulomb case.

\section{A simple picture for the Fourier component}\label{app:ONMC_S0}

We consider a phenomenological density ansatz for the orbital-nematic
molecular crystal,
\begin{equation}
    n(\bm r)
    =
    \sum_j n_0(\bm r-\bm R_j;\bm R_j),
\end{equation}
where \(\bm R_j\) denotes the position of the \(j\)-th moir\'e minimum.
The local density profile around each minimum is assumed to be
\begin{equation}
    n_0(\bm r;\bm R_j)
    =
    r^2 e^{-2\alpha r^2}
    \cos^2\!\left[
        \varphi+\gamma(\bm R_j)
    \right].
\end{equation}
Here \(\alpha^{-1/2}\) characterizes the spatial size of the molecular
orbital, while $\gamma(\bm R_j)$ parametrizes the angular dependence of the local $p$-wave density. With the convention above, the corresponding molecular director is oriented along $-\gamma(\bm R_j)$ modulo $\pi$. In general,
\(\gamma(\bm R_j)\) may depend on the lattice site.

In this simplified discussion, we retain only the orientational \(p\)-wave component and omit the isotropic \(s\)-wave contribution. Since the \(s\)-wave component is identical on all moir\'e sites, it contributes only at reciprocal vectors of the original moir\'e lattice and does not generate the half-shifted Fourier components associated with the ON--MC unit-cell doubling. The \(p\)-wave-only ansatz is therefore sufficient for identifying the additional symmetry-allowed Umklapp resonances. The \(s\)-wave component is restored in the quantitative calculation, where it affects the resonance energies and the distribution of optical weight.

We define the Fourier transform as $\tilde n(\bm k)=\int d^2r\,
    n(\bm r)e^{i\bm k\cdot\bm r}$.
Substituting the density ansatz gives
\begin{equation}
\begin{aligned}
    \tilde n(\bm k)
    &=
    \sum_j
    \int d^2r\,
    n_0(\bm r-\bm R_j;\bm R_j)
    e^{i\bm k\cdot\bm r}=
    \sum_j
    e^{i\bm k\cdot\bm R_j}
    \tilde n_0(\bm k;\bm R_j),
\end{aligned}
\end{equation}
where $\tilde n_0(\bm k;\bm R_j) =\int d^2r\,n_0(\bm r;\bm R_j)e^{i\bm k\cdot\bm r}$.

Writing $\bm k=k(\cos\varphi_{\bm k},\sin\varphi_{\bm k})$,
one obtains
\begin{equation}
\begin{aligned}
    \tilde n_0(\bm k;\bm R_j)
    =
    \pi e^{-k^2/(8\alpha)}
    \left[
        \frac{1}{8\alpha^2}
        -
        \frac{k^2}{32\alpha^3}
        \cos^2\!\left(
            \varphi_{\bm k}+\gamma(\bm R_j)
        \right)
    \right].
\end{aligned}
\end{equation}
Equivalently, this can be written as $\tilde n_0(\bm k;\bm R_j)= I_0(k)+I_1(\bm k;\bm R_j)$, with
\begin{equation}
    I_0(k)
    =
    \pi e^{-k^2/(8\alpha)}
    \left(
        \frac{1}{8\alpha^2}
        -
        \frac{k^2}{32\alpha^3}
    \right),\qquad I_1(\bm k;\bm R_j)
    =
    \pi e^{-k^2/(8\alpha)}
    \frac{k^2}{32\alpha^3}
    \sin^2\!\left[
        \varphi_{\bm k}+\gamma(\bm R_j)
    \right].
\end{equation}

Therefore,
\begin{equation}
    \tilde n(\bm k)
    =
    I_0(k)
    \sum_j e^{i\bm k\cdot\bm R_j}
    +
    \sum_j
    I_1(\bm k;\bm R_j)
    e^{i\bm k\cdot\bm R_j}.
\end{equation}

The first term is independent of the molecular orientation. For a finite
periodic lattice, we have $\sum_j e^{i\bm k\cdot\bm R_j}= N\sum_{\bm G_M}\delta_{\bm k,\bm G_M}$,
where \(N\) is the number of moir\'e unit cells and \(\bm G_M\) denotes
a moir\'e reciprocal lattice vector. Hence the orientation-independent
contribution is
\begin{equation}
    I_0(k)N
    \sum_{\bm G_M}
    \delta_{\bm k,\bm G_M}.
\end{equation}
This term gives Bragg peaks at the moir\'e reciprocal lattice vectors.
The peak heights are controlled by the Gaussian envelope
\(\exp[-k^2/(8\alpha)]\). Thus, a larger \(\alpha\), corresponding to a
more localized real-space molecular orbital, leads to a broader momentum-space
profile and relatively stronger finite-\(\bm G_M\) peaks.

We next discuss the orientation-dependent contribution. If the orientation is
site independent, namely $\gamma(\bm R_j)=\gamma$,
then
\begin{equation}
\begin{aligned}
    \sum_j I_1(\bm k;\bm R_j)
    e^{i\bm k\cdot\bm R_j}
    =
    \frac{\pi e^{-k^2/(8\alpha)} k^2 \sin^2\!\left(
        \varphi_{\bm k}+\gamma
    \right)
    N}{32\alpha^3}    
    \sum_{\bm G_M}
    \delta_{\bm k,\bm G_M}.
\end{aligned}
\end{equation}
Therefore the Bragg peaks remain located at the ordinary reciprocal lattice
vectors \(\bm G_M\), but their weights acquire an angular modulation. 
In particular, for \(\gamma=0\), the anisotropic contribution \(I_1(\bm k;\bm R_j)\)
is maximal for momenta along the transverse \(y\)-direction and vanishes for
momenta along the \(x\)-direction. However, the total Fourier amplitude is given
by \(I_0(k)+I_1(\bm k;\bm R_j)\), so the direction of the enhanced
\(|\tilde n_0(\bm k)|\) is determined by the competition between the isotropic
and anisotropic parts. For fixed \(k\), the transverse enhancement is most clearly
realized in the localized limit of large \(\alpha\), where the relevant momenta
satisfy \(k^2\ll 8\alpha\). In this regime, a molecule elongated along the
\(x\)-direction in real space gives a larger Fourier weight along the transverse
\(y\)-direction. 

Now consider a triangular moir\'e lattice, $    \bm R_j=m_{jx}\bm a_1+m_{jy}\bm a_2$,
with
\begin{equation}
    \bm a_1=a_M(1,0),
    \qquad
    \bm a_2=
    a_M
    \left(
        \frac{1}{2},
        \frac{\sqrt{3}}{2}
    \right).
\end{equation}
The corresponding reciprocal lattice vectors are
\begin{equation}
    \bm b_1
    =
    \frac{2\pi}{a_M}
    \left(
        1,
        -\frac{1}{\sqrt{3}}
    \right),
    \qquad
    \bm b_2
    =
    \frac{2\pi}{a_M}
    \left(
        0,
        \frac{2}{\sqrt{3}}
    \right).
\end{equation}
We take a staggered orientational pattern, $\gamma(\bm R_j)=\gamma_0+(-1)^{m_{jy}}\Delta$.
The orientation-dependent part then becomes
\begin{equation}
\begin{aligned}
\sum_j I_1(\bm k;\bm R_j)
e^{i\bm k\cdot\bm R_j}
&=
\pi e^{-k^2/(8\alpha)}
\frac{k^2}{32\alpha^3}
\sum_j
\sin^2\!\left[
    \varphi_{\bm k}
    +\gamma_0
    +(-1)^{m_{jy}}\Delta
\right]
e^{i\bm k\cdot\bm R_j}\\
&=
\frac{k^2 \pi e^{-k^2/(8\alpha)}}{64\alpha^3}
\sum_j e^{i\bm k\cdot\bm R_j}
\left\{
1-
\cos\!\left[
    2(\varphi_{\bm k}
    +\gamma_0)
    +(-1)^{m_{jy}}2\Delta
\right]
\right\}\\
&=\frac{k^2 \pi e^{-k^2/(8\alpha)}}{64\alpha^3}\left\{
\left[1-\cos(2\varphi_{\bm k}+2\gamma_0)\cos(2\Delta)
\right]
\sum_j e^{i\bm k\cdot\bm R_j}+
\sin(2\varphi_{\bm k}+2\gamma_0)\sin(2\Delta)
\sum_j
(-1)^{m_{jy}}
e^{i\bm k\cdot\bm R_j}\right\}.
\end{aligned}
\end{equation}

The staggered factor can be written as $(-1)^{m_{jy}}=e^{i\pi m_{jy}}=e^{i(\bm b_2/2)\cdot\bm R_j}$.
Therefore
\begin{equation}
\begin{aligned}
    \sum_j
    (-1)^{m_{jy}}
    e^{i\bm k\cdot\bm R_j}=
    \sum_j
    e^{i(\bm k+\bm b_2/2)\cdot\bm R_j}=
    N
    \sum_{\bm G_M}
    \delta_{\bm k+\bm b_2/2,\bm G_M}.
\end{aligned}
\end{equation}
Thus the staggered orientational pattern generates additional Bragg peaks at $\bm k=\bm G_M-\frac{\bm b_2}{2}$.

Combining the above results, the orientation-dependent contribution is
\begin{equation}
\begin{aligned}
\sum_j I_1(\bm k;\bm R_j)
e^{i\bm k\cdot\bm R_j}
&=
\frac{k^2 \pi e^{-k^2/(8\alpha)} N}{64\alpha^3}
\left\{
\left[1-
    \cos(2\varphi_{\bm k}+2\gamma_0)\cos(2\Delta)
\right]
\sum_{\bm G_M}
\delta_{\bm k,\bm G_M}+
\sin(2\varphi_{\bm k}+2\gamma_0)\sin(2\Delta)
\sum_{\bm G_M}
\delta_{\bm k+\bm b_2/2,\bm G_M}\right\}.
\end{aligned}
\end{equation}

Finally, adding the orientation-independent part, we obtain
\begin{equation}
\begin{aligned}
&\frac{\tilde n(\bm k)}{N}=
\pi e^{-k^2/(8\alpha)}
\Bigg\{\left[
    \frac{1}{8\alpha^2}
    -
    \frac{k^2}{64\alpha^3}
    \left(
        1
        +
        \cos(2\varphi_{\bm k}+2\gamma_0)\cos(2\Delta)
    \right)
\right]
\sum_{\bm G_M}
\delta_{\bm k,\bm G_M}+\frac{k^2\sin(2\varphi_{\bm k}+2\gamma_0)\sin(2\Delta)}{64\alpha^3}
\sum_{\bm G_M}
\delta_{\bm k+\bm b_2/2,\bm G_M}
\Bigg\}.
\end{aligned}
\end{equation}
This expression contains two sets of peaks. The first set appears at the
ordinary reciprocal lattice vectors \(\bm G_M\). The second set originates
from the staggered molecular orientation and is shifted by \(-\bm b_2/2\).
Therefore, the staggered ON-MC pattern produces additional folded Bragg peaks
at $\bm k=\bm G_M-\frac{\bm b_2}{2}$.

As a simple example, consider $\gamma_0=\frac{\pi}{2},\ \Delta=\frac{\pi}{4}$.
Then $\cos(2\Delta)=0,\  \sin(2\Delta)=1,$ and $    \sin(2\varphi_{\bm k}+2\gamma_0)=-\sin(2\varphi_{\bm k})$.

The Fourier component reduces to
\begin{equation}
\begin{aligned}
\frac{1}{N}\tilde n(\bm k)
&=
\pi e^{-k^2/(8\alpha)}
\Bigg\{
\left(
    \frac{1}{8\alpha^2}
    -
    \frac{k^2}{64\alpha^3}
\right)
\sum_{\bm G_M}
\delta_{\bm k,\bm G_M}
-
\frac{k^2}{64\alpha^3}
\sin(2\varphi_{\bm k})
\sum_{\bm G_M}
\delta_{\bm k+\bm b_2/2,\bm G_M}
\Bigg\}.
\end{aligned}
\end{equation}

For a finite periodic system, \(\delta_{\bm k,\bm G_M}\) should be understood
as a Kronecker delta for the allowed momenta. In the thermodynamic limit, the
Bragg peaks become Dirac-comb delta functions in momentum space.

Since the bare structure factor can be decomposed as $S_0(\bm k) = S(\bm k) + \frac{1}{N_e} \left|\tilde n(\bm k)\right|^2$, and since the connected part \(S(\bm k)\) does not exhibit sharp Bragg peaks, as shown in the main text, the sharp resonance peaks in \(S_0(\bm k)\) are controlled by the elastic contribution \(\left|\tilde n(\bm k)\right|^2/N_e\). Therefore, the simple analytical picture developed above for \(\tilde n(\bm k)\) directly explains the peak positions and angular intensity modulation observed in the bare structure factor.

\section{An effective theory for the orientational order of ON-MC}
\label{app:classical_ONMC}

In the orbital-nematic molecular crystal (ON-MC) phase, the charge density
around each moir\'e minimum has a localized two-lobe structure. As a simple
classical model, we approximate the two density lobes of the molecule at
moir\'e site $\bm R_j$ by two point-like charge packets located at $\bm r_{j,a} =\bm R_j+a d \hat{\bm n}_j,\ a=\pm 1$,
where $d$ characterizes the molecular size and $\hat{\bm n}_j=(\cos\theta_j,\sin\theta_j)$
is the orientational director of the molecule. Since the two lobes are
related by inversion, the configurations with $\theta_j$ and $\theta_j+\pi$
are physically equivalent. Therefore the orientational variable is a
headless director rather than a polar vector.

For a general central interaction $V(\bm r)=V(r)$, the interaction energy
between two molecules at $\bm R_i$ and $\bm R_j$ is
\begin{equation}
    E_{ij}
    =
    \sum_{a,b=\pm1}
    V\left[
        \bm R_{ij}
        +
        d(a\hat{\bm n}_i-b\hat{\bm n}_j)
    \right],
    \qquad
    \bm R_{ij}\equiv \bm R_i-\bm R_j .
\end{equation}
Here the overall charge factors are absorbed into the definition of $V$.

Since the molecular size is much smaller than the moir\'e lattice spacing,
$d\ll R_{ij}$, we can expand the interaction in powers of $d/R_{ij}$.
Introduce the directional derivatives
\begin{equation}
    D_i
    \equiv
    \hat{\bm n}_i\cdot\nabla_{\bm R_{ij}},
    \qquad
    D_j
    \equiv
    \hat{\bm n}_j\cdot\nabla_{\bm R_{ij}} .
\end{equation}
Then
\begin{equation}
    E_{ij}
    =
    \sum_{a,b=\pm1}
    \exp\left[
        d(aD_i-bD_j)
    \right]
    V(\bm R_{ij})=4\cosh(dD_i)\cosh(dD_j)V(\bm R_{ij}).
\end{equation}

Expanding to fourth order in $d$ gives
\begin{equation}
\begin{aligned}
    E_{ij}
    &=
    4V(R_{ij})
    +
    2d^2
    \left(
        D_i^2+D_j^2
    \right)
    V(R_{ij})
    +
    \frac{d^4}{6}
    \left(
        D_i^4+D_j^4
    \right)
    V(R_{ij})
    +
    d^4D_i^2D_j^2V(R_{ij})
    +
    O(d^6).
\end{aligned}
\label{eq:Eij_expansion}
\end{equation}
All odd powers of $d$ vanish because each molecule is inversion symmetric.
This is the classical origin of the absence of an electric dipole moment.

The leading term, $4V(R_{ij})$ is the monopole-monopole interaction and is independent of the molecular
orientations. The second-order term contains the interaction between the
monopole moment of one molecule and the quadrupole moment of the other.
To make this explicit, let $\phi_{ij}$ be the polar angle of $\bm R_{ij}$.
For a central potential, one has
\begin{equation}
\begin{aligned}
    D_i^2 V(R_{ij})
    &=
    V''(R_{ij})\cos^2(\theta_i-\phi_{ij})
    +
    \frac{V'(R_{ij})}{R_{ij}}
    \sin^2(\theta_i-\phi_{ij})
    \\
    &=
    \frac{1}{2}
    \left[
        V''(R_{ij})
        +
        \frac{V'(R_{ij})}{R_{ij}}
    \right]
    +
    \frac{1}{2}
    \left[
        V''(R_{ij})
        -
        \frac{V'(R_{ij})}{R_{ij}}
    \right]
    \cos 2(\theta_i-\phi_{ij}) .
\end{aligned}
\label{eq:D2V}
\end{equation}
Thus the orientational dependence appears through $\cos 2\theta_i$, rather
than through $\cos\theta_i$. This reflects the headless nature of the
$p$-wave molecular director and shows that the lowest nontrivial multipole
of a localized two-lobe molecule is a quadrupole rather than a dipole.

However, the $O(d^2)$ term does not directly couple the orientations of two
different molecules. The leading orientational coupling between two
molecules appears at order $d^4$, through $d^4D_i^2D_j^2V(R_{ij})$.
For a general central potential, this term can be decomposed as
\begin{equation}
\begin{aligned}
    D_i^2D_j^2V(R_{ij})
    &=
    C_0(R_{ij})+
    C_1(R_{ij})
    \left[
        \cos 2(\theta_i-\phi_{ij})
        +
        \cos 2(\theta_j-\phi_{ij})
    \right]
    +
    J_1(R_{ij})
    \cos 2(\theta_i-\theta_j)
    +
    J_2(R_{ij})
    \cos 2(\theta_i+\theta_j-2\phi_{ij}) .
\end{aligned}
\label{eq:D2D2_decomposition}
\end{equation}
Here
\begin{equation}
    C_0(R)
    =
    \frac{1}{4}
    \left[
        V^{(4)}(R)
        +
        \frac{2V^{(3)}(R)}{R}
        -
        \frac{V''(R)}{R^2}
        +
        \frac{V'(R)}{R^3}
    \right],\qquad C_1(R)
    =
    \frac{1}{4}
    \left[
        V^{(4)}(R)
        -
        \frac{3V''(R)}{R^2}
        +
        \frac{3V'(R)}{R^3}
    \right],
\end{equation}
and the genuine quadrupole-quadrupole couplings are
\begin{equation}
    J_1(R)
    =
    \frac{1}{8}
    \left[
        V^{(4)}(R)
        +
        \frac{2V^{(3)}(R)}{R}
        -
        \frac{V''(R)}{R^2}
        +
        \frac{V'(R)}{R^3}
    \right],
\end{equation}
\begin{equation}
    J_2(R)
    =
    \frac{1}{8}
    \left[
        V^{(4)}(R)
        -
        \frac{6V^{(3)}(R)}{R}
        +
        \frac{15V''(R)}{R^2}
        -
        \frac{15V'(R)}{R^3}
    \right].
\end{equation}
The term proportional to $J_1$ depends only on the relative orientation of
the two molecular directors, whereas the term proportional to $J_2$ couples
the molecular orientations to the bond direction $\phi_{ij}$. Therefore,
even for an isotropic central interaction, the lattice geometry generates a
bond-direction-dependent quadrupolar interaction.

In addition to this
two-molecule quadrupole-quadrupole coupling, the fourth-order expansion also
contains the single-molecule terms \(D_i^4V\) and \(D_j^4V\). For a central
potential, these terms can be written as
\begin{equation}
\begin{aligned}
    D_i^4 V(R_{ij})
    &=
    K_0(R_{ij})
    +
    K_2(R_{ij})
    \cos 2(\theta_i-\phi_{ij})
    +
    K_4(R_{ij})
    \cos 4(\theta_i-\phi_{ij}) ,
\end{aligned}
\label{eq:D4V}
\end{equation}
and similarly for \(D_j^4V(R_{ij})\) with \(\theta_i\) replaced by
\(\theta_j\). The radial coefficients are
\begin{equation}
\begin{aligned}
    K_0(R)
    &=
    \frac{3}{8}
    \left[
        V^{(4)}(R)
        +
        \frac{2V^{(3)}(R)}{R}
        -
        \frac{V''(R)}{R^2}
        +
        \frac{V'(R)}{R^3}
    \right],\\
    K_2(R)
    &=
    \frac{1}{2}
    \left[
        V^{(4)}(R)
        -
        \frac{3V''(R)}{R^2}
        +
        \frac{3V'(R)}{R^3}
    \right],\\
    K_4(R)
    &=
    \frac{1}{8}
    \left[
        V^{(4)}(R)
        -
        \frac{6V^{(3)}(R)}{R}
        +
        \frac{15V''(R)}{R^2}
        -
        \frac{15V'(R)}{R^3}
    \right].
\end{aligned}
\end{equation}

Therefore,
\begin{equation}
\begin{aligned}    &D_i^4V(R_{ij})+D_j^4V(R_{ij})=
    2K_0(R_{ij})
    +
    K_2(R_{ij})
    \left[
        \cos 2(\theta_i-\phi_{ij})
        +
        \cos 2(\theta_j-\phi_{ij})
    \right]
    +
    K_4(R_{ij})
    \left[
        \cos 4(\theta_i-\phi_{ij})
        +
        \cos 4(\theta_j-\phi_{ij})
    \right].
\end{aligned}
\label{eq:D4_sum}
\end{equation}
These terms describe the coupling between the monopole moment of one
molecule and the fourth multipole moment of the other molecule. They still
depend on the orientation of each molecule relative to the bond direction,
but they do not directly couple \(\theta_i\) and \(\theta_j\). The genuine
intermolecular orientational coupling at order \(d^4\) is therefore encoded
in \(D_i^2D_j^2V\), through the coefficients \(J_1\) and \(J_2\).

This effective description also suggests that the orientational ordering observed in the ON-MC phase is not a peculiarity of a specific interaction potential. Rather, it reflects a generic low-energy mechanism of the fully polarized molecular insulating regime. Once the electrons are localized near the moir\'e minima, the Pauli principle and repulsive interaction favor an anisotropic two-electron molecular state with an internal quadrupolar degree of freedom. The remaining low-energy physics is then governed by the interactions between these local nematic directors on the moir\'e lattice. Different microscopic interactions, such as the Coulomb and PT potentials considered here, mainly modify the effective quadrupolar couplings and therefore select different orientational patterns. In this sense, the ON-MC phase can be viewed as a generic quadrupolar molecular crystal, whose pattern is determined by a bond-dependent nematic-director model.

\subsection{Local quadrupolar pattern for Coulomb interaction}
\label{app:classical_ONMC_Coulomb}

We now specialize the classical expansion to the Coulomb interaction $V(R)={r_s}/{R}$ (here we do not include the correction from Ewald sum for simplicity).
The radial derivatives are
\begin{equation}
\begin{aligned}
    V'(R)&=-\frac{r_s}{R^2},\ \ \ \ 
    V''(R)=\frac{2r_s}{R^3},\ \ \ \ V^{(3)}(R)=-\frac{6r_s}{R^4},\ \ \ \ 
    V^{(4)}(R)=\frac{24r_s}{R^5}.    
\end{aligned}
\end{equation}

Let us consider two molecules at the same bond $R_{ij}$, for notational simplicity, we write
\(R=R_{ij}\) and \(\phi=\phi_{ij}\).  We also define the angular
factors
\begin{equation}
    c_i=\cos 2(\theta_i-\phi),
    \qquad
    c_j=\cos 2(\theta_j-\phi), \qquad
    q_i=\cos 4(\theta_i-\phi),
    \qquad
    q_j=\cos 4(\theta_j-\phi),
\end{equation}
and
\begin{equation}
    c_{-,ij}=\cos 2(\theta_i-\theta_j),
    \qquad
    c_{+,ij}=\cos 2(\theta_i+\theta_j-2\phi).
\end{equation}

The second-order contribution to the intermolecular energy is
\begin{equation}
    2d^2
    \left(
        D_i^2+D_j^2
    \right)V(R)
    =
    \frac{r_s d^2}{R^3}
    \left[
        2+3(c_i+c_j)
    \right].
\end{equation}

At fourth order, there are two types of terms.  The first type is the
single-molecule fourth-derivative contribution,
\begin{equation}
\begin{aligned}
    \frac{d^4}{6}
    \left(
        D_i^4+D_j^4
    \right)V(R)
    &=
    \frac{r_s d^4}{R^5}
    \left[
        \frac{9}{8}
        +
        \frac{5}{4}(c_i+c_j)
        +
        \frac{35}{16}(q_i+q_j)
    \right].
\end{aligned}
\end{equation}

The second type of fourth-order term is the genuine two-molecule
quadrupole-quadrupole contribution,
for the Coulomb interaction, the general decomposition gives
\begin{equation}
\begin{aligned}
    d^4D_i^2D_j^2V(R)
    =
    \frac{r_s d^4}{R^5}
    \Bigg[
        &
        \frac{9}{4}
        +
        \frac{15}{4}(c_i+c_j)
        +
        \frac{9}{8}c_{-,ij}
        +
        \frac{105}{8}c_{+,ij}
    \Bigg].
\end{aligned}
\end{equation}
Equivalently, the genuine quadrupole-quadrupole couplings are
\begin{equation}
    J_1(R)
    =
    \frac{9r_s}{8R^5},
    \qquad
    J_2(R)
    =
    \frac{105r_s}{8R^5}.
\end{equation}
Thus $J_1>0,\ J_2>0,
    \ {J_2}/{J_1}={35}/{3}$.
The dominant fourth-order orientational coupling is therefore the
bond-direction-dependent term proportional to \(J_2\).

Combining all terms up to order \(d^4\), the Coulomb interaction energy
between two molecules is
\begin{equation}
\begin{aligned}
    E_{ij}
    &=
    \frac{4r_s}{R}
    +
    \frac{r_s d^2}{R^3}
    \left[
        2+3(c_i+c_j)
    \right]
    +
    \frac{r_s d^4}{R^5}
    \Bigg[
        \frac{27}{8}
        +
        5(c_i+c_j)
        +
        \frac{35}{16}(q_i+q_j)+
        \frac{9}{8}c_{-,ij}
        +
        \frac{105}{8}c_{+,ij}
    \Bigg]
    +
    O(d^6).
\end{aligned}
\label{eq:Coulomb_full_pair_energy}
\end{equation}
The terms proportional to \(c_i,c_j,q_i,q_j\) describe the coupling of an
individual molecular director to the bond direction.  They do not directly
couple \(\theta_i\) and \(\theta_j\).  The genuine intermolecular
orientational coupling is instead
\begin{equation}
    E_{ij}^{QQ}
    =
    d^4
    \left[
        J_1(R)\cos 2(\theta_i-\theta_j)
        +
        J_2(R)\cos 2(\theta_i+\theta_j-2\phi)
    \right].
\end{equation}
Within this minimal quadrupole-quadrupole part, a single isolated bond
would prefer
\begin{equation}
    \cos 2(\theta_i-\theta_j)=-1,
    \qquad
    \cos 2(\theta_i+\theta_j-2\phi)=-1.
\end{equation}
Equivalently,
\begin{equation}
    \theta_i-\theta_j=\frac{\pi}{2}\quad \mathrm{mod}\ \pi ,\qquad \theta_i+\theta_j-2\phi
    =
    \frac{\pi}{2}\quad \mathrm{mod}\ \pi .
\end{equation}
Thus the minimal quadrupole-quadrupole interaction favors two neighboring
molecular directors that are approximately perpendicular to each other, with
their average orientation locked to the bond direction.

On a triangular lattice, however, each molecule is connected to several
bonds with different bond directions.  The above single-bond condition
cannot be simultaneously satisfied on all bonds.  This is the origin of
frustration in the classical quadrupole model.

To make this explicit, consider a small hexagonal cluster of the triangular
lattice, consisting of one central site and six nearest-neighbor sites.  Let
the central site be denoted by \(c\), and let the six outer sites be
\(m=0,\ldots,5\), with positions
\begin{equation}
    \bm R_m
    =
    a_M
    \left(
        \cos\phi_m,
        \sin\phi_m
    \right),
    \qquad
    \phi_m=\frac{m\pi}{3}.
\end{equation}
We keep the twelve nearest-neighbor bonds of this cluster.  These include
six radial bonds between the center and the outer sites, with bond angles $\phi_{cm}=\phi_m$,
and six bonds along the outer hexagon, with bond angles $\psi_m=\phi_m+\frac{2\pi}{3}$.

For this symmetric twelve-bond cluster, the single-director anisotropies
cancel after summing over bonds: $\sum_{\langle ij\rangle}
    (c_i+c_j)
    =
    0,
    \ 
    \sum_{\langle ij\rangle}
    (q_i+q_j)
    =
    0.$
This cancellation follows from the fact that the bond directions meeting
each site are related by the triangular lattice symmetry.  Therefore, for
this cluster, the orientational part of the full Coulomb energy is governed
by the same \(J_1\)-\(J_2\) quadrupole-quadrupole model, up to an
orientation-independent constant.

The nearest-neighbor quadrupolar energy is
\begin{equation}
\begin{aligned}
    &E_{\mathrm{hex}}^{QQ}=
    d^4
    \left\{\sum_{m=0}^{5}
    \left[
        J_1\cos 2(\theta_c-\theta_m)
        +
        J_2\cos 2(\theta_c+\theta_m-2\phi_m)
    \right]+
    \sum_{m=0}^{5}
    \left[
        J_1\cos 2(\theta_m-\theta_{m+1})
        +
        J_2\cos 2(\theta_m+\theta_{m+1}-2\psi_m)
    \right]\right\},
\end{aligned}
\label{eq:hexagon_QQ_energy_Coulomb}
\end{equation}
where \(m+1\) is understood modulo \(6\).

A uniform nematic configuration, $\theta_c=\theta_m=\theta_0,$
has
\begin{equation}
    \sum_{\langle ij\rangle}
    \cos 2(\theta_i-\theta_j)
    =
    12,\qquad \sum_{\langle ij\rangle}
    \cos 2(\theta_i+\theta_j-2\phi_{ij})
    =
    0.
\end{equation}
Hence
\begin{equation}
    E_{\mathrm{hex}}^{QQ,\mathrm{uniform}}
    =
    12d^4J_1
    =
    \frac{27}{2}
    \frac{r_s d^4}{a_M^5}.
\end{equation}
The large \(J_2\) contribution averages to zero in the uniform state.

A representative low-energy \(C_2\) configuration is
\begin{equation}
    \theta_c=105^\circ,\qquad     (\theta_0,\theta_1,\theta_2,\theta_3,\theta_4,\theta_5)
    =
    (15^\circ,105^\circ,15^\circ,15^\circ,105^\circ,15^\circ).
\end{equation}

\begin{figure}[h]
\includegraphics[trim=0 0 0 0, clip,width=0.2\columnwidth]{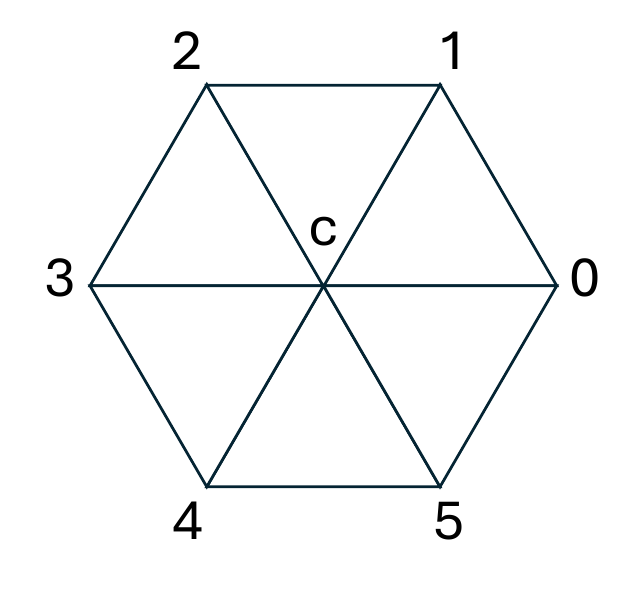}
\caption{
Schematic illustration of the small hexagonal cluster. 
}
\label{parallel_quadrupole_picture}
\end{figure}

For this configuration, the twelve bonds give
\begin{equation}
    \sum_{\langle ij\rangle}
    \cos 2(\theta_i-\theta_j)
    =
    -4,\qquad     \sum_{\langle ij\rangle}
    \cos 2(\theta_i+\theta_j-2\phi_{ij})
    =
    -8.
\end{equation}
Therefore
\begin{equation}
    E_{\mathrm{hex}}^{QQ,C_2}
    =
    d^4
    \left(
        -4J_1-8J_2
    \right)=
    -\frac{219}{2}
    \frac{r_s d^4}{a_M^5}.
\end{equation}
This is much lower than the uniform nematic energy because the \(C_2\)
configuration partially satisfies the frustrated bond-dependent \(J_2\)
constraints.

Note that for an isolated bond, the lowest possible value of the minimal
quadrupole-quadrupole energy would be
\(-d^4(J_1+J_2)\).  If all twelve bonds could be optimized independently,
one would obtain $-12d^4(J_1+J_2)$.
The actual \(C_2\) cluster energy is higher than this ideal bound, reflecting
the frustration imposed by the triangular geometry.

The configuration above breaks the sixfold rotational symmetry of the local
triangular environment down to a twofold pattern.  It therefore provides a
simple classical picture for the emergence of orbital-nematic order in the
ON-MC phase:
\begin{equation}
    \mathcal{C}_6
    \quad\longrightarrow\quad
    \mathcal{C}_2 .
\end{equation}

There are symmetry-related domains.  For example, another equivalent
configuration can be written using the two perpendicular directions
\(45^\circ\) and \(135^\circ\):
\begin{equation}
    \theta_c=45^\circ,\qquad     (\theta_0,\theta_1,\theta_2,\theta_3,\theta_4,\theta_5)
    =
    (45^\circ,135^\circ,135^\circ,45^\circ,135^\circ,135^\circ).
\end{equation}

This configuration has the same quadrupolar energy,
\begin{equation}
    E_{\mathrm{hex}}^{QQ,C_2}
    =
    -4d^4J_1
    -
    8d^4J_2,
\end{equation}
because it is related to the previous one by a lattice symmetry operation.
However, one should not arbitrarily rotate all directors by an angle that is
not a symmetry of the triangular lattice while keeping the lattice fixed.
The \(J_2\) term explicitly depends on the bond direction \(\phi_{ij}\), and
therefore the absolute orientations of the molecular directors are locked to
the triangular lattice. We also verified the above energy-minimizing configuration through numerical calculations.

This analysis should be interpreted as a local classical picture.  It shows
that the Coulomb interaction naturally favors a frustrated quadrupolar
arrangement on the triangular lattice, and that the resulting local pattern
tends to break \(\mathcal{C}_6\) down to \(\mathcal{C}_2\). 

In the effective Hamiltonian of the main text and in the numerical estimates quoted below, we absorb the common factor $d^4$ into the definitions of $J_1$ and $J_2$. For the parameters used in the main text, taking $d/a_M\sim 0.25$, we obtain
\begin{equation}
    J_1\sim 0.14\,\mathrm{meV}\approx 1.6\,\mathrm{K},\qquad J_2\sim 1.61\,\mathrm{meV}\approx 18.7\,\mathrm{K}.
\end{equation}

\subsection{Local quadrupolar pattern for P\"oschl--Teller interaction}
\label{app:classical_ONMC_PT}

We next consider the short-range P\"oschl--Teller interaction
\begin{equation}
    V(R)
    =
    V_{\mathrm{PT}}
    \frac{\hbar^2\kappa^2}{m^*}
    f(\kappa R),
\end{equation}
where $f(x)=\mathrm{\rm sech}^2 x={1}/{{\rm cosh}^2(x)}$.

The $n$-th radial derivative has the form
\begin{equation}
    V^{(n)}(R)
    =
    V_{\mathrm{PT}}
    \frac{\hbar^2\kappa^{n+2}}{m^*}
    f^{(n)}(\kappa R).
\end{equation}

Substituting this potential into the general quadrupole-quadrupole couplings $J_1$ and $J_2$, we obtain
\begin{equation}
    J_1(R)
    =
    V_{\mathrm{PT}}
    \frac{\hbar^2\kappa^6}{8m^*}
    F_1(\kappa R),
\qquad
    J_2(R)
    =
    V_{\mathrm{PT}}
    \frac{\hbar^2\kappa^6}{8m^*}
    F_2(\kappa R),
\end{equation}
where
\begin{equation}
    F_1(x)
    =
    f^{(4)}(x)
    +
    \frac{2f^{(3)}(x)}{x}
    -
    \frac{f''(x)}{x^2}
    +
    \frac{f'(x)}{x^3},
\qquad
    F_2(x)
    =
    f^{(4)}(x)
    -
    \frac{6f^{(3)}(x)}{x}
    +
    \frac{15f''(x)}{x^2}
    -
    \frac{15f'(x)}{x^3}.
\end{equation}

For the parameter used in our calculations, $\kappa^{-1}=a\approx0.37a_M, \ \ \ \ x=\kappa a_M\simeq 2.7027.$
At this value, one finds $F_1(2.7027)\simeq 0.1374,
\ F_2(2.7027)\simeq 0.7135.$
Therefore, for $\kappa^{-1}=0.37a_M$, the nearest-neighbor quadrupolar
couplings of the P\"oschl--Teller interaction have the same sign structure
as those of the Coulomb interaction:
\begin{equation}
\begin{aligned}
    J_1(a_M)
    &\simeq
    6.69\,
    V_{\mathrm{PT}}
    \frac{\hbar^2}{m^*a_M^6}>0,  \qquad J_2(a_M)
    \simeq
    34.76\,
    V_{\mathrm{PT}}
    \frac{\hbar^2}{m^*a_M^6}>0,\qquad
    \frac{J_2(a_M)}{J_1(a_M)}
    \simeq 5.19.
\end{aligned}
\end{equation}

For a single nearest-neighbor bond, the positive $J_1$ term favors $\theta_i-\theta_j=\frac{\pi}{2}\  \mathrm{mod}\ \pi,$
namely perpendicular neighboring directors. The positive $J_2$ term favors
$\theta_i+\theta_j-2\phi_{ij}=\frac{\pi}{2}\ \mathrm{mod}\ \pi,$
which locks the molecular orientations to the bond direction. Therefore,
at the nearest-neighbor level, the P\"oschl--Teller interaction with
$\kappa^{-1}=0.37a_M$ favors a local orientational tendency similar to that
of the Coulomb interaction: on an isolated bond, one molecule tends to be
approximately parallel to the bond while the other is approximately
perpendicular to it.
However, the relative strength of the bond-direction-dependent coupling is
smaller than in the Coulomb case. 
Thus the P\"oschl--Teller interaction still favors a Coulomb-like frustrated
quadrupolar pattern, but the dominance of the bond-dependent $J_2$ term is
weaker.

For the seven-site hexagonal cluster discussed above, the same two-director
$C_2$ pattern gives
\begin{equation}
    E_{\mathrm{hex}}^{C_2}
    =
    -4d^4J_1(a_M)
    -
    8d^4J_2(a_M)\simeq
    -304.84\,
    V_{\mathrm{PT}}
    \frac{\hbar^2}{m^*a_M^2}
    \left(\frac{d}{a_M}\right)^4.
\end{equation}
This analysis suggests that the P\"oschl--Teller interaction with
$\kappa^{-1}=0.37a_M$ can still generate a local $C_2$-like quadrupolar
tendency. Nevertheless, compared with the Coulomb case, the interaction is
shorter ranged and the ratio $J_2/J_1$ is smaller. Consequently, the
classical orientational locking is expected to be weaker, and the resulting
orientational order can be more easily affected by molecular-shape
renormalization, kinetic energy, exchange effects, and many-body
correlations. Moreover, owing to the high tunability of ultracold-atom experiments, the
detailed form of the short-range interaction can in principle be engineered,
which may allow other ON-MC orientational patterns beyond the Coulomb-like
$C_2$ texture considered here.

\subsection{Candidate phase diagram}

\begin{figure}[h]
\centering
\includegraphics[trim=0 0 0 0, clip,width=0.3\columnwidth]{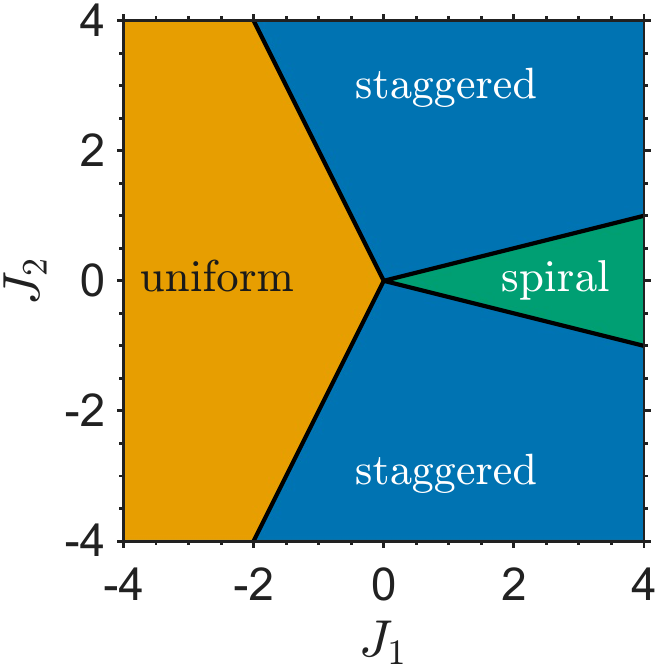}
\caption{
Candidate zero-temperature phase diagram of the classical effective orientational model in the $J_1$-$J_2$ plane.
The colors denote the lowest classical energy among three high-symmetry candidate states: uniform nematic order, staggered stripe order, and $K$-point spiral order of $\psi_i=e^{2i\theta_i}$.
The black lines show the analytic boundaries obtained by comparing the classical energies
$E_{\rm uni}/N=3J_1$,
$E_{\rm stag}/N=-J_1-2|J_2|$,
and
$E_{\rm spi}/N=-3J_1/2$.
Thermal and quantum fluctuations of the molecular directors are not included in this classical zero-temperature analysis.
}
\label{J1_J2_phase_diagram}
\end{figure}

In Fig.~\ref{J1_J2_phase_diagram}, we plot a candidate zero-temperature phase diagram of the effective orientational model in the $J_1$-$J_2$ plane. The classical orientational Hamiltonian is
\begin{equation}
H_{J_1,J_2}[\{\theta_i\}]
=
\sum_{\langle ij\rangle}
\left[
J_1\cos 2(\theta_i-\theta_j)
+
J_2\cos 2(\theta_i+\theta_j-2\phi_{ij})
\right],
\end{equation}
where $\theta_i$ denotes the molecular director angle at site $i$, with the identification $\theta_i\equiv \theta_i+\pi$, and $\phi_{ij}$ is the angle of the nearest-neighbor bond. It is therefore natural to introduce the complex nematic variable $\psi_i=e^{2i\theta_i}$.
For the triangular lattice, we count each nearest-neighbor bond once by keeping the three positive bond directions $\phi_1=0,\ 
\phi_2=\frac{\pi}{3},\ 
\phi_3=\frac{2\pi}{3},$
so that the total number of nearest-neighbor bonds is $3N$.

We compare three high-symmetry candidate states in the ON-MC phase. The first one is a uniform nematic state, $\theta_i=\alpha,\ \psi_i=e^{2i\alpha}$.

For this state, $\cos 2(\theta_i-\theta_j)=1$ on every bond, giving ${E_{J_1}^{\rm uni}}/{N}=3J_1$.
The $J_2$ contribution vanishes after summing over the three bond orientations, $\sum_{\mu=1}^3
\cos(4\alpha-4\phi_\mu)=0.$
Therefore,
\begin{equation}
\frac{E_{\rm uni}}{N}=3J_1.
\end{equation}

For the staggered stripe state shown here, we choose a representative stripe orientation in which the director alternates along the $\bm a_1$ direction. Writing the triangular-lattice sites as $\bm R_{x,y}=x\bm a_1+y\bm a_2,\ 
x,y\in\mathbb Z,$
the staggered configuration is $\theta_{x,y}=\alpha+\frac{\pi}{2}(x\bmod 2).$
Equivalently, in terms of the nematic variable $\psi_{x,y}=e^{2i\theta_{x,y}}$, this becomes $\psi_{x,y}=
e^{2i\alpha}e^{i\pi x}=e^{2i\alpha}e^{i\bm Q_M\cdot\bm R_{x,y}},$
where the corresponding $M$-point wave vector is $\bm Q_M={\bm b_1}/{2}.$

For a representative stripe configuration, the three bond directions give the $J_1$ factors $-1,\ +1,\ -1$, and hence ${E_{J_1}^{\rm stag}}/{N}=-J_1.$
The $J_2$ contribution depends on the global director angle $\alpha$. For the same representative stripe, the three bond contributions combine to ${E_{J_2}^{\rm stag}}/{N}
=
-2J_2\cos\left(4\alpha-\frac{\pi}{3}\right).$
Since $\alpha$ is a variational parameter, the $J_2$ energy can be minimized for either sign of $J_2$, giving $\min_\alpha {E_{J_2}^{\rm stag}}/{N}
=
-2|J_2|.$
Thus the optimized staggered-stripe energy is
\begin{equation}
\frac{E_{\rm stag}}{N}
=
-J_1-2|J_2|.
\end{equation}

The third candidate is a classical $K$-point spiral state of the nematic variable $\psi_i$, $\psi_i=e^{i(\mathbf K\cdot\mathbf R_i+2\alpha)},\ \theta_i=\alpha+\frac{1}{2}\mathbf K\cdot\mathbf R_i,$
where $\mathbf K=\frac{\bm b_1-\bm b_2}{3}$ is a corner of the triangular-lattice Brillouin zone. This is a $120^\circ$ order of $\psi_i=e^{2i\theta_i}$, corresponding to a $60^\circ$ rotation of the director angle $\theta_i$ between neighboring sublattices. 

For this state, the nearest-neighbor phase difference of $\psi_i$ is $\pm 2\pi/3$, and hence $\cos 2(\theta_i-\theta_j)
=
\cos\left(\pm\frac{2\pi}{3}\right)
=
-\frac{1}{2}.$
The $J_1$ contribution is ${E_{J_1}^{\rm spi}}/{N}=-\frac{3}{2}J_1.$
For a lattice commensurate with the $K$ point, the $J_2$ contribution averages to zero because it contains spatial phases proportional to $e^{2i\mathbf K\cdot\mathbf R_i}$. Therefore,
\begin{equation}
\frac{E_{\rm spi}}{N}
=
-\frac{3}{2}J_1.
\end{equation}

Comparing the three candidate energies,
\begin{equation}
\frac{E_{\rm uni}}{N}=3J_1,
\qquad
\frac{E_{\rm stag}}{N}=-J_1-2|J_2|,
\qquad
\frac{E_{\rm spi}}{N}=-\frac{3}{2}J_1,
\end{equation}
yields the phase boundaries
\begin{equation}
\begin{aligned}
\mathrm{uniform-staggered}:\quad J_1&=-\frac{|J_2|}{2},\qquad\mathrm{staggered-spiral}:\quad J_1=4|J_2|.    
\end{aligned}
\end{equation}

The candidate phase diagram is symmetric under $J_2\to -J_2$. This symmetry follows from an exact change of variables in the orientational configuration space. Specifically, one has
\begin{equation}
H_{J_1,J_2}[{\theta_i}]
=
H_{J_1,-J_2}
\left[
\left\{\theta_i+\frac{\pi}{4}\right\}
\right].
\end{equation}
Therefore, the two parameter points $(J_1,J_2)$ and $(J_1,-J_2)$ have the same set of classical energies, and in particular the same ground-state energy. This is not a spatial rotation symmetry of the triangular lattice, but a duality between two Hamiltonians with opposite signs of $J_2$. Although in Fig.~\ref{J1_J2_phase_diagram} we compare only three high-symmetry candidate states, the above symmetry argument is general for the effective orientational model.

The structure of the candidate phase diagram can be understood from three simple limits. For $J_1\to -\infty$ and $J_2=0$, the $J_1$ term favors parallel alignment of the nematic variables, and the uniform nematic state has the lowest energy. For $J_1\to +\infty$ and $J_2=0$, the $J_1$ term is frustrated on the triangular lattice, leading to the classical $K$-point spiral state. Finally, for $J_1=0$ and $|J_2|\to\infty$, the $J_2$ term dominates. After optimizing the global director angle, the staggered stripe state minimizes the energy for either sign of $J_2$, which is consistent with the $J_2\to -J_2$ symmetry.

We note that the above discussion treats the molecular directors as classical variables and neglects their intrinsic quantum fluctuations. The role of such quantum fluctuations will be investigated in future work. Another interesting limit is $J_2=0$, where the effective model recovers a continuous $U(1)$ symmetry associated with rotations of the nematic order parameter $\psi_i=e^{2i\theta_i}$. In this case, although true long-range orientational order is forbidden at any finite temperature in two dimensions, the system may instead exhibit a Berezinskii-Kosterlitz-Thouless transition. Across this transition, the high-temperature disordered phase gives way to a low-temperature phase with quasi-long-range nematic order, characterized by algebraically decaying correlations of $\psi_i$ and bound vortex-antivortex pairs. Thus, in realistic finite-temperature experiments, the $J_2=0$ or weak-$J_2$ regime may provide a natural setting to observe BKT physics associated with the molecular directors.

\section{Mean-field theory of the exciton absorption spectrum}
\label{app:exciton-absorption}

\subsection{Effective exciton--electron interaction}

We describe the optically created exciton as a structureless quasiparticle
whose internal electron--hole degrees of freedom are encoded in an
effective exciton--electron interaction. The exciton center-of-mass
coordinate is denoted by \(\bm R\), while \(\bm r\) denotes the position
of an electron. We employ the rotationally symmetric interaction~\cite{PhysRevX.15.011049}
\begin{equation}
V_{e-X}(\bm R-\bm r)
=
-\frac{W_0}
{\left(a_X^2+|\bm R-\bm r|^2\right)^2}
+
P_0\exp\left(-\frac{|\bm R-\bm r|}{a_X}\right),
\label{eq:app-real-space-interaction}
\end{equation}
where \(W_0>0\) controls the attractive component, \(P_0>0\) determines
the strength of the short-range repulsive component, and \(a_X\) sets
their characteristic interaction range. The competition between these
two contributions allows the interaction to capture both the attractive
exciton--electron coupling and the short-distance repulsive core.

We use the two-dimensional Fourier-transform convention
\begin{equation}
\widetilde V_{e-X}(\bm q)
=
\int d^2\rho\,
V_{e-X}(\bm \rho)e^{-i\bm q\cdot\bm \rho},
\qquad
\bm \rho=\bm R-\bm r.
\label{eq:app-fourier-convention}
\end{equation}
Because the interaction is rotationally invariant, its Fourier transform
depends only on \(q=|\bm q|\) and can be written as
\begin{equation}
\widetilde V_{e-X}(q)
=
2\pi\int_0^\infty d\rho\,\rho\,
V_{e-X}(\rho)J_0(q\rho),
\label{eq:app-radial-fourier}
\end{equation}
where \(J_0\) is the Bessel function of the first kind and \(\rho=|\bm\rho|\). Using
\begin{align}
\int_0^\infty d\rho\,
\frac{\rho J_0(q\rho)}
{\left(a_X^2+\rho^2\right)^2}
&=
\frac{q}{2a_X}K_1(a_Xq),\quad
\int_0^\infty d\rho\,
\rho e^{-\rho/a_X}J_0(q\rho)
=
\frac{a_X^2}
{\left[1+(a_Xq)^2\right]^{3/2}},
\end{align}
we obtain
\begin{equation}
\widetilde V_{e-X}(q)
=
\begin{cases}
\displaystyle
-\frac{\pi W_0}{a_X}\,
qK_1(a_Xq)
+
\frac{2\pi P_0a_X^2}
{\left[1+(a_Xq)^2\right]^{3/2}},
&
q\neq0,
\\[4mm]
\displaystyle
-\frac{\pi W_0}{a_X^2}
+
2\pi P_0a_X^2,
&
q=0,
\end{cases}
\label{eq:app-interaction-fourier}
\end{equation}
where \(K_1\) is the modified Bessel function of the second kind. The
\(q=0\) expression follows continuously from
\(qK_1(a_Xq)\rightarrow 1/a_X\) as \(q\rightarrow0\). In constructing
the periodic exciton mean-field potential, Eq.~\eqref{eq:app-interaction-fourier}
is evaluated at the reciprocal vectors of the enlarged ON--MC unit cell,
\(q=|\bm G|\).

\subsection{Static mean-field approximation}

Before applying the mean-field approximation, the exciton--electron
interaction Hamiltonian can be written as
\begin{equation}
\hat H_{e-X}
=
\int d^2R\,d^2r\,
\hat X^\dagger(\bm R)\hat X(\bm R)
V_{e-X}(\bm R-\bm r)
\hat n_e(\bm r),
\label{eq:app-full-exciton-electron}
\end{equation}
where \(\hat X^\dagger(\bm R)\) creates an exciton centered at \(\bm R\),
and \(\hat n_e(\bm r)\) is the electronic density operator. Within the
static mean-field approximation, the density operator is replaced by its
expectation value,
\begin{equation}
\hat n_e(\bm r)
\longrightarrow
n(\bm r)
=
\langle\hat n_e(\bm r)\rangle.
\end{equation}
The exciton then experiences the effective periodic potential
\begin{equation}
U_X(\bm R)
=
\int d^2r\,
V_{e-X}(\bm R-\bm r)n(\bm r).
\label{eq:exciton-mean-field-potential}
\end{equation}

For a periodic electronic density, we write
\(n(\bm r)=\sum_{\bm G}\tilde n_{\bm G}e^{i\bm G\cdot\bm r}\), where
\(\tilde n_{\bm G}=\mathcal A_{\mathrm{uc}}^{-1}
\int_{\mathcal A_{\mathrm{uc}}}d^2r\,
n(\bm r)e^{-i\bm G\cdot\bm r}\), and
\(\mathcal A_{\mathrm{uc}}\) is the area of the unit cell of the ordered
electronic state. The effective potential can then be expressed as
\(U_X(\bm R)=\sum_{\bm G}U_{\bm G}e^{i\bm G\cdot\bm R}\), with
\begin{equation}
U_{\bm G}
=
\widetilde V_{e-X}(G)\tilde n_{\bm G}.
\label{eq:app-mean-field-fourier}
\end{equation}
Thus, each Fourier component of the electronic density generates a
corresponding Fourier component of the exciton mean-field potential.

The uniform component is
\begin{equation}
U_{\bm 0}
=
\widetilde V_{e-X}(0)\tilde n_{\bm 0}
=
\left(
-\frac{\pi W_0}{a_X^2}
+
2\pi P_0a_X^2
\right)\overline n,
\label{eq:app-uniform-shift}
\end{equation}
where \(\overline n=\tilde n_{\bm 0}\) is the mean electronic density.
This term produces a common shift of all exciton energies. The nonuniform
components \(U_{\bm G\neq\bm 0}\), by contrast, mix exciton plane waves
with different momenta and reconstruct the absorption spectrum.

For the ON--MC state considered here, the unit cell is doubled along
\(\bm a_2\). Its reciprocal vectors can therefore be written as
\(\bm G=m\bm b_1+n\bm b_2/2\), with \(m,n\in\mathbb Z\). Here, \(\bm a_{1,2}\) and \(\bm b_{1,2}\) denote the primitive real-
and reciprocal-lattice vectors of the original moir\'e lattice,
respectively. Components with
even \(n\) are reciprocal vectors of the original moir\'e lattice, whereas
those with odd \(n\) belong to the folded reciprocal-vector sector
generated by the enlarged ON--MC unit cell.

The resulting single-exciton mean-field Hamiltonian is
\begin{equation}
\hat H_X^{\mathrm{MF}}
=
-\frac{\hbar^2\nabla_{\bm R}^2}{2M_X}
+
U_X(\bm R),
\label{eq:app-exciton-hamiltonian}
\end{equation}
where \(M_X\) is the exciton center-of-mass mass. For a Bloch momentum \(\bm k\) in the reduced Brillouin zone, we label
the exciton eigenstates by \(s\) and expand them as
\begin{equation}
\psi_{s\bm k}(\bm R)
=
\frac{1}{\sqrt{\mathcal A}}
\sum_{\bm G}
c_{s,\bm k}(\bm G)
e^{i(\bm k+\bm G)\cdot\bm R},
\label{eq:app-bloch-expansion}
\end{equation}
where \(\mathcal A\) is the total system area. The corresponding
eigenvalue equation in the reciprocal-vector basis is
\begin{equation}
\sum_{\bm G'}
H_{\bm G\bm G'}(\bm k)
c_{s,\bm k}(\bm G')
=
E_s(\bm k)c_{s,\bm k}(\bm G),
\label{eq:app-eigenvalue-equation}
\end{equation}
with
\begin{equation}
H_{\bm G\bm G'}(\bm k)
=
\frac{\hbar^2|\bm k+\bm G|^2}{2M_X}
\delta_{\bm G\bm G'}
+
U_{\bm G-\bm G'}.
\label{eq:app-plane-wave-hamiltonian}
\end{equation}
Diagonalizing this matrix yields the exciton eigenenergies \(E_s(\bm k)\)
and the corresponding plane-wave coefficients
\(c_{s,\bm k}(\bm G)\).

\subsection{Optical matrix elements}

To describe optical excitation, we introduce the photon annihilation operator $\hat a_{\bm q\lambda}$, where $\bm q$ and $\lambda$ denote the photon momentum and polarization, respectively. We also define the momentum-space exciton operator as
\begin{equation}
\hat X_{\bm p}^{\dagger}=\frac{1}{\sqrt{\mathcal A}}
\int d^2R\,
e^{i\bm p\cdot\bm R}
\hat X^\dagger(\bm R).
\end{equation}
Within the rotating-wave approximation, the exciton--photon coupling can be written in the explicitly Hermitian form
\begin{equation}
\hat H_{X-\gamma}
=
-\sum_{\bm q,\lambda}
\left[
g_{\bm q\lambda}
\hat X_{\bm q_{\parallel}}^\dagger
\hat a_{\bm q\lambda}
+
g_{\bm q\lambda}^{*}
\hat a_{\bm q\lambda}^{\dagger}
\hat X_{\bm q_{\parallel}}
\right],
\label{eq:app-light-matter-hamiltonian}
\end{equation}
where \(g_{\bm q\lambda}\) is the exciton--photon coupling matrix element, and \(\bm q_\parallel\) is the in-plane component of the photon momentum. The first term describes photon absorption and exciton creation, while the second term describes the reverse process.

At normal incidence, \(\bm q_{\parallel}=\bm 0\). The photon therefore
directly creates the zero-momentum bare exciton,
\(\hat X_{\bm 0}^{\dagger}|\mathrm{vac}\rangle\), where
\(|\mathrm{vac}\rangle\) denotes the exciton vacuum.
At $\bm k=\bm 0$, the eigenstates obtained by diagonalizing Eq.~\eqref{eq:app-plane-wave-hamiltonian} are
\begin{equation}
|s,\bm 0\rangle
=
\sum_{\bm G}
c_{s,\bm 0}(\bm G)
\hat X_{\bm G}^{\dagger}
|\mathrm{vac}\rangle.
\label{eq:app-exciton-eigenstate}
\end{equation}

Consider an initial state with one photon of momentum \(\bm q\) and
polarization \(\lambda\), and no exciton,
\(|i\rangle=|\mathrm{vac}\rangle\otimes
|1_{\bm q\lambda}\rangle_\gamma\), and a final state with the \(s\)th
exciton eigenstate at \(\bm k=\bm 0\) and no photon,
\(|f_s\rangle=|s,\bm 0\rangle\otimes|0\rangle_\gamma\).
Here, \(|\mathrm{vac}\rangle\) and \(|0\rangle_\gamma\) denote the
exciton and photon vacuum states, respectively, while
\(|1_{\bm q\lambda}\rangle_\gamma\) denotes a single-photon state.
The corresponding absorption matrix element is
\begin{align}
\mathcal M_s
&=
\langle f_s|
\hat H_{X-\gamma}
|i\rangle
\nonumber=
-g_{\bm q\lambda}
\langle s,\bm 0|
\hat X_{\bm 0}^{\dagger}
|\mathrm{vac}\rangle
\nonumber=
-g_{\bm q\lambda}
c_{s,\bm 0}^{*}(\bm G=\bm 0).
\label{eq:app-optical-matrix-element}
\end{align}
After removing the common microscopic prefactor $|g_{\bm q\lambda}|^2$, the normalized optical weight is therefore $w_s=|c_{s,\bm 0}(\bm G=\bm 0)|^2.$

\subsection{Absorption spectrum and its relation to the mean-field eigenstates}

According to Fermi's golden rule, the absorption spectrum before broadening is proportional to
\begin{equation}
I(E)
\propto
\sum_s
|\mathcal M_s|^2
\delta\left[E-E_s(\bm 0)\right],
\label{eq:app-golden-rule-spectrum}
\end{equation}
where \(E\) denotes the photon energy.
Replacing each delta function by a Lorentzian with phenomenological half-width at half-maximum $\Gamma$, and removing the common light--matter coupling prefactor, gives
\begin{equation}
I(E)
=
\sum_s
w_s
\frac{\Gamma/\pi}
{\left[E-E_s(\bm 0)\right]^2+\Gamma^2}.
\label{eq:app-absorption-spectrum}
\end{equation}

Finally, because the electronic density operator has been replaced by its static expectation value, the present treatment includes only the reconstruction of the exciton spectrum by the periodic mean-field potential. Dynamical density fluctuations, exciton-polaron branches, particle--hole continua, and interaction-induced linewidths are not included; the linewidth $\Gamma$ in Eq.~\eqref{eq:app-absorption-spectrum} is introduced phenomenologically. For the calculations presented in Fig.~\ref{exciton_scan}, we use
\(W_0=296.7~{\rm meV\,nm^4}\), \(P_0=114~{\rm meV}\),
\(a_X=1.1~{\rm nm}\), \(M_X=1.2m_e\), and
\(\Gamma=1~{\rm meV}\).

\clearpage

\end{document}